\documentclass[12pt,twoside,pointlessnumbers,nofootinbib,smallheadings]{revtex4}

\usepackage[CJKbookmarks]{hyperref}
\usepackage{CJK}
\usepackage{amsmath}
\usepackage{graphics}
\usepackage{slashed}
\usepackage{graphicx}
\usepackage{indentfirst}
\usepackage{amssymb}
\usepackage{cancel}
\usepackage{leftidx}
\usepackage{simplewick}
\usepackage[page,title,titletoc,header]{appendix}

\begin{document}

\begin{CJK*}{GBK}{song}

\title{Lightness of Higgs Boson and Spontaneous CP Violation in Lee Model}

\author{Ying-nan Mao $^1$ and Shou-hua
Zhu $^{1,2,3}$  }

\affiliation{
$ ^1$ Institute of Theoretical Physics $\&$ State Key Laboratory of
Nuclear Physics and Technology, Peking University, Beijing 100871,
China \\
$ ^2$ Collaborative Innovation Center of Quantum Matter, Beijing 100871, China \\
$ ^3$ Center for High Energy Physics, Peking University,
Beijing 100871, China
}

\begin{abstract}

We proposed a mechanism in which the lightness of Higgs boson and the smallness of CP-violation are correlated
based on the Lee model, namely the spontaneous CP-violation two-Higgs-doublet-model.
In this model, the mass of the lightest Higgs boson $m_h$ as well as the quantities $K$
and $J$ are $\propto t_{\beta}s_{\xi}$
in the limit $t_{\beta}s_{\xi}\rightarrow0$ (see text for definitions of $t_\beta$ and $\xi$), namely the CP conservation limit. Here $K$
and $J$ are the measures for CP-violation effects in scalar and Yukawa sectors respectively.
It is a new way to understand why the Higgs boson discovered at the LHC is light.
We investigated the important constraints from both high
energy LHC data and numerous low energy experiments, especially the measurements of EDMs of electron and neutron as well as the quantities of B-meson and kaon.
Confronting all data, we found that this model is still viable. It should be emphasized that there is no standard-model limit for this scenario, thus it is always
testable for future experiments.
In order to pin down Lee model, it is important to discover the extra neutral and charged Higgs bosons and measure their CP properties and
the flavor-changing decays.
At the LHC with $\sqrt{s}=14\textrm{TeV}$, this scenario is favored if there is significant
suppression in the $b\bar{b}$ decay channel or any vector boson fusion (VBF), V+H production channels. On the contrary, it will be disfavored if
the signal strengths are  standard-model-like more and more. It can be easily excluded at $(3\sim5)\sigma$ level with several $\textrm{fb}^{-1}$ at future $e^+e^-$ colliders,
via the accurately measuring the Higgs boson production cross sections.

\end{abstract}

\date{\today}

\maketitle

\newpage

\section{Introduction}

How to realize the electro-weak gauge symmetry breaking and CP violation are important topics in the
standard model (SM) and beyond the SM (BSM) in particle physics. In order to induce the spontaneous gauge symmetry breaking,
the Higgs mechanism was proposed in 1964 \cite{higgs}. Meanwhile in the SM the CP violation is put by hand via the complex
Yukawa couplings among Higgs field and fermions, namely Kobayashi and Maskawa (KM) mechanism \cite{KM}.
In 1973, Kobayashi and Maskawa \cite{KM} proposed that if there are three generations of fermions,
there would be a nontrivial phase which leads to CP violation in the fermion mixing matrix (CKM matrix \cite{KM}\cite{cabbibo}).
In a word, one single scalar field plays the two-fold roles.
In the SM, only one doublet Higgs field is introduced. After spontaneous symmetry breaking, there exists one physical scalar, the Higgs boson.
It is essential to discover and measure the properties of the Higgs boson, in order to test the SM or discover the BSM.

\subsection{Status of experimental measurements on new scalar boson}

Experimentally, in July 2012, both CMS \cite{disc1} and ATLAS \cite{disc2} discovered a new boson with the mass around $125.7\textrm{GeV}$
in $\gamma\gamma$ and $ZZ^*$ final states with the luminosity of about 10$\textrm{fb}^{-1}$. At the LHC, the SM Higgs boson
can be produced through the following three processes: (1)gluon-gluon fusion (ggF); (2)vector boson fusion (VBF);
(3) associated production with a vector boson (V+H). It can also be produced associated with a pair of top
quarks due to the large $m_t$, but the cross section is suppressed by its phase space and parton distribution function (PDF) of proton.
A SM Higgs boson would mainly decay to fermion pairs ($b\bar{b},\tau^+\tau^-$, or $t\bar{t}$ if heavier than $2m_t$),
massive gauge boson pair ($W^+W^-, Z^0Z^0$), massless gauge boson pair ($gg, \gamma\gamma$), etc.
The decay properties for a $125.7\textrm{GeV}$ SM higgs boson are listed in \autoref{smp}, for the production and
decay properties, see also the reviews \cite{pro1} and \cite{pro2}.

\begin{table}[h]
\caption{Table for the SM prediction for the decay branching ratios of a 125.7GeV Higgs boson, the numbers are from \cite{pro2}.}\label{smp}
\begin{tabular}{|c|c|c|}
\hline
Decay Channel & Branching Ratio ($\%$)& Relative Uncertainty ($\%$)\\
\hline
$b\bar{b}$ & $56.6$ & $\pm3.3$\\
\hline
$c\bar{c}$ & $2.85$ & $\pm12.2$\\
\hline
$\tau^+\tau^-$ & $6.21$ & $\pm5.6$\\
\hline
$gg$ & $8.51$ & $^{+10.2}_{-9.9}$\\
\hline
$WW^*$ & $22.6$ & $^{+4.2}_{-4.1}$\\
\hline
$ZZ^*$ & $2.81$ & $^{+4.2}_{-4.1}$\\
\hline
$\gamma\gamma$ & $0.228$ & $\pm4.9$\\
\hline
$Z\gamma$ & $0.16$ & $^{+8.9}_{-8.8}$\\
\hline
Total Width & $4.17\textrm{MeV}$ & $\pm3.9$\\
\hline
\end{tabular}
\end{table}

The updated searches by CMS \cite{updc1,updc2,updc3,updc4} and ATLAS \cite{upda1,upda2,upda3,upda4}
with the luminosity of about 25$\textrm{fb}^{-1}$ till the end of 2012 \footnote{Some new analysis updated
in 2014 are used as well which modify the old results a little bit.}
gave the significance $s$ and signal strengths $\mu$ (defined as the ratios between observed
$\sigma\cdot Br$ and the corresponding SM prediction) for some channels. Because the measurements will be utilized to constrain the new model in this paper, we list the results in
\autoref{sgs1} for CMS and \autoref{sgs2} for ATLAS.\footnote{The VBF events are usually easy to tag with
two jets which have large invariant mass, while sometimes it is difficult to tag a gluon fusion event.}
\begin{table}
\caption{Signal strengths for some production and decay channels of the new boson at CMS (with combined significance over $3\sigma$).}\label{sgs1}
\begin{tabular}{|c|c|c|c|c|}
\hline
 & $\mu$(VBF/V+H) & $\mu$(ggF) & $\mu$(combined) & significance \\
\hline
$\gamma\gamma$ & $1.58^{+0.77}_{-0.68}$ & $1.12^{+0.37}_{-0.32}$ & $1.14^{+0.26}_{-0.23}$ & $5.7\sigma$ \\
\hline
$ZZ^*$ & $1.7^{+2.2}_{-2.1}$ & $0.80^{+0.46}_{-0.36}$ & $0.93^{+0.29}_{-0.25}$ & $6.8\sigma$ \\
\hline
$WW^*$ & $0.60^{+0.57}_{-0.46}$ & $0.74^{+0.22}_{-0.20}$ & $0.72^{+0.20}_{-0.18}$ & $4.3\sigma$ \\
\hline
$\tau^+\tau^-$ & $0.94\pm0.41$ & & $0.78\pm0.27$ & $3.2\sigma$ \\
\hline
\end{tabular}
\end{table}
\begin{table}
\caption{Signal strengths for some production and decay channels of the new boson at ATLAS (with combined significance over $3\sigma$).}\label{sgs2}
\begin{tabular}{|c|c|c|c|c|}
\hline
 & $\mu$(VBF/V+H) & $\mu$(ggF) & $\mu$(combined) & significance \\
\hline
$\gamma\gamma$ & $0.8\pm0.7$ & $1.32\pm0.38$ & $1.17\pm0.27$ & $5.2\sigma$ \\
\hline
$ZZ^*$ & $0.26^{+1.64}_{-0.94}$& $1.66^{+0.51}_{-0.44}$ & $1.43^{+0.40}_{-0.33}$ & $8.1\sigma$ \\
\hline
$WW^*$ & $1.28^{+0.53}_{-0.45}$ & $1.01^{+0.28}_{-0.26}$ & $1.09^{+0.23}_{-0.20}$ & $6.1\sigma$ \\
\hline
$\tau^+\tau^-$ & $1.24^{+0.58}_{-0.54}$ & $1.93^{+1.45}_{-1.15}$ & $1.42^{+0.44}_{-0.38}$ & $4.5\sigma$ \\
\hline
\end{tabular}
\end{table}
The new boson has a combined mass $125.7$GeV and it is also favored as a $0^+$ particle in spin and parity
by the data \cite{updc2,CPP,CPP2} if we assume that there is no CP violation induced by this boson.

The experimental measurements of the new particle are in agreement with the SM predictions within the current accuracy. In the SM  the electro-weak fitting results \cite{Fit} also favors a light one.
It allows a SM Higgs boson lighter than 145GeV at $95\%$C.L. inferred from the
oblique parameters \cite{obl} with fixed $U=0$. However there are still spacious room for the BSM. For example, if we assume that
the new particle is
a CP-mixing state, the general effective interaction for $hZZ$ can be written as \cite{updc2,CPmix,CPmix2}
\begin{equation}
\label{hZZ}
\mathcal{L}_{hZZ}=\frac{h}{v}\left(a_1m^2_ZZ^{\mu}Z_{\mu}
+\frac{1}{2}a_2Z_{\mu\nu}Z^{\mu\nu}+\frac{1}{2}a_3Z_{\mu\nu}\tilde{Z}^{\mu\nu}\right)
\end{equation}
with $\tilde{Z}_{\mu\nu}=(1/2)\epsilon_{\mu\nu\alpha\beta}Z^{\alpha\beta}$. Define
\begin{equation}
\label{fa3}
f_{a3}=\frac{(a_3/a_1)^2}{(a_3/a_1)^2+\sigma_1/\sigma_3}
\end{equation}
where $\sigma_{1(3)}$ are the partial width for pure CP even (odd) state with $a_{1(3)}=1$. Direct search by CMS gives
$f_{a3}<0.47$ at $95\%$ C.L. which leads to $|a_3/a_1|<2.4$ \cite{updc2}. In a renormalized theory, $a_2$ and $a_3$ which
are loop induced are expected to behave as $a_{2,3}/a_1\ll\mathcal{O}(1)$, so they are still not constrained by current LHC data.

\subsection{The issue of lightness of new scalar boson in the SM and BSM}

BSM is well motivated because SM can't account for the matter-dominant universe and provide the suitable dark matter candidate. However BSM scale is usually pushed to a much higher value than that of weak interaction, given
the great success of the SM. In such circumstance, the $125.7$ GeV scalar boson is unnatural. In other word, the lightness of the new scalar must link to certain mechanism.
The issue of the lightness of the new scalar differs in the SM and the BSM.
In the SM, we cannot predict the mass
of Higgs boson, and the Higgs boson with the mass $125.7$ GeV simply implies that the interactions are in the weak regime. For example the Higgs boson self-coupling
\begin{equation}
\lambda=\frac{m^2_h}{2v^2}=0.13\ll1.
\end{equation}
Compared with the strong interactions at low energy, the mass of $\sigma$
particle (or we call it $f_0(600)$ which plays a similar role as the Higgs boson) $m_{\sigma}\gg f_{\pi}
(\approx93\textrm{MeV})$ appears at a typical scale $\Lambda\sim4\pi f_{\pi}\sim\mathcal{O}(1)\textrm{GeV}$.
Thus we can argue that the new boson with mass 125.7GeV is rather light compared with the strong interaction. As a side remark,
the pion mass $m_{\pi}\sim\mathcal{O}(f_{\pi})$ is light compared with $\sigma$
due to the approximate chiral symmetry. This has motivated the idea that new scalar boson may be the pseudo-Nambu-Goldstone boson for certain unknown symmetry breaking.

Theoretically, in some BSM models there exists a light scalar naturally. For example, (1) in the minimal
super-symmetric model, the lightest Higgs boson should be lighter than 140GeV including higher-order corrections \cite{susy}
(at tree level it should be lighter than the mass of $Z_0$ boson); (2) in the little higgs model, a Higgs boson which is
treated as a pseudo-Nambu-Goldstone boson must be light due to classical global symmetry and it acquires mass
through quantum effects only \cite{LH}; (3)similarly, anomalous in scale invariance
can also generate a light Higgs boson as well \cite{scale}; (4) the lightness of Higgs boson can intimately connect with the spontaneous CP violation \cite{zhu}.
While the first three approaches base on the conjectured symmetry, the last one utilizes the observed approximate CP symmetry.
Historically Lee proposed the spontaneous CP violation in 1973 \cite{Lee} as an alternative way to induce CP violation. For the fourth approach, Lee's idea is extended to account for the lightness of
the observed Higgs boson.

\subsection{The lightness of new scalar boson and spontaneous CP violation}

CP violation was first discovered in neutral K-meson in 1964 \cite{CP}. Experimentally people have already measured
several kinds of CP violated effects in neutral K- and B-meson, and charged B meson systems \cite{PDG}. These CP violation can
be successfully accounted for by
the CKM matrix, which is usually parameterized as the Wolfenstein formalism \cite{wolf}
\begin{equation}
V_{\textrm{CKM}}=\left(\begin{array}{ccc}1-\lambda^2/2&\lambda&A\lambda^3(\rho-i\eta)\\
-\lambda&1-\lambda^2/2&A\lambda^2\\A\lambda^3(1-\rho-i\eta)&-A\lambda^2&1\end{array}\right)+\mathcal{O}(\lambda^4)
\end{equation}
The Jarlskog invariant \cite{PDG}\cite{Jarskog}
\begin{equation}
J=A^2\lambda^6\eta=(3.06^{+0.21}_{-0.20})\times10^{-5}
\end{equation}
measures the CP violation in flavor sector. The smallness of $J$ means the smallness of CP-violation
in the real world in SM. Another possible explicit CP-violation comes from the $\theta$ term
\begin{equation}
\mathcal{L}_{\textrm{str}}=\frac{\theta\alpha_s}{8\pi}G_{\mu\nu}\tilde{G}^{\mu\nu}
\end{equation}
in the QCD lagrangian \cite{str}\cite{str2}. The parameter $\theta$ is strongly constrained by the neutron
electric dipole moment (EDM) measurement \cite{dncal}\cite{dn}, namely $|\theta|\lesssim10^{-10}$.
Why $\theta$ is extremely small is known as the strong CP problem. It is often interesting, necessary and useful to
search for other sources of CP violation beyond the KM-mechanism.
As a common reason, for example, CP-violation is one of the conditions to produce
the matter-antimatter asymmetry in the universe today \cite{sakh}, but SM itself cannot provide the first order
electro-weak phase transition and large enough CP-violation to get the right asymmetry between matter
and anti-matter \cite{PDG,asy,bar,cohen}.

In 1973, Lee proposed a 2HDM (Lee model) \cite{Lee} in which all parameters in the scalar potential are real but it is possible
to leave a nontrivial phase $\xi$ between the vacuum expectation values (VEV) of the two Higgs doublets. CP can be spontaneously
broken in this model. Chen et. al. \cite{slc} proposed the possibility that the complex vacuum could lead to a correct
CKM matrix, which means that we can set all Yukawa couplings real thus the complex vacuum would become the only source of
CP-violation. It is also a possible way to solve strong CP problem, for example, in spontaneous CP-violation scenarios,
$\theta$ arises only from the determinant of quark mass matrix. Assuming $\theta\equiv0$ at tree level, the loop corrections
can generate naturally small $\theta$ \cite{Gino}\cite{Barr}, the so-called ``calculable $\theta$" \cite{str}.
Without imposing symmetry \cite{GW}, the Yukawa couplings are arbitrary which will generate the flavor changing neutral currents (FCNC)
at tree level. FCNC is severely constrained by experiments. Cheng and Sher proposed  an ansatz \cite{CS} that the flavor changing
couplings should be $\propto\sqrt{m_im_j}$ for two fermions with mass $m_i$ and $m_j$.
One of the authors of this paper had proposed a mechanism \cite{zhu} to understand the lightness of Higgs boson in the $\xi\rightarrow0$
limit. In this paper, we will explore the relation between the smallness of CP-violation and the lightness of Higgs
boson in a similar way in Lee model further. Specifically we will study the full phenomenology of the Lee model and to see
whether this model is still viable confronting LHC data and numerous low energy measurements.

We should mention that there are also cosmological implication for Lee model. In this model, CP is a
spontaneously broken discrete symmetry thus it may face the domain wall problem \cite{Dom} during the electro-weak phase transition. It is argued that if there is
a small initial bias thus one of the vacuum states is favored, the domain walls would disappear soon \cite{Dom}\cite{domsol1},
for example, if there is small explicit CP-violation \cite{expli}. In the soft CP breaking
model, the electro-weak baryogenesis effects is estimated by Cohen et. al. \cite{cohen} at early time, and was estimated
again by Shu and Zhang \cite{shu} after including LHC data. They found that the observed matter-anti-matter asymmetry can be
explained. It is also discussed numerically that an inflation during
the symmetry breaking would forbid the domain wall production \cite{domsol2}.

This paper is organized as following. Section II presents the Lee model and the scenario that lightness of Higgs boson and smallness of CP violation are
correlated. Section III and IV contain the constraints on Lee model from high energy and low energy data respectively. Section V studies the
perspectives for Lee model for future experiments. The last section collects our conclusions and discussions.

\section{The Lee model: mass spectrum and couplings}

We begin with the description of Lee model \cite{Lee} assuming that in the whole lagrangian there are no explicit CP-violation terms,
which means all the CP-violation effects come from a complex vacuum \footnote{For a review on two Higgs doublet models (2HDM), the interested reader can read Ref. \cite{2HDM}}. For the Lee model, the interactions of scalar fields read \cite{Lee}
\begin{equation}
\mathcal{L}=(D_{\mu}\phi_1)^{\dag}(D^{\mu}\phi_1)+(D_{\mu}\phi_2)^{\dag}(D^{\mu}\phi_2)-V(\phi_1,\phi_2).
\end{equation}
Here
\begin{equation}
\label{vev}
\phi_1=\left(\begin{array}{c}\phi_1^+\\ \frac{v_1+R_1+iI_1}{\sqrt{2}}\end{array}\right),\quad\quad
\phi_2=\left(\begin{array}{c}\phi_2^+\\ \frac{v_2e^{i\xi}+R_2+iI_2}{\sqrt{2}}\end{array}\right)
\end{equation}
are the two higgs doublets. We can get the masses of gauge bosons
\begin{equation}
m_W=\frac{g\sqrt{v^2_1+v^2_2}}{2},\quad\quad m_Z=\frac{\sqrt{(g^2+g'^2)(v^2_1+v^2_2)}}{2}
\end{equation}
by setting $v=\sqrt{v_1^2+v^2_2}=246\textrm{GeV}$. Defining $R(I)_{ij}$ as
the real(imaginary) part of $\phi_i^{\dag}\phi_j$, we can write a general potential as
\begin{eqnarray}
V&=&V_2+V_4\nonumber\\
&=&\mu_{1}^2R_{11}+\mu_{2}^2R_{22}\nonumber\\
&&+\lambda_1R_{11}^2+\lambda_2R_{11}R_{12}+\lambda_3R_{11}R_{22}\nonumber\\
\label{pot}&&+\lambda_4R_{12}^2+\lambda_5R_{12}R_{22}+\lambda_6R_{22}^2+\lambda_7I_{12}^2;
\end{eqnarray}
in which we can always perform a rotation between $\phi_1$ and $\phi_2$ to keep the coefficient of $R_{12}$ term zero in $V_2$.
We can also write the general Yukawa couplings as
\begin{equation}
\mathcal{L}_y=-\bar{Q}_{Li}(Y_{1d}\phi_1+Y_{2d}\phi_2)_{ij}D_{Rj}
-\bar{Q}_{Li}(Y_{1u}\tilde{\phi}_1+Y_{2u}\tilde{\phi}_2)_{ij}U_{Rj},
\end{equation}
in which $\tilde{\phi}_i=\textrm{i}\sigma_2\phi_i^*$ and all Yukawa couplings are real.

Minimizing the higgs potential, and for some parameter choices, we can get a nonzero phase difference $\xi$
between two higgs VEVs, which would induce spontaneous CP violation. We can always perform a gauge transformation
to get at least one of the VEVs real like in (\ref{vev}). When $v_1,v_2,\xi\neq0$, we can express
\begin{eqnarray}
\mu_{1}^2&=&-\lambda_1v_1^2-\frac{\lambda_3+\lambda_7}{2}v_2^2-\frac{\lambda_2}{2}v_1v_2\cos\xi;\\
\mu_{2}^2&=&-\frac{\lambda_3+\lambda_7}{2}v_1^2-\lambda_6v_2^2-\frac{\lambda_5}{2}v_1v_2\cos\xi.
\end{eqnarray}
$\tan\beta$ is identified as $v_2/v_1$ as usual. We also have an equation about $\xi$
\begin{equation}
\label{cpv}\frac{\lambda_2}{2}v_1^2+\frac{\lambda_5}{2}v_2^2+(\lambda_4-\lambda_7)v_1v_2\cos\xi=0,
\end{equation}
which requires $\lambda_2v_1^2+\lambda_5v^2_2<2|\lambda_4-\lambda_7|v_1v_2$. Of course, the couplings
$\lambda_i$ must keep the vacuum stable, for the conditions see \autoref{vs} for details.

All the CP-violation effects in the real world are small (see the data in \cite{PDG}) corresponding to the
smallness of the off-diagonal elements in the CKM-matrix which leads to the smallness of the Jarlskog
invariant. As a limit, when $t_{\beta}\equiv\tan\beta\rightarrow0$,\footnote{We write
$s_{\alpha}\equiv\sin\alpha, c_{\alpha}\equiv\cos\alpha, t_{\alpha}\equiv\tan\alpha$ for short in this paper.}
or we may write $t_{\beta}s_{\xi}\rightarrow0$ instead since $|s_{\xi}|<1$ always holds,
there would be no CP-violation in the scalar sector. The CKM-matrix would be real thus there would be no
CP-violation in flavor sector as well. In this paper we will consider the small $t_{\beta}$ limit,
in which all CP-violation effects tends to zero as $t_{\beta} \rightarrow0$. We treat the whole world as an
expansion around the point without CP-violation.

The two higgs doublets contain 8 degrees of freedom, 3 of which should be eaten by massive gauge bosons as
Goldstones. So there are 5 physical scalars left, 2 of which are charged and 3 of which are neutral. If CP is
a good symmetry, there will be 2 CP even and 1 CP odd scalars among the 3 neutral ones. However, when CP is
spontaneously breaking, the CP eigenstates will mix with each other thus the neutral scalars have no certain
CP charge. We have the Goldstones as
\begin{eqnarray}
G^{\pm}&=&c_{\beta}\phi_1^{\pm}+e^{\mp i\xi}\phi_2^{\pm};\\
G^0&=&c_{\beta}I_1+s_{\beta}c_{\xi}I_2-s_{\beta}s_{\xi}R_2.
\end{eqnarray}

The charged Higgs boson is the orthogonal state of the charged Goldstone as
\begin{equation}
H^{\pm}=-e^{\pm i\xi}s_{\beta}\phi_1^{\pm}+c_{\beta}\phi_2^{\pm}
\end{equation}
and its mass square should be
\begin{equation}
m^2_{H^{\pm}}=-\frac{\lambda_7v^2}{2}.
\end{equation}
While for the neutral part, we write the mass square matrix as $\tilde{m}v^2/2$
in the basis $(-s_{\beta}I_1+c_{\beta}c_{\xi}I_2-c_{\beta}s_{\xi}R_2,R_1,s_{\xi}I_2+c_{\xi}R_2)^T$.
The symmetric matrix $\tilde{m}$ is
\begin{equation}
\label{higmass}
\left(\begin{array}{ccc}(\lambda_4-\lambda_7)s^2_{\xi}&
\begin{array}{c}-((\lambda_4-\lambda_7)s_{\beta}c_{\xi}\\+\lambda_2c_{\beta})s_{\xi}\end{array}&
\begin{array}{c}-((\lambda_4-\lambda_7)c_{\beta}c_{\xi}\\+\lambda_5s_{\beta})s_{\xi}\end{array}\\
&&\\
&\begin{array}{c}4\lambda_1c^2_{\beta}+2\lambda_2c_{\beta}s_{\beta}c_{\xi}\\+(\lambda_4-\lambda_7)s^2_{\beta}c^2_{\xi}\end{array}&
\begin{array}{c}(2(\lambda_3+\lambda_7)+(\lambda_4-\lambda_7)c^2_{\xi})s_{\beta}c_{\beta}\\
+\lambda_2c^2_{\beta}c_{\xi}+\lambda_5s^2_{\beta}c_{\xi}\end{array}\\
&&\\
&&\begin{array}{c}(\lambda_4-\lambda_7)c^2_{\beta}c^2_{\xi}\\+2\lambda_5s_{\beta}c_{\beta}c_{\xi}+4\lambda_6s^2_{\beta}\end{array}\end{array}\right)
\end{equation}
and its three eigenvalues correspond to the masses of three neutral bosons.

We expand the matrix $\tilde{m}$ in series of $t_{\beta}(s_{\xi})$ as
\begin{equation}
\tilde{m}=\tilde{m}_0+(t_{\beta}s_{\xi})\tilde{m}_1+(t_{\beta}s_{\xi})^2\tilde{m}_2+\cdots
\end{equation}
to get the approximate analytical behavior of its eigenvalues and eigenstates. Certainly we have
\begin{equation}
\mathop{\lim}_{t_{\beta}s_{\xi}\rightarrow0}\det(\tilde{m})=\det(\tilde{m}_0)=0
\end{equation}
which means a zero eigenvalue of $\tilde{m}_0$ thus there must be a light neutral scalar when $t_{\beta}s_{\xi}$ is small.
To the leading order of $t_{\beta}s_{\xi}$, for the lightest scalar $h$, we have
\begin{eqnarray}
\label{mh}
m^2_{h}&=&\frac{v^2t^2_{\beta}s^2_{\xi}}{2}\left((\tilde{m}_2)_{11}-\frac{(\tilde{m}_1)^2_{12}}{(\tilde{m}_0)_{22}}
-\frac{(\tilde{m}_1)^2_{13}}{(\tilde{m}_0)_{33}}\right)\nonumber\\
&=&\frac{v^2t^2_{\beta}s^2_{\xi}}{2}\bigg[4\lambda_6+2\lambda_5(\lambda_3+\lambda_7)s_{2\theta}
\left(\frac{1}{(\tilde{m}_0)_{22}}-\frac{1}{(\tilde{m}_0)_{33}}\right)\nonumber\\
&&-4(\lambda_3+\lambda_7)^2\left(\frac{c^2_{\theta}}{(\tilde{m}_0)_{22}}+
\frac{s^2_{\theta}}{(\tilde{m}_0)_{33}}\right)-\lambda^2_5\bigg(\frac{s^2_{\theta}}{(\tilde{m}_0)_{22}}
+\frac{c^2_{\theta}}{(\tilde{m}_0)_{33}}\bigg)\bigg];\\
h&=&I_2-t_{\beta}s_{\xi}\left(\frac{(\tilde{m}_1)_{12}}{(\tilde{m}_0)_{22}}(c_{\theta}R_1+s_{\theta}R_2)
+\frac{(\tilde{m}_1)_{13}}{(\tilde{m}_0)_{33}}(c_{\theta}R_2-s_{\theta}R_1)+\frac{I_1}{t_{\xi}}\right)\nonumber\\
&=&I_2-t_{\beta}s_{\xi}\bigg[\bigg(2(\lambda_3+\lambda_7)\bigg(\frac{c^2_{\theta}}{(\tilde{m}_0)_{22}}+
\frac{s^2_{\theta}}{(\tilde{m}_0)_{33}}\bigg)+\frac{\lambda_5s_{2\theta}}{2}\bigg(\frac{1}{(\tilde{m}_0)_{22}}
-\frac{1}{(\tilde{m}_0)_{33}}\bigg)\bigg)R_1\nonumber\\
&&+\bigg((\lambda_3+\lambda_7)s_{2\theta}\bigg(\frac{1}{(\tilde{m}_0)_{22}}
-\frac{1}{(\tilde{m}_0)_{33}}\bigg)+\lambda_5\bigg(\frac{s^2_{\theta}}{(\tilde{m}_0)_{22}}+
\frac{c^2_{\theta}}{(\tilde{m}_0)_{33}}\bigg)\bigg)R_2+\frac{I_1}{t_{\xi}}\bigg].
\end{eqnarray}
While for the two heavier neutral Higgs, we have
\begin{equation}
m^2_{2(3)}=\frac{v^2}{2}\left((\tilde{m}_0)_{22(33)}+\mathcal{O}(t_{\beta}s_{\xi}) \right),
\end{equation}
in which $(\tilde{m}_0)_{22(33)}$ are the other two eigenvalues of $\tilde{m}_0$ and
\begin{equation}
(\tilde{m}_0)_{22(33)}=\frac{4\lambda_1+\lambda_4-\lambda_7}{2}\pm
\left(\frac{4\lambda_1-(\lambda_4-\lambda_7)}{2}c_{2\theta}+\lambda_2s_{2\theta}\right)
\end{equation}
where $\theta=(1/2)\arctan(2\lambda_2/(4\lambda_1-\lambda_4+\lambda_7))$. The physical states are
\begin{equation}
\left(\begin{array}{c}h_2\\h_3\end{array}\right)=\left(\begin{array}{cc}c_{\theta}&s_{\theta}\\
-s_{\theta}&c_{\theta}\end{array}\right)\left(\begin{array}{c}R_1\\R_2\end{array}\right)
+\mathcal{O}(t_{\beta}s_{\xi}).
\end{equation}
For all the details about scalar spectra and its small $t_{\beta}s_{\xi}$ expansion series, the interested reader can see \autoref{mass}.

From the Yukawa couplings we will get the mass matrixes for fermions as
\begin{eqnarray}
\label{m1}(M_U)_{ij}=\frac{v}{\sqrt{2}}(Y_{1u}c_{\beta}+Y_{2u}s_{\beta}e^{-i\xi})_{ij},\\
\label{m2}(M_D)_{ij}=\frac{v}{\sqrt{2}}(Y_{1d}c_{\beta}+Y_{2d}s_{\beta}e^{i\xi})_{ij}.
\end{eqnarray}
We can always perform the diagonalization for $M_{U(D)}$ with matrixes $U(D)_L$ and $U(D)_R$ as
\begin{equation}
U_LM_UU_R^{\dag}=\left(\begin{array}{ccc}m_u&&\\&m_c&\\&&m_t\end{array}\right),\quad
D_LM_DD_R^{\dag}=\left(\begin{array}{ccc}m_d&&\\&m_s&\\&&m_b\end{array}\right).
\end{equation}
And $V_{\textrm{CKM}}=U_LD_L^{\dag}$ is the CKM matrix.

In this scenario, the couplings for the discovered light Higgs boson should be modified from SM by a factor as
\begin{eqnarray}
\mathcal{L}_{h,\textrm{eff}}&=&c_V\left(\frac{2m^2_W}{v}W^+_{\mu}W^{\mu-}+\frac{m^2_Z}{v}Z_{\mu}Z^{\mu}\right)h-c_{\pm}vH^+H^-h\nonumber\\
&&-\mathop{\sum}_{i}\left(c_{Ui}\bar{U}_{Li}U_{Ri}+c_{Di}\bar{D}_{Li}D_{Ri}+\textrm{h.c.}\right)h,
\end{eqnarray}
where the factors $c_{\pm}$ and $c_V$ must be real, but $c_{Ui}$ and $c_{Di}$ may be complex.
According to (\ref{c1})-(\ref{c4}) in \autoref{Fr}, to the leading order of $t_{\beta}s_{\xi}$, we straightforwardly have
\begin{eqnarray}
\label{c1*}
c_V&=&t_{\beta}s_{\xi}(1-\eta_1);\\
\frac{m_{D_i}}{v}c_{Di}&=&\frac{i(Y_{2d}')_{ii}}{\sqrt{2}}-\frac{t_{\beta}s_{\xi}}{\sqrt{2}}(\eta_1(Y_{1d}')_{ii}+\eta_2(Y_{2d}')_{ii});\\
\label{c4*}
\frac{m_{U_i}}{v}c_{Ui}&=&-\frac{i(Y_{2u}')_{ii}}{\sqrt{2}}-\frac{t_{\beta}s_{\xi}}{\sqrt{2}}(\eta_1(Y_{1u}')_{ii}+\eta_2(Y_{2u}')_{ii});
\end{eqnarray}
and the coupling including charged higgs should be
\begin{equation}
\label{c5*}
c_{\pm}=t_{\beta}s_{\xi}\left((2\lambda_6-\lambda_7)-\lambda_3\eta_1-\frac{\lambda_5\eta_2}{2}\right);
\end{equation}
where
\begin{eqnarray}
\eta_1&=&c_{\theta}\frac{(\tilde{m}_1)_{12}}{(\tilde{m}_0)_{22}}-s_{\theta}\frac{(\tilde{m}_1)_{13}}{(\tilde{m}_0)_{33}}\nonumber\\
&=&2(\lambda_3+\lambda_7)\left(\frac{c^2_{\theta}}{(\tilde{m}_0)_{22}}+\frac{s^2_{\theta}}{(\tilde{m}_0)_{33}}\right)
+\frac{\lambda_5s_{2\theta}}{2}\left(\frac{1}{(\tilde{m}_0)_{22}}-\frac{1}{(\tilde{m}_0)_{33}}\right);\\
\eta_2&=&s_{\theta}\frac{(\tilde{m}_1)_{12}}{(\tilde{m}_0)_{22}}+c_{\theta}\frac{(\tilde{m}_1)_{13}}{(\tilde{m}_0)_{33}}\nonumber\\
&=&(\lambda_3+\lambda_7)s_{2\theta}\left(\frac{1}{(\tilde{m}_0)_{22}}-\frac{1}{(\tilde{m}_0)_{33}}\right)
+\lambda_5\left(\frac{c^2_{\theta}}{(\tilde{m}_0)_{22}}+\frac{s^2_{\theta}}{(\tilde{m}_0)_{33}}\right).
\end{eqnarray}

We choose all the nine free parameters as nine observables in Higgs sector: masses of four scalars $m_h,m_2,m_3$ and $m_{H^{\pm}}$;
vacuum expected values $v_1,v_2,\xi$ and two mixing angles for neutral bosons. The mixing angles are represented as $c_1$ and $c_2$.
\begin{equation}
\mathcal{L}_{h_iVV}=c_ih_i\left(\frac{2m^2_W}{v}W^+_{\mu}W^{\mu-}+\frac{m^2_Z}{v}Z_{\mu}Z^{\mu}\right).
\end{equation}
The $c_i$ just stands for the $h_iVV$ vertex strength ratio comparing with that in SM\footnote{There is a sum rule
$c_1^2+c_2^2+c_3^2=1$ due to spontaneous electro-weak symmetry broken, thus only two of the $c_i$ are free,
and $c_1$ here is just the $c_V$ in (\ref{c1*}).}. In the scalar sector, for non-degenerate neutral Higgs
bosons, a quantity $K=c_1c_2c_3$ measures the CP violation effects \cite{2HDM}\cite{K}\footnote{If at least two of the neutral
bosons have degenerate mass, we can always perform a rotation among the neutral fields to keep $K=0$.},
while in Yukawa sector, the Jarlskog invariant $J$ \cite{Jarskog} measures that. In this scenario,
to the leading order of $t_{\beta}s_{\xi}$, we have
\begin{equation}
\label{K}K=c_1c_2c_3=-s_{\theta}c_{\theta}(1+\eta_1)t_{\beta}s_{\xi}\propto t_{\beta}s_{\xi}
\end{equation}

In order to calculate J, we define matrix $\hat{C}$ as
\begin{equation}
\hat{C}\equiv \left[M_UM_U^{\dag},M_DM_D^{\dag}\right].
\end{equation}
We can always choose a basis in which the diagonal elements of $\hat{C}$ are zero. Thus
\begin{equation}
\hat{C}=\left(\begin{array}{ccc}0&C_3&-C_2\\-C_3&0&C_1\\C_2&-C_1&0\end{array}\right)+
i\left(\begin{array}{ccc}0&C_3^*&C_2^*\\C_3^*&0&C_1^*\\C_2^*&C_1^*&0\end{array}\right)=
\left(\textrm{Re}\hat{C}+i\textrm{Im}\hat{C}\right)
\end{equation}
in which using equations (\ref{m1}) and (\ref{m2}), to the leading order of $t_{\beta}s_{\xi}$, we have
\begin{eqnarray}
\textrm{Re}\hat{C}&=&\frac{v^4c^4_{\beta}}{4}\left[Y_{u1}Y_{u1}^{\dag},Y_{d1}Y_{d1}^{\dag}\right];\\
\textrm{Im}\hat{C}&=&\frac{v^4c^4_{\beta}}{4}\left(\left[Y_{u1}Y_{u2}^{\dag}-Y_{u2}Y_{u1}^{\dag},Y_{d1}Y_{d1}^{\dag}\right]\right.\nonumber\\
&&\quad+\left.\left[Y_{u1}Y_{u1}^{\dag},Y_{d2}Y_{d1}^{\dag}-Y_{d1}Y_{d2}^{\dag}\right]\right)t_{\beta}s_{\xi}\propto t_{\beta}s_{\xi}.
\end{eqnarray}
To the leading order of $t_{\beta}s_{\xi}$, the determinant
\begin{eqnarray}
\det\left(i\hat{C}\right)&\equiv&2J\mathop{\prod}_{i<j}\left(m^2_{U_i}-m^2_{U_j}\right)\mathop{\prod}_{i<j}\left(m^2_{D_i}-m^2_{D_j}\right)\nonumber\\
&=&C_1C_2C_3\left(\frac{C_1^*}{C_1}+\frac{C_2^*}{C_2}+\frac{C_3^*}{C_3}\right),
\end{eqnarray}
where $C_i^*\propto t_{\beta}s_{\xi}$, thus
\begin{equation}
\label{J}J=\frac{\prod C_i\sum(C_i^*/C_i)}{\prod\left(m^2_{U_i}-m^2_{U_j}\right)\prod\left(m^2_{D_i}-m^2_{D_j}\right)}\propto t_{\beta}s_{\xi}.
\end{equation}
According to the equations (\ref{K}), (\ref{J}), and (\ref{mh}), we propose that the lightness of the Higgs
boson and the smallness of CP-violation effects could be correlated through small $t_{\beta}s_{\xi}$ since both
the Higgs mass $m_h$ and the quantities $K$ and $J$ to measure CP-violation effects are proportional to
$t_{\beta}s_{\xi}$ at the small $t_{\beta}s_{\xi}$ limit.

In the following two sections, we will study whether the Lee model is still viable confronting the current numerous high and low energy measurements.  From Eq. (\ref{c1*}), it is quite clear that couplings of discovered scalar
boson differ from those in the SM, namely Lee model does not have SM limit. Provided that LHC obtained only a small portion of its designed integrated luminosity, there would be spacious room for Lee model. In the long run, LHC and future facilities have the great potential to discover/exclude Lee model. We will discuss this part in \autoref{disco}.

\section{Constraints from High Energy Phenomena}

In this model there are two more neutral bosons and one more charged boson pair comparing with SM, these
degree of freedoms may affect on the physics at electro-weak scale, and they could also be constrained
by direct searches at the LHC. For the discovered boson, SM predicts the decay branching ratios for a Higgs
boson with mass 125.7GeV in \autoref{smp}. However in Lee model, the modified couplings will change the
total width and branching ratios due to equations (\ref{c1*})-(\ref{c5*}), together with the
production cross sections modified by (\ref{c4*}) for gluon fusion and (\ref{c1*}) for vector boson
fusion and the associated production with vector bosons. Of course, this model may also affect top
physics because the couplings between Higgs boson and top quark are not suppressed and it may also change the
flavor changing couplings especially for top quark. Thus it is necessary to
discuss the constraints to this model from high energy phenomena.

\subsection{Constraints on heavy neutral bosons}

A heavy Higgs boson may decay to $W^+W^-$, $2Z_0$, $2h$, $t\bar{t}$ (for neutral bosons heavier than
$2m_t\approx346\textrm{GeV}$), or $H^+H^-$(for light charged Higgs and a neutral boson heavier than $2m_{H^{\pm}}$).
Based on the searches for the SM Higgs boson using diboson final state \cite{HH}, masses and couplings of
the other two heavier neutral Higgs bosons should be constrained by the data. For a neutral Higgs boson
heavier than 350GeV, the $t\bar{t}$ resonance search \cite{tt} may also give some constraints.

In this scenario, the totol width of a heavy boson can be expressed as
\begin{equation}
\Gamma_i=\Gamma_{i,VV}+\Gamma_{i,\pm}+\Gamma_{i,2h}+\Gamma_{i,t\bar{t}},
\end{equation}
where $\Gamma_{i,VV}$, $\Gamma_{i,\pm}$, $\Gamma_{i,2h}$, $\Gamma_{i,t\bar{t}}$ correspond to massive gauge boson pairs, charged Higgs pair, neutral Higgs pair and top quark pair final states respectively.
The partial decay width for a heavy neutral Higgs with mass $m_i$ are
\begin{eqnarray}
\label{gamma1}
\frac{\Gamma_{i,VV}}{m_i}&=&\frac{3c_i^2}{16\pi}\left(\frac{m_i}{v}\right)^2\quad\quad(m_i\gg m_V);\\
\frac{\Gamma_{i,t\bar{t}}}{m_i}&=&\frac{3|c_{t,i}|^2}{8\pi}\left(\frac{m_t}{v}\right)^2\sqrt{1-\frac{4m_t^2}{m_i^2}}
\left(1-\frac{4m_t^2\cos^2(\arg(c_{t,i}))}{m_i^2}\right);\\
\frac{\Gamma_{i,\pm}}{m_i}&=&\frac{\lambda^2_{i,\pm}}{16\pi}\left(\frac{v}{M}\right)^2\sqrt{1-\frac{4m_{H^{\pm}}^2}{m_i^2}};\\
\label{gamma4}
\frac{\Gamma_{i,hh}}{m_i}&=&\frac{\lambda^2_{i,hh}}{32\pi}\left(\frac{v}{M}\right)^2\sqrt{1-\frac{4m_{h}^2}{m_i^2}}.
\end{eqnarray}
in unit of its mass. Here we have the vertices
\begin{equation}
\lambda_{i,\pm}=\frac{1}{v}\frac{\partial^3V}{\partial h_i\partial H^+\partial H^-},\quad\quad
\lambda_{i,hh}=\frac{1}{v}\frac{\partial^3V}{\partial h_i\partial h^2}.
\end{equation}
The couplings $\lambda_{i,\pm},\lambda_{i,hh}\sim\mathcal{O}(1)$.

The signal strength is defined as
\begin{equation}
\mu=\frac{\sigma}{\sigma_{\textrm{SM}}}\cdot\frac{\Gamma_{i,VV}}{\Gamma_i}\cdot\frac{1}{Br_{\textrm{SM}(VV)}}
\end{equation}
for a production channel. The $\sigma/\sigma_{\textrm{SM}}\lesssim\mathcal{O}(1)$ for different channels.
For a heavy Higgs with $m_i\leq2m_t$, $Br_{\textrm{SM}}(VV)$ is very close to 1; while for $m_i>2m_t$,
$Br_{\textrm{SM}}(VV)$ has a minimal value of about $0.8$ when $m_i\sim500\textrm{GeV}$. According to (\ref{gamma1})
-(\ref{gamma4}), we can estimate that for both $m_i\sim v$ and $m_i\gg v$, $\mu\sim\mathcal{O}(0.1\sim1)$.

Thus according to the figures in \cite{HH}, we have three types of typical choices for the mass of two heavy
neutral higgs particles in \autoref{cases}. (Here we write the mass of the lighter boson $m_2$ and
the heavier one $m_3$.)
\begin{table}[h]
\caption{Typical choices for the masses of the two heavy neutral scalars.}\label{cases}
\begin{tabular}{|c|c|c|}
\hline
Case & Allowed $m_2$(GeV) & Allowed $m_3$(GeV) \\
\hline
I & $\lesssim300$ & $\lesssim300$ \\
\hline
II & $\lesssim300$ & $\gtrsim700$ \\
\hline
III & $\gtrsim700$ & $\gtrsim700$ \\
\hline
\end{tabular}
\end{table}

\subsection{Constraints due to Oblique Parameters}

After the discovery of the new boson, there are new electro-weak fit for the standard model \cite{Fit}.
Choosing $m_{t,\textrm{ref}}=173\textrm{GeV}$ and $m_{h,\textrm{ref}}=126\textrm{GeV}$, the
oblique parameters \cite{obl} are
\begin{eqnarray}
&S=0.03\pm0.10,\quad\quad T=0.05\pm0.12,\quad\quad U=0.03\pm0.10,&\nonumber\\
&R_{ST}=+0.89,\quad\quad R_{SU}=-0.54,\quad\quad R_{TU}=-0.83,&
\end{eqnarray}
with $R$ the correlation coefficient between two quantities; or
\begin{equation}
S=0.05\pm0.09,\quad\quad T=0.08\pm0.07, \quad\quad R=+0.91,
\end{equation}
with fixed $U=0$, where R is the correlation coefficient between S and T. The basic mathematica code to
draw the S-T ellipse can be found on the webpage \cite{STcode}\footnote{Assuming Gaussian distribution,
the second $\Delta\chi^2$ should be 6.0 instead of 6.8 in the code. See the 36th chapter (statistics)
of the reviews in PDG \cite{PDG}, in its 2014 updated version please see the 38th chapter instead.}. The contribution
to S and T parameters due to multi-higgs doublets were calculated in \cite{ST} (see the formulae in \cite{2HDM}).
\begin{eqnarray}
\Delta T&=&\frac{1}{16\pi s^2_Wm^2_W}\bigg[\mathop{\sum}_{i=1}^3(1-c_i^2)F(m^2_{H^{\pm}},m^2_i)-c_1^2F(m^2_2,m_3^2)
-c_2^2F(m^3_3,m_1^2)\nonumber\\
&&-c_3^2F(m_1^2,m_2^2)+3\mathop{\sum}_{i=1}^3c^2_i(F(m^2_Z,m^2_i)-F(m^2_W,m^2_i))\nonumber\\
&&-3(F(m^2_Z,m^2_{h,\textrm{ref}})-F(m^2_W,m^2_{h,\textrm{ref}}))\bigg];\\
\Delta S&=&\frac{1}{24\pi}\Bigg[(1-2s^2_W)^2G(z_{\pm},z_{\pm})+c_1^2G(z_2,z_3)+c_2^2G(z_3,z_1)+c_3^2G(z_1,z_2)\nonumber\\
&&+\mathop{\sum}_{i=1}^3\left(c^2_iH(z_i)+\ln\left(\frac{m^2_i}{m^2_{H^{\pm}}}\right)\right)
-H\left(\frac{m^2_{h,\textrm{ref}}}{m^2_Z}\right)-\ln\left(\frac{m^2_{h,\textrm{ref}}}{m^2_{H^{\pm}}}\right)\Bigg];
\end{eqnarray}
where $c_i$ is the rate of the $h_iV_{\mu}V^{\mu}$ coupling to that in SM ($c_1$ represents above-mentioned $c_V$) and $\sum c^2_i=1$.
$m_{h,\textrm{ref}}=m_1=126\textrm{GeV}$ is the reference point for Higgs Boson, $z_{\pm}=(m_{H^{\pm}}/m_Z)^2$
and $z_i=(m_i/m_Z)^2$. The functions
$F,G,H$ read  (following the fomulae in \cite{2HDM})
\begin{eqnarray}
F(x,y)&=&\frac{x+y}{2}-\frac{xy}{x-y}\ln\left(\frac{x}{y}\right);\\
G(x,y)&=&-\frac{16}{3}+5(x+y)-2(x-y)^2\nonumber\\
&&+3\left(\frac{x^2+y^2}{x-y}+y^2-x^2+\frac{(x-y)^3}{3}\right)\ln\left(\frac{x}{y}\right)+(1-2(x+y)\nonumber\\
&&+(x-y)^2)f(x+y-1,1-2(x+y)+(x-y)^2);\\
H(x)&=&-\frac{79}{3}+9x-2x^2+\left(-10+18x-6x^2+x^3-9\frac{x+1}{x-1}\right)\ln x\nonumber\\
&&+(12-4x+x^2)f(x,x^2-4x);
\end{eqnarray}
where
\begin{equation}
f(x,y)=\left\{\begin{array}{l}\sqrt{y}\ln\left|\frac{x-\sqrt{y}}{x+\sqrt{y}}\right|,\quad\quad y\geq0;\\
2\sqrt{-y}\arctan\left(\frac{\sqrt{-y}}{x}\right),\quad\quad y<0.\end{array}\right.
\end{equation}
At removable singularities the functions are defined as the limit.

\begin{figure}
\caption{S-T ellipse for case I, $m_2=280\textrm{GeV}$ and $m_3=300\textrm{GeV}$.}\label{I}
\includegraphics[scale=0.7]{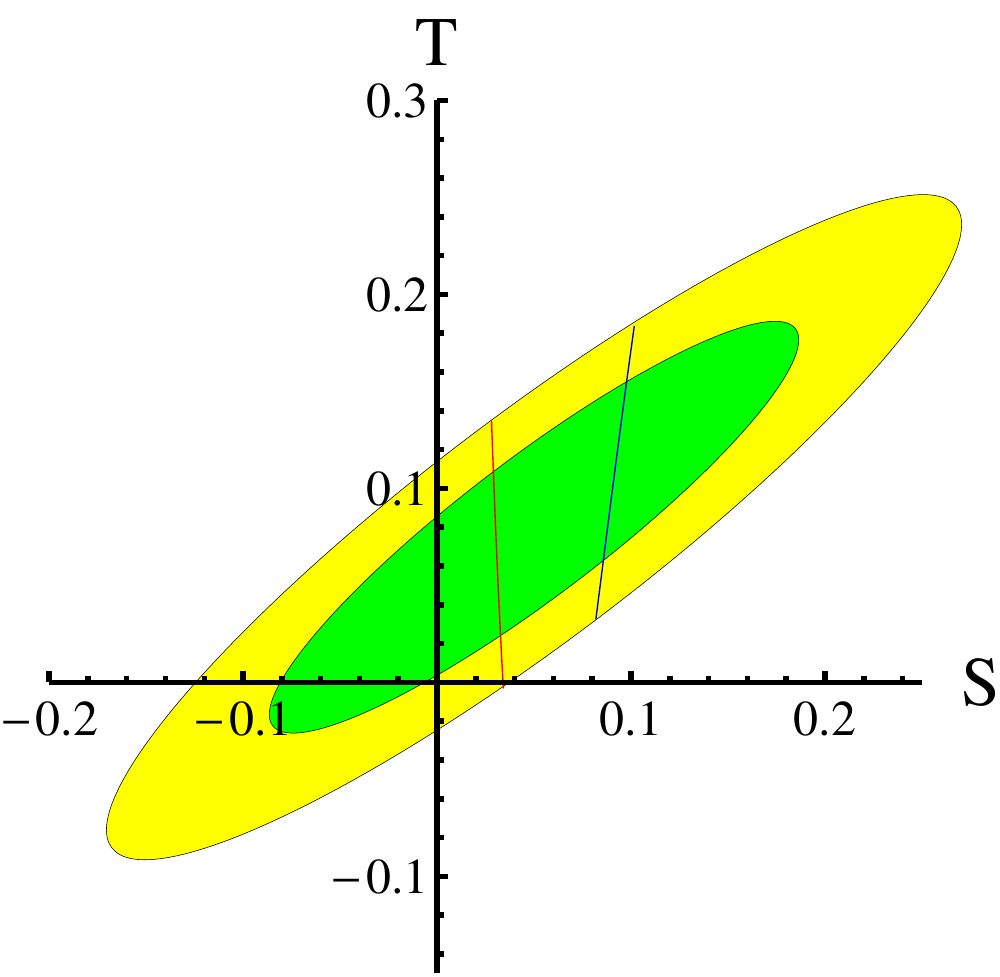}\quad\includegraphics[scale=0.7]{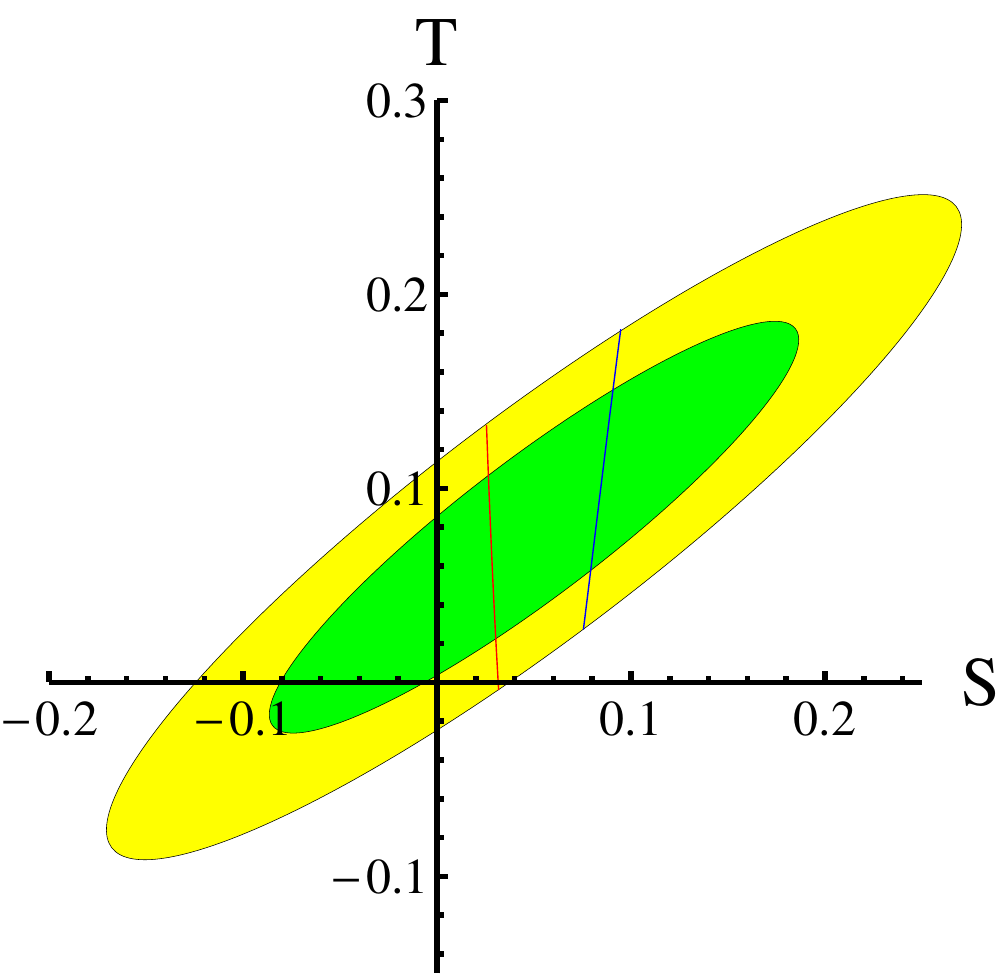}
\end{figure}
\begin{figure}
\caption{S-T ellipse for case II, $m_2=300\textrm{GeV}$ and $m_3=700\textrm{GeV}$.}\label{II}
\includegraphics[scale=0.7]{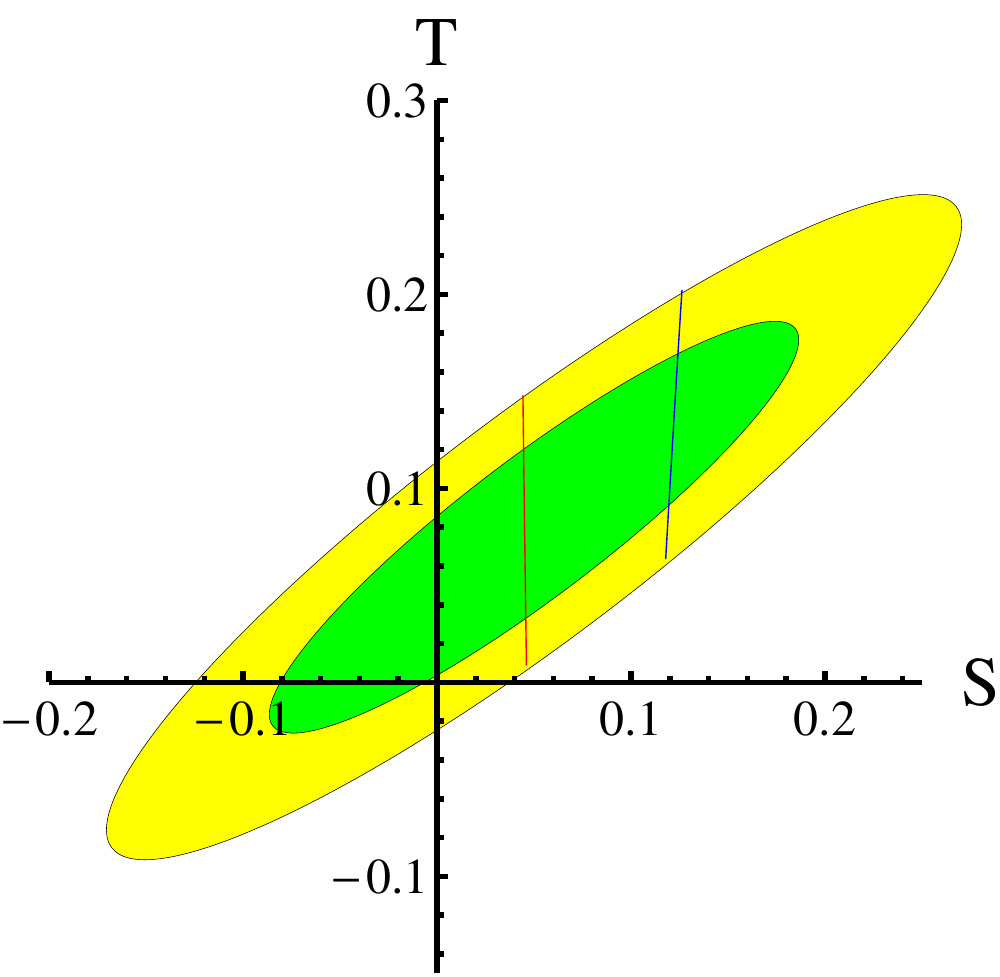}\quad\includegraphics[scale=0.7]{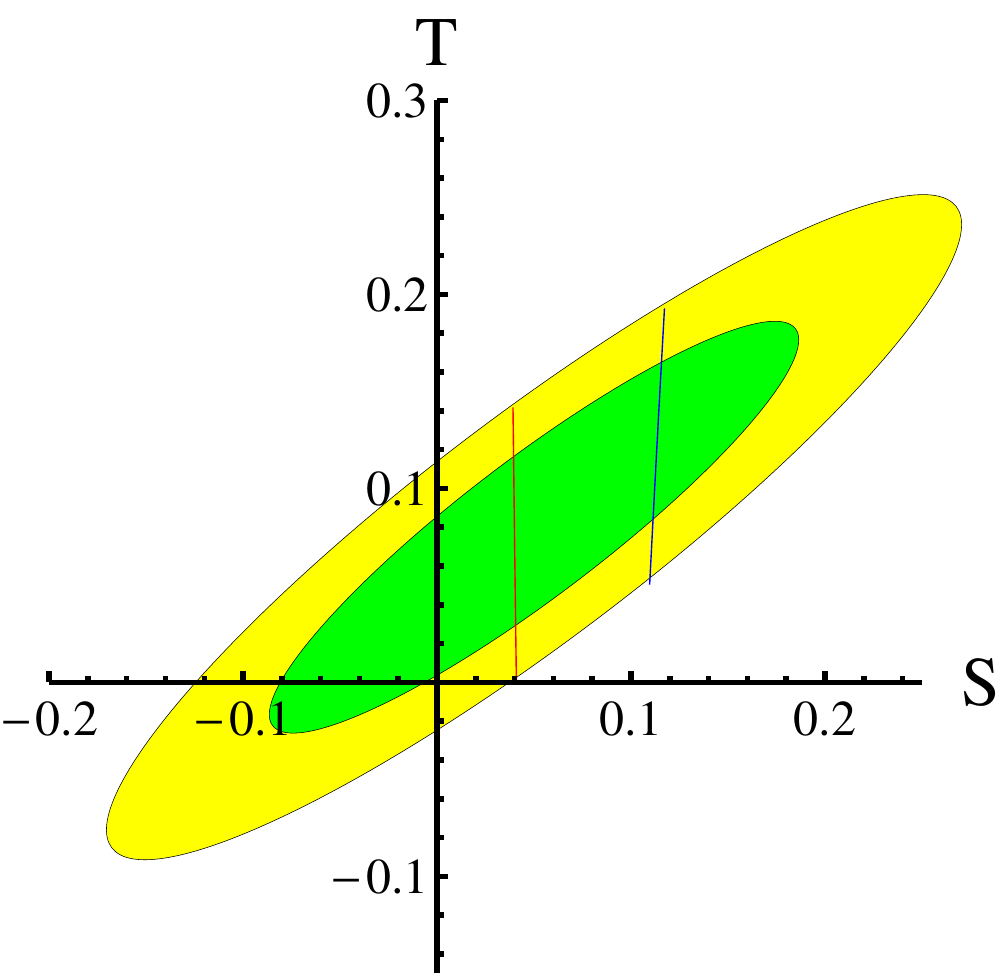}
\end{figure}
\begin{figure}
\caption{S-T ellipse for case III, $m_2=700\textrm{GeV}$ and $m_3=750\textrm{GeV}$.}\label{III}
\includegraphics[scale=0.7]{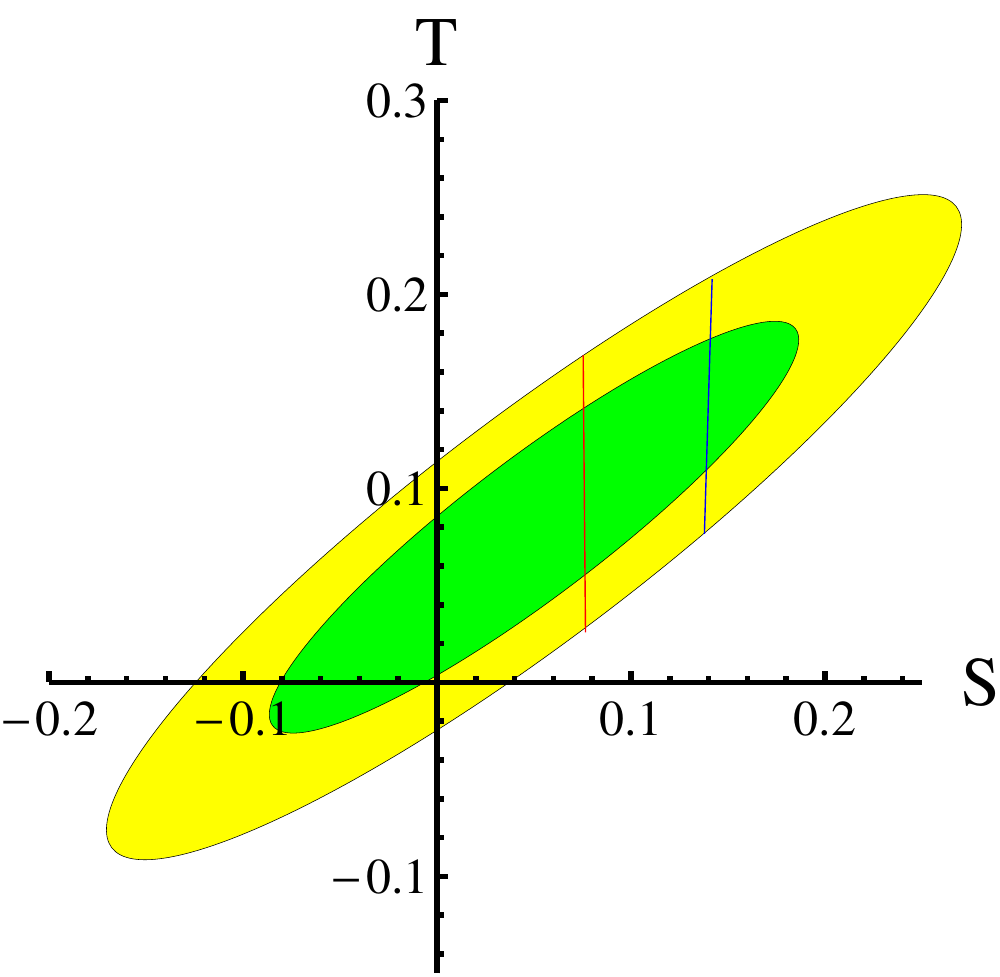}\quad\includegraphics[scale=0.7]{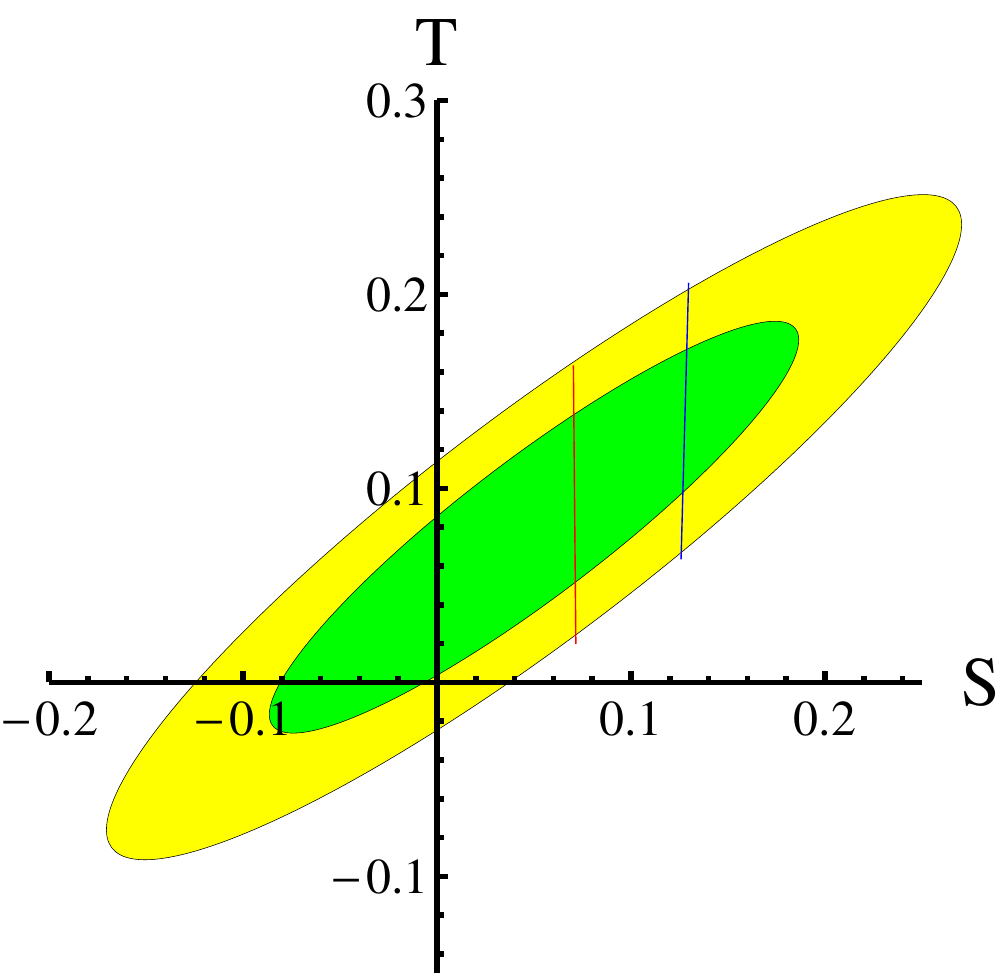}
\end{figure}
The parameter $U$ is usually small so that we fix $U=0$ from now on. We take the benchmark points according to
the cases in \autoref{cases}. We show the contours in \autoref{I}-\autoref{III} for the cases listed in
\autoref{cases} in last section. Throughout the  paper, the region outside the green area is excluded at $68\%$ C.L. and
the region outside the yellow area is excluded at $95\%$ C.L.
Firstly, for case I, we take $m_2=280\textrm{GeV}$ and $m_3=300\textrm{GeV}$. The typical values for the left diagram in \autoref{I} are $c_1^2=0.2$, $c_2^2=c_3^2=0.4$. Here the blue and red lines refer to
$86\textrm{GeV}<m_{H^{\pm}}<126\textrm{GeV}$ and $312\textrm{GeV}<m_{H^{\pm}}<350\textrm{GeV}$ respectively.
For the right diagram, $c_1^2=0.25$, $c_2^2=0.4$, $c_3^2=0.35$, and the blue and red lines refer to
$94\textrm{GeV}<m_{H^{\pm}}<136\textrm{GeV}$ and  $312\textrm{GeV}<m_{H^{\pm}}<351\textrm{GeV}$ respectively.

Secondly, for case II, we take $m_2=300\textrm{GeV}$ and $m_3=700\textrm{GeV}$. The typical values for the left diagram in \autoref{II} are $c_1^2=0.2$, $c_2^2=0.5$, $c_3^2=0.3$. Here the blue and red lines refer to
$127\textrm{GeV}<m_{H^{\pm}}<149\textrm{GeV}$ and  $580\textrm{GeV}<m_{H^{\pm}}<600\textrm{GeV}$ respectively.
For the right diagram, $c_1^2=c^2_3=0.25$, $c_2^2=0.5$, and the blue and red lines refer to
$141\textrm{GeV}<m_{H^{\pm}}<163\textrm{GeV}$ and $598\textrm{GeV}<m_{H^{\pm}}<618\textrm{GeV}$ respectively.

Thirdly, for case III, we take $m_2=700\textrm{GeV}$ and $m_3=750\textrm{GeV}$. The typical values for the left diagram  in \autoref{III} are $c_1^2=0.2$, $c_2^2=c_3^2=0.4$. Here the blue and red lines refer to
$218\textrm{GeV}<m_{H^{\pm}}<235\textrm{GeV}$ and  $748\textrm{GeV}<m_{H^{\pm}}<765\textrm{GeV}$ respectively.
For the right diagram, $c_1^2=0.25$, $c_2^2=0.4$, $c_3^2=0.35$, and the blue and red lines refer to
$250\textrm{GeV}<m_{H^{\pm}}<269\textrm{GeV}$ and  $749\textrm{GeV}<m_{H^{\pm}}<767\textrm{GeV}$ respectively.

In type II 2HDM the charged Higgs should be heavier than 360GeV \cite{chm}\cite{chm2} mainly due to the constraint from inclusive
$b\rightarrow s\gamma$ process. However in other models, there is no such strict constraints. Direct searches
by LEP told us that the charged Higgs boson should be heavier than 78.6GeV \cite{LEP}. In case I and II above,
a light (around $100\sim200$GeV) charged Higgs boson is allowed, while in case III the charged Higgs boson cannot be
lighter than about 250GeV. In case I and III, a charged higgs boson with the mass near the heavy neutral bosons
is allowed, while in case II a heavy charged higgs boson must be lighter than the heaviest neutral scalar.

\subsection{Constraints due to Signal Strengths}
In \autoref{sgs1} and \autoref{sgs2}, for a certain channel, the signal strength is defined as
\begin{equation}
\mu_f=\frac{\sigma\cdot Br_f}{(\sigma\cdot Br_f)_{\textrm{SM}}}=\frac{\sigma}{\sigma_{\textrm{SM}}}
\cdot\frac{\Gamma_f}{\Gamma_{f,\textrm{SM}}}\cdot\frac{\Gamma_{\textrm{tot,SM}}}{\Gamma_{\textrm{tot}}},
\end{equation}
in which $\sigma/\sigma_{\textrm{SM}}=|c_t'|^2$ for gluon fusion processes and $\sigma/\sigma_{\textrm{SM}}=c_V^2$
for vector boson fusion (VBF) processes and associated productions with a gauge boson. For decays
without interference, we simply have $\Gamma_f/\Gamma_{f,\textrm{SM}}=|c_f|^2$ such as for $f=V,b,\tau$.
While for the two photons final state, we have \cite{pro1}\cite{susy}
\begin{equation}
\label{rr}\frac{\Gamma_{\gamma\gamma}}{\Gamma_{\gamma\gamma,\textrm{SM}}}=\left|\frac{(4/3)c_t'\mathcal{A}_{1/2}(x_t)
+c_V\mathcal{A}_1(x_W)+(c_{\pm}v^2/2m^2_{H^{\pm}})\mathcal{A}_0(x_{\pm})}
{(4/3)\mathcal{A}_{1/2}(x_t)+\mathcal{A}_{1}(x_W)}\right|^2,
\end{equation}
in which $x_i=m^2_h/4m^2_i$ for $i=t,W,H^{\pm}$. The loop integration functions are
\begin{eqnarray}
\mathcal{A}_0(x)&=&\frac{1}{x^2}(x-f(x))\\
\mathcal{A}_{1/2}(x)&=&-\frac{2}{x^2}(x+(x-1)f(x))\\
\mathcal{A}_1(x)&=&\frac{1}{x^2}(2x^2+3x+3(2x-1)f(x))
\end{eqnarray}
for scalar, fermion and vector boson loop respectively and
\begin{equation}
f(x)=\left\{\begin{array}{l}\arcsin^2\sqrt{x},\quad\quad x\leq1;\\
-\frac{1}{4}\left(\ln\left(\frac{1+\sqrt{1-1/x}}{1-\sqrt{1-1/x}}\right)-\pi i\right)^2,\quad\quad x>1.\end{array}\right.
\end{equation}

In a spontaneous CP-violation model, $c_t'$ (together with other $c_f$ for fermions) can be complex while
$c_V$ and $c_{\pm}$ must be real. Notice all the $c_V,c_{\pm},c_b$ and $c_{\tau}$ are the same with those in
(\ref{c1*})-(\ref{c5*}), but $c_t$ should be modified to $c_t'$ as
\begin{equation}
c_t'=\textrm{Re}(c_t)+i\frac{\mathcal{B}_{1/2}(m^2_h/4m^2_t)}{\mathcal{A}_{1/2}(m^2_h/4m^2_t)}\textrm{Im}(c_t)
\end{equation}
in which the function
\begin{equation}
\mathcal{B}_{1/2}(x)=-\frac{2f(x)}{x}.
\end{equation}
Thus defining $\alpha_t\equiv\arg(c_t)$ and $\alpha_t'\equiv\arg(c_t')$, numerically we have
\begin{equation}
\alpha_t'=\arctan(1.52\tan\alpha_t),\quad\quad|c_t'|=|c_t|\sqrt{1+1.31\sin^2\alpha_t}.
\end{equation}
Assuming there is no unknown decay channel which contributes several percentages or
more to the total width, we can estimate that
\begin{equation}
\frac{\Gamma_{\textrm{tot}}}{\Gamma_{\textrm{tot,SM}}}=0.57|c_b|^2+0.25c_V^2+0.06|c_{\tau}|^2+0.03|c_c|^2+0.09|c_t'|^2
\end{equation}
according to \autoref{smp}.

Define the $\chi^2$
\begin{equation}
\label{chi2}
\chi^2=\mathop{\sum}_{i,f}\left(\frac{\mu_{i,f,\textrm{obs}}-\mu_{i,f,\textrm{pre}}}{\sigma_{i,f}}\right)^2
\end{equation}
where $i=\textrm{VBF},\textrm{ggF},\textrm{VH}$ and $f=\gamma\gamma,WW^*,ZZ^*,\tau^+\tau^-$ at a detector (CMS or ATLAS). The
$\mu_{i,f,\textrm{obs(pre)}}$ are the observed (predicted) signal strength for the production channel
$i$ and final state $f$. We ignored all correlation coefficients between channels since they are small.

Numerically we find that the minimal $\chi^2$ is not sensitive to the charged Higgs mass since the scalar
loop contributes less than the top and $W$ loop in $\gamma\gamma$ decay channel. Thus we take the benchmark
point as $m_{H^{\pm}}=150\textrm{GeV}$. For six degrees of freedom, parameter space with $\chi^2\leq7.0$ is
allowed at $68\%$C.L. and $\chi^2\leq12.6$ is allowed at $95\%$C.L. For both CMS and ATLAS data, the
minimal $\chi^2$ is very sensitive to $c_V$ and $c_t'$, since they give dominant contributions to most production
cross sections and partial decay widths; it is sensitive to $c_b$ as well since the total width is sensitive to $|c_b|$.
With the CMS data, we have $c_V\geq0.22$; and with the ATLAS data, we have $c_V\geq0.31$, both at $95\%$C.L.
So $c_V=0.5$ is a good benchmark point as we have chosen in the last section, and it will also be taken around this point in later analysis.
The $\chi^2$ is not very sensitive to $|c_{\tau}|$ and $c_{\pm}$, as both of them contribute to only one channel, and
the charged Higgs loop contributes less in the $\gamma\gamma$ decay channel. Thus for most analysis we don't discuss
these two parameters carefully.

\begin{figure}
\caption{Allowed $|c^\prime_t|-\alpha^\prime_t$ contour when taking $c_V=0.4$, $c_{\pm}=0.2$ and $|c_{\tau}|=0.8$ for CMS data.
$|c_b|=0.1$ for the left figure and $|c_b|=0.4$ for the right figure.}\label{cmsct1}
\includegraphics[scale=0.7]{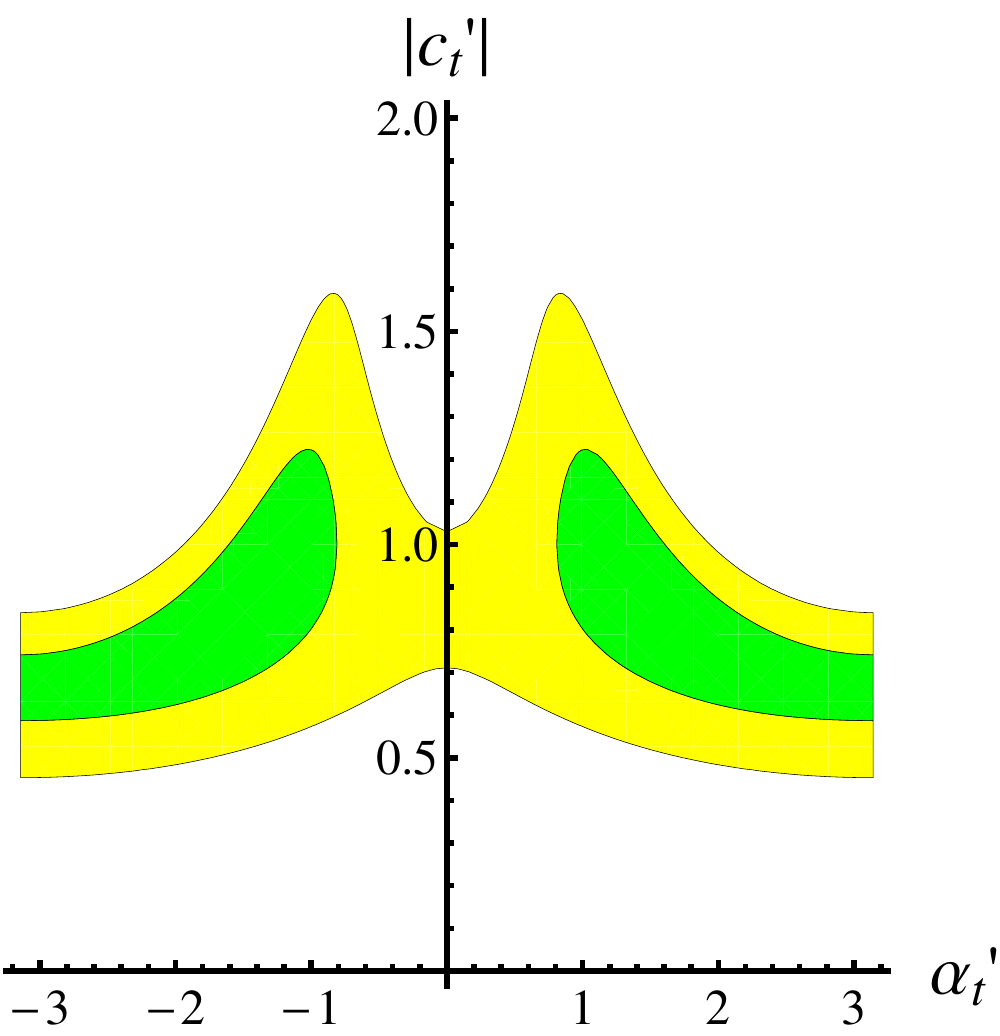}\quad\includegraphics[scale=0.7]{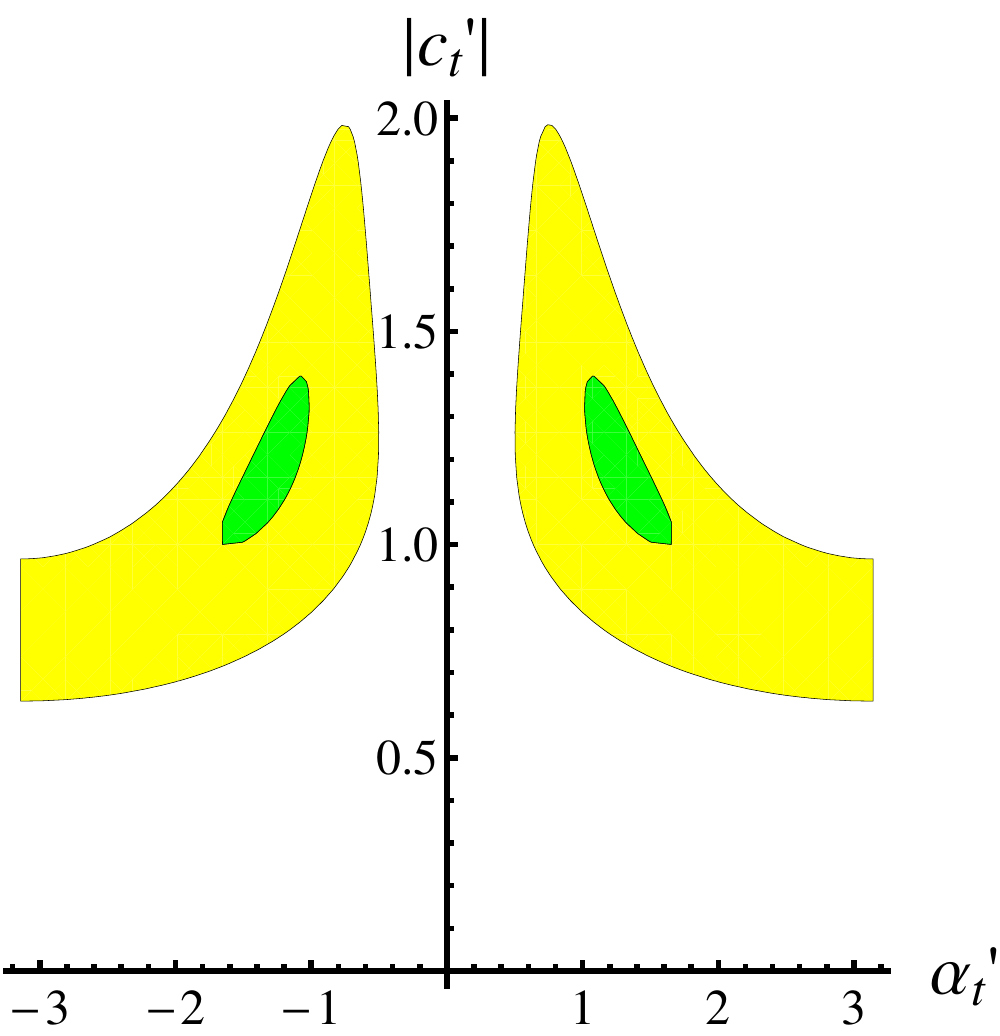}
\end{figure}

\begin{figure}
\caption{Allowed $|c^\prime_t|-\alpha^\prime_t$ contour  when taking $c_V=0.4$, $c_{\pm}=0.2$ and $|c_b|=0.7$ for CMS data.
$|c_{\tau}|=0.8$ for the left figure and $|c_{\tau}|=1.3$ for the right figure.}\label{cmsct2}
\includegraphics[scale=0.7]{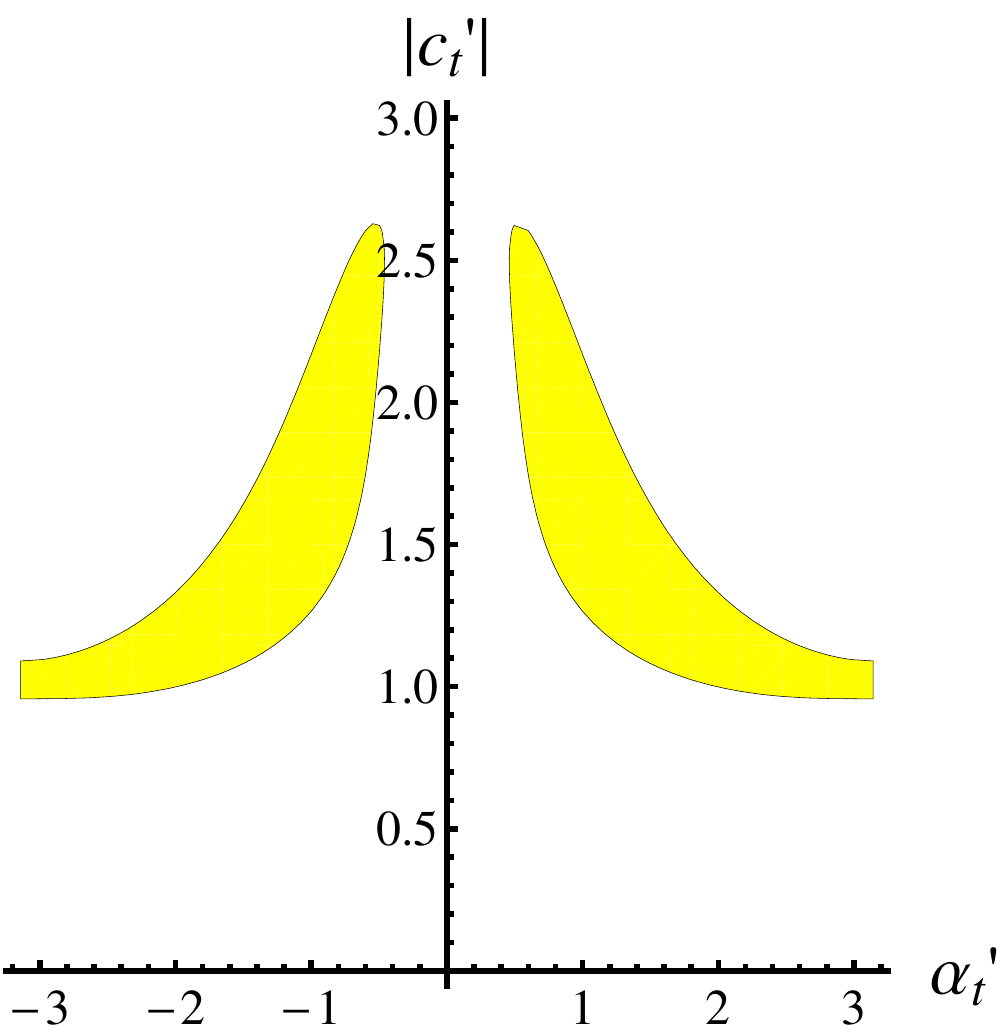}\quad\includegraphics[scale=0.7]{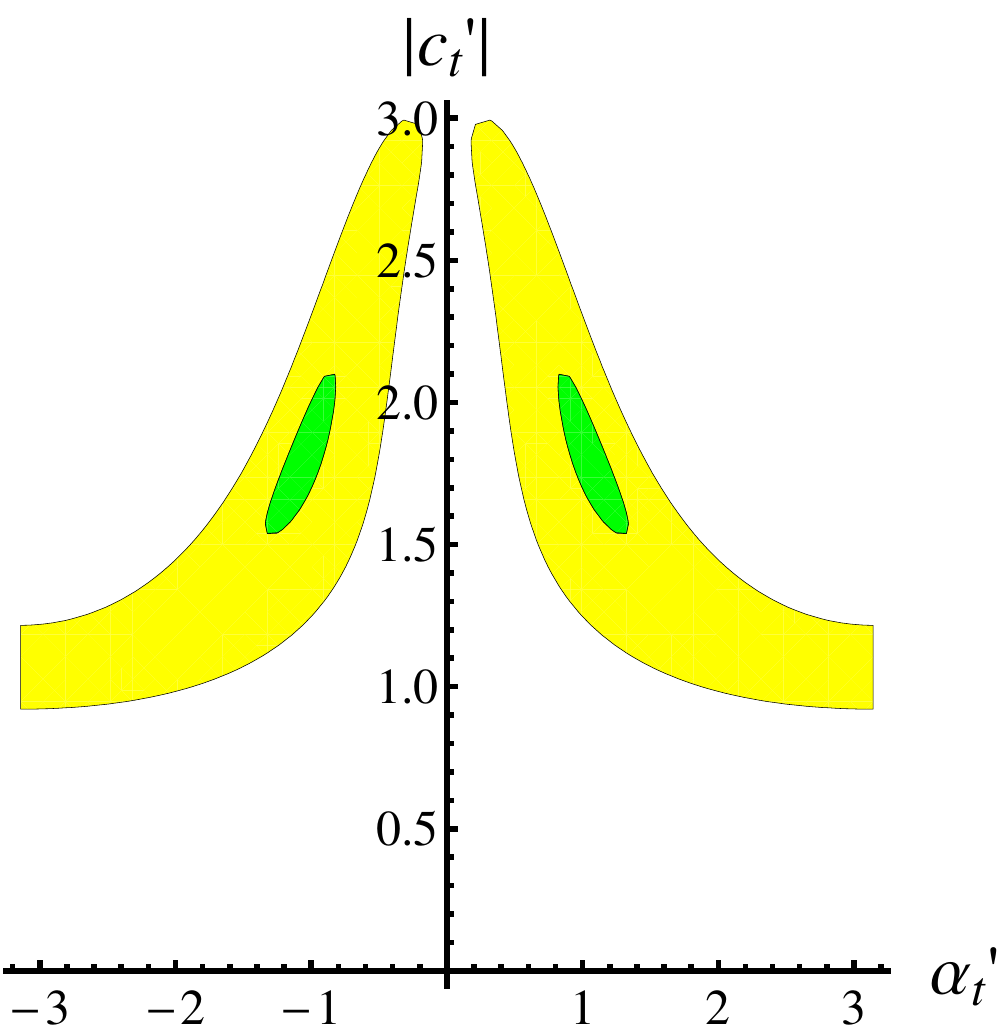}
\end{figure}

\begin{figure}
\caption{Allowed $|c^\prime_t|-\alpha^\prime_t$ contour  when taking $c_V=0.5$, $c_{\pm}=0.2$ and $|c_{\tau}|=1$ for CMS data.
The four figures correspond to $|c_b|=0.1,0.3,0.5,0.7$ respectively.}\label{cmsct3}
\includegraphics[scale=0.7]{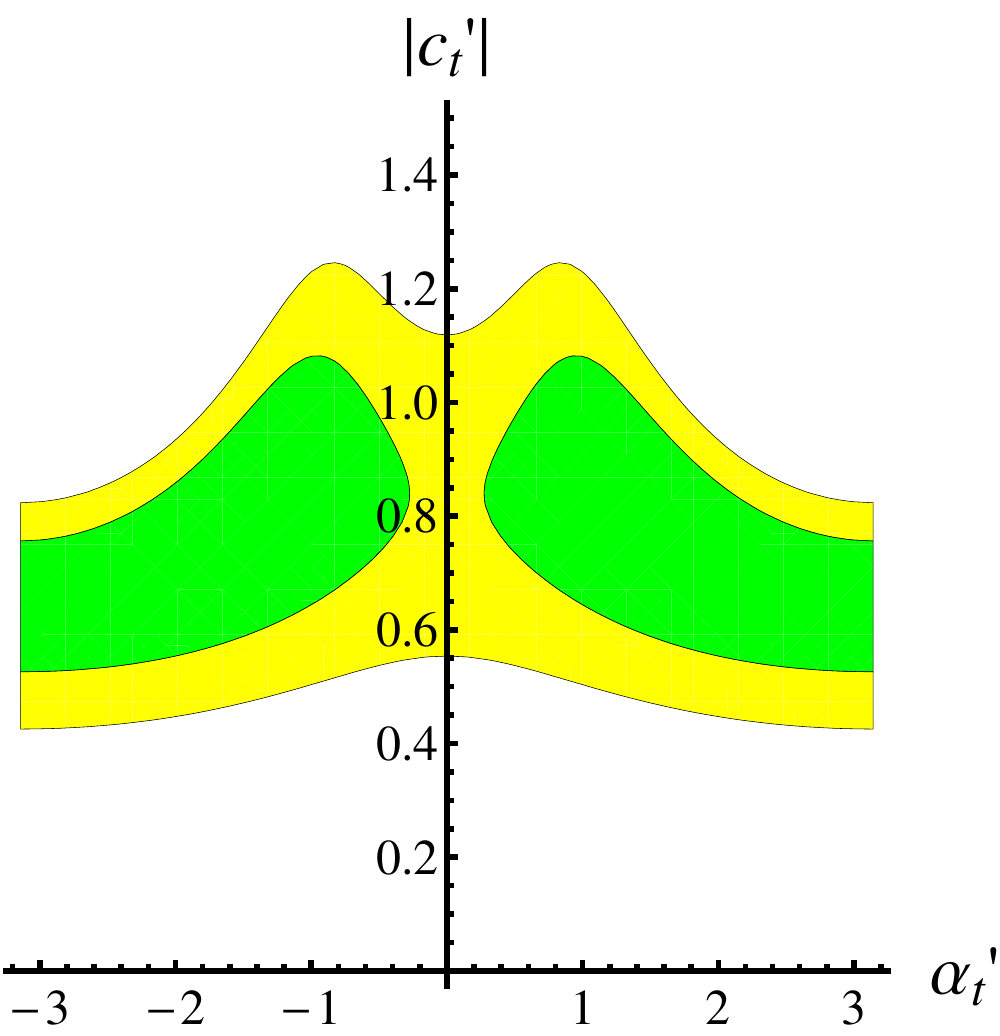}\quad\includegraphics[scale=0.7]{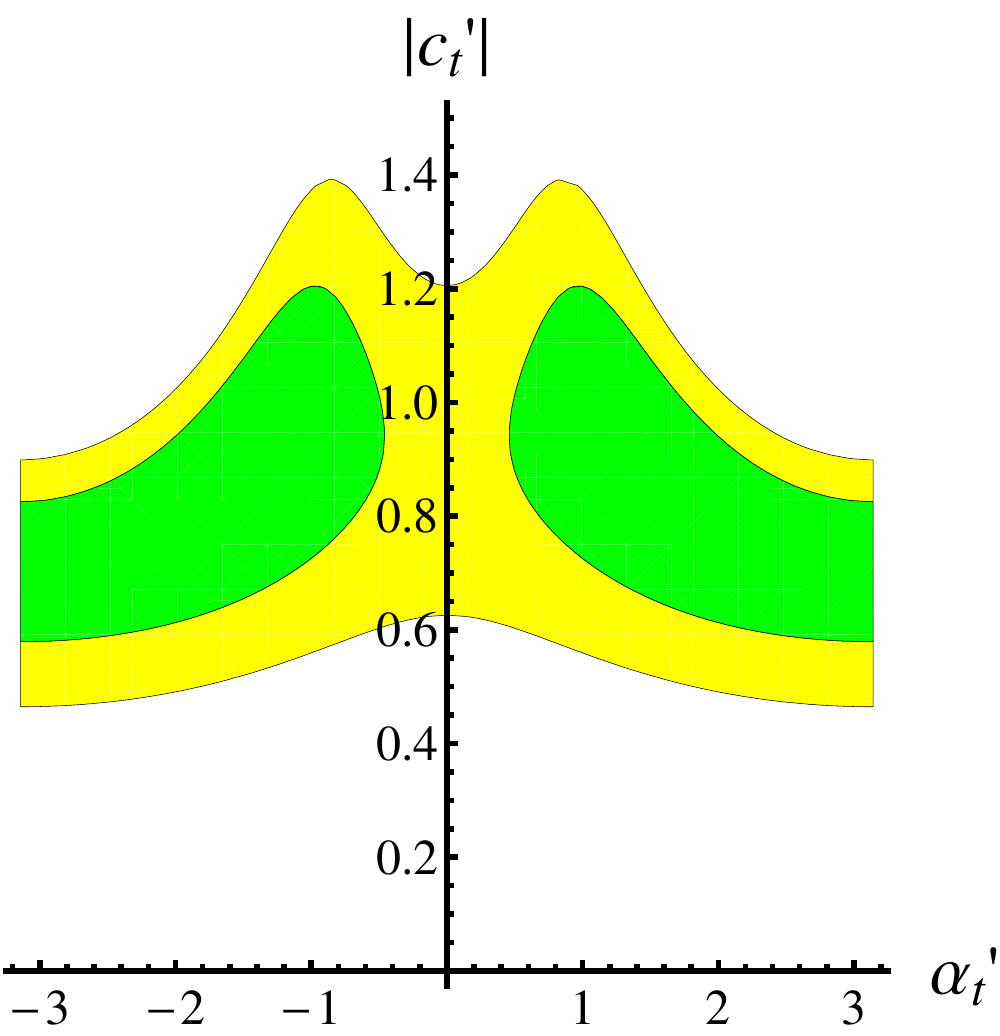}\\
\includegraphics[scale=0.7]{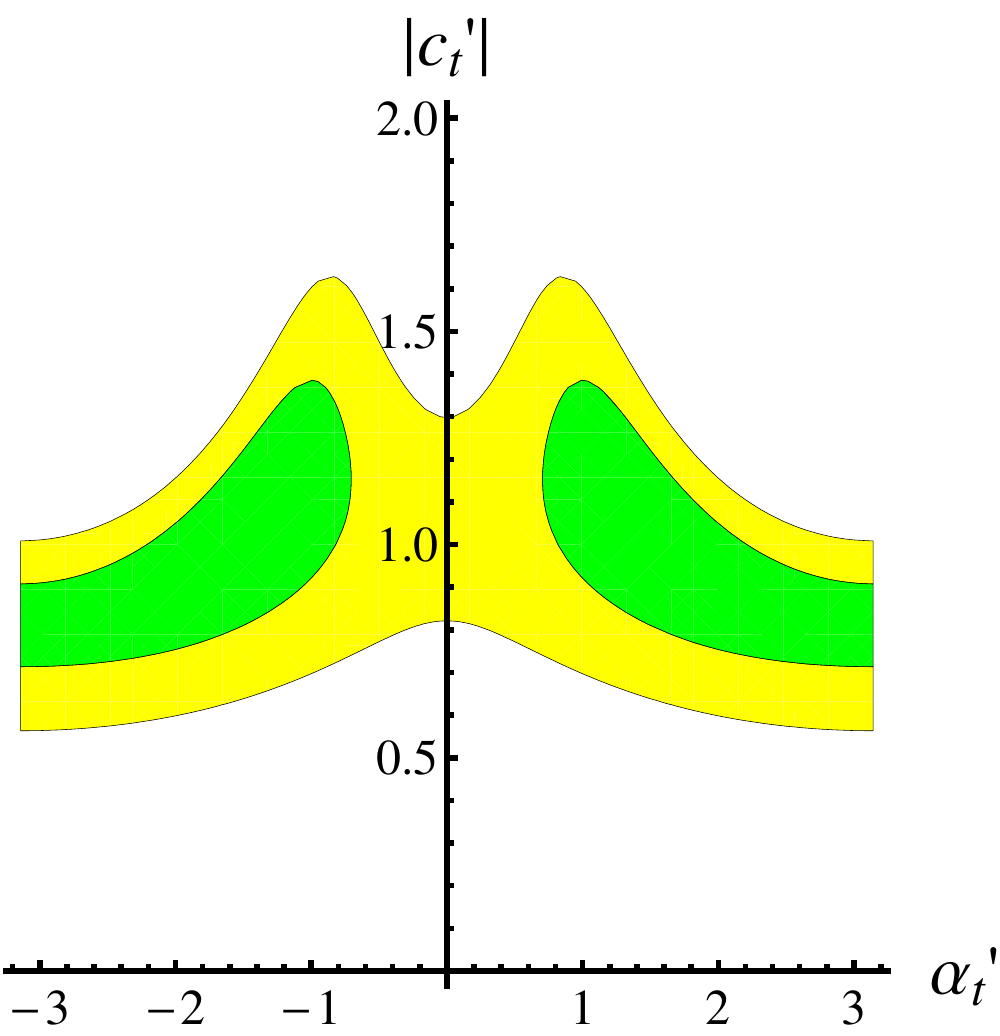}\quad\includegraphics[scale=0.7]{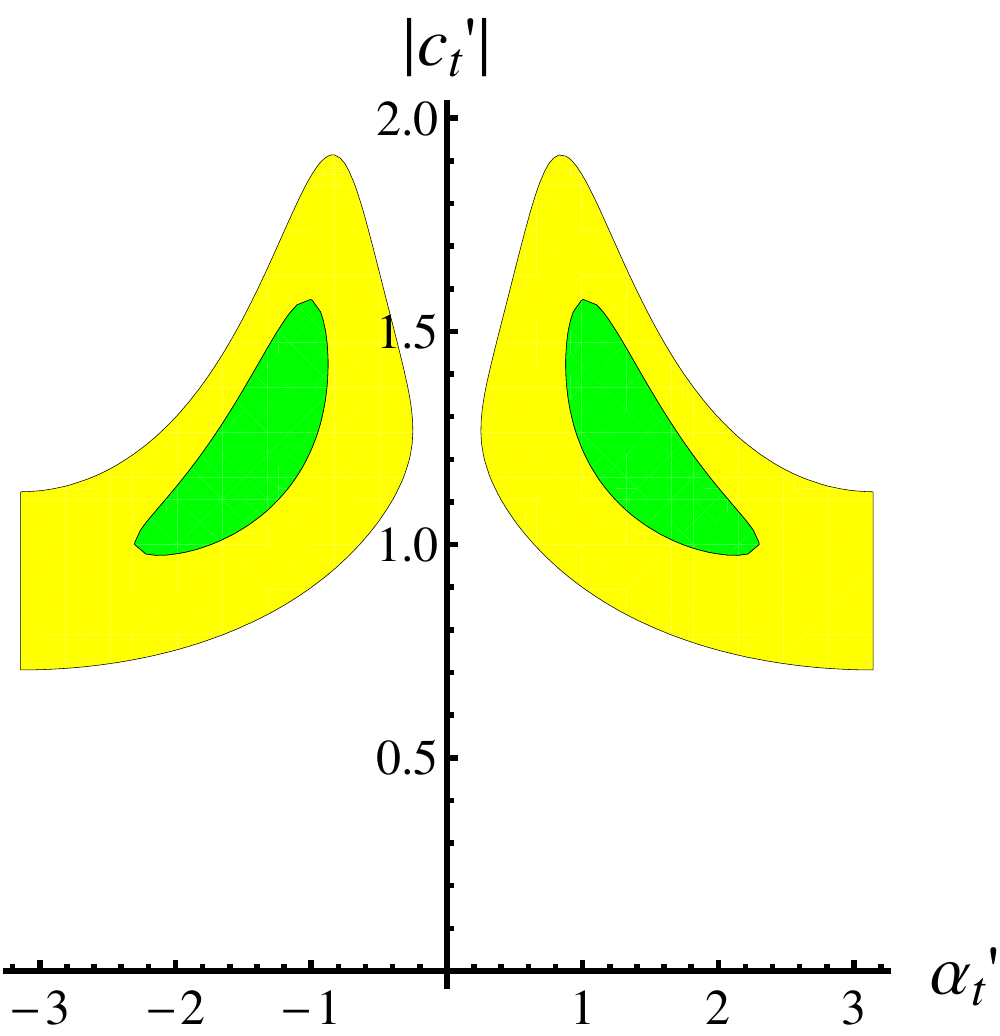}
\end{figure}

For the CMS data, when $c_V\sim0.5$, the $\chi^2_{\textrm{min}}\approx2$. The data favors smaller $|c_b|$ but the minimal
value of $\chi^2$ changes little as $c_b$ varies,  since the points are far away from the $95\%$ allowed boundary $c_V=0.22$.
\autoref{cmsct1}-\autoref{cmsct3} show CMS allowed $|c_t'|$ and $\alpha_t'\equiv\arg(c_t')$ for
some benchmark points\footnote{In this paper, the benchmark points are close to the best fit points for a certain case
thus the allowed regions are typical enough.}.


In \autoref{cmsct1} and \autoref{cmsct2}, we choose $c_V=0.4$. Fixing $c_{\pm}=0.2$ and $|c_{\tau}|=0.8$,
and taking $|c_b|=0.1,0,4,0.7$, we have the three figures in \autoref{cmsct1} and \autoref{cmsct2}.
The best fit point for $|c_t|$ has positive
correlation with $|c_b|$. For larger $|c_b|$, the best fit point for $|c_{\tau}|$ increases as well,
thus in the right figure in \autoref{cmsct2} we set $|c_{\tau}|=1.3$ and get better fitting result.

In \autoref{cmsct3}, we have $c_V=0.5$. Fixing $c_{\pm}=0.2$ and $|c_{\tau}|=0.9$,
and taking $|c_b|=0.1,0,3,0.5,0.7$, we get the four figures. The fitting results are less sensitive to $|c_{\tau}|$
than in \autoref{cmsct1} and \autoref{cmsct2}, and the best fit point for $|c_t'|$ has positive correlation with
$|c_b|$ as well. Usually $\alpha_t'\sim0$ is disfavored while for smaller $|c_b|$ and
larger $c_V$ any $\alpha_t'$ is allowed.  For each case, the best fit point is about $|\alpha_t'|\sim1.2$.

\begin{figure}
\caption{Allowed $|c^\prime_t|-\alpha^\prime_t$ contour  when taking $c_V=0.5$, $c_{\pm}=0.4$ and $|c_{\tau}|=0.7$ for ATLAS data.
The left and right figures correspond to $|c_b|=0.2$ and $|c_b|=0.4$ respectively.}\label{atlasct1}
\includegraphics[scale=0.7]{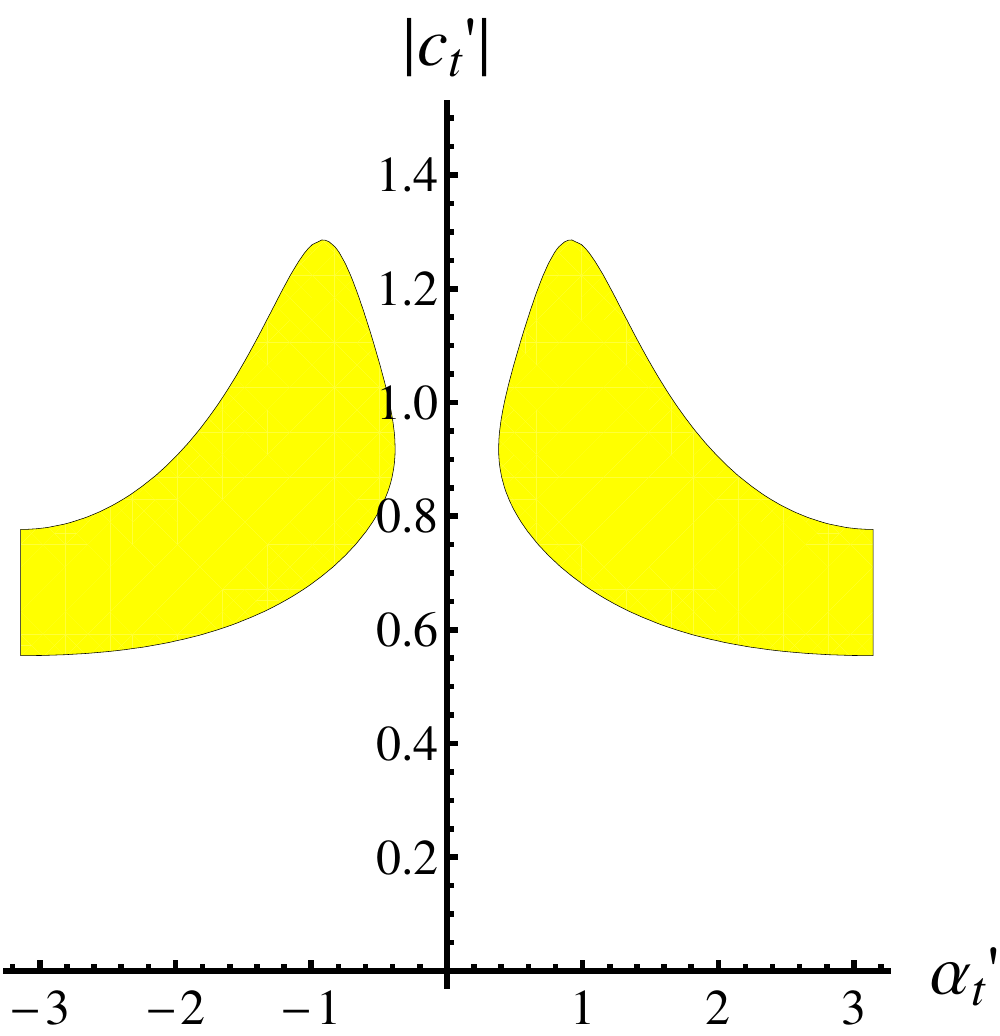}\quad\includegraphics[scale=0.7]{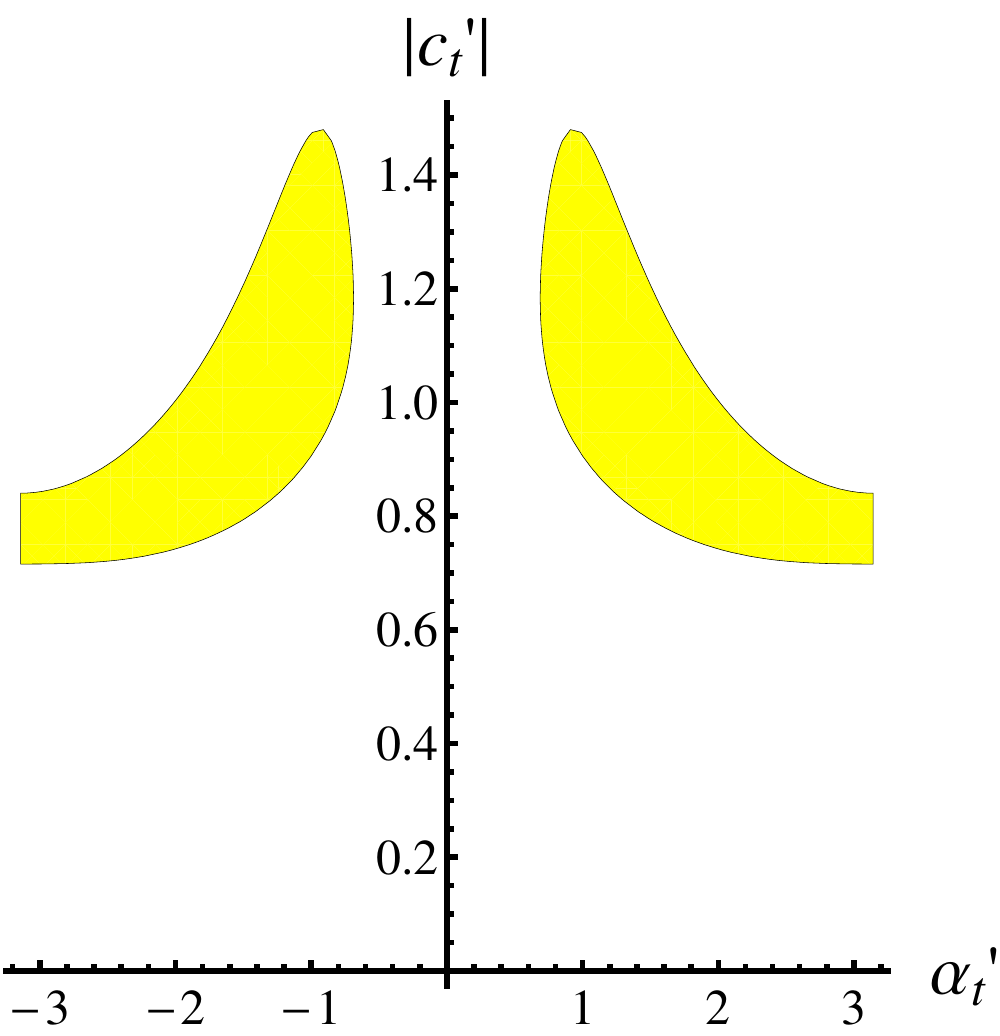}
\end{figure}

\begin{figure}
\caption{Allowed $|c^\prime_t|-\alpha^\prime_t$ contour  when taking $c_V=0.6$, $c_{\pm}=0.4$ and $|c_{\tau}|=0.8$ for ATLAS data.
The four figures correspond to $|c_b|=0.1,0.3,0.5,0.7$ respectively.}\label{atlasct2}
\includegraphics[scale=0.7]{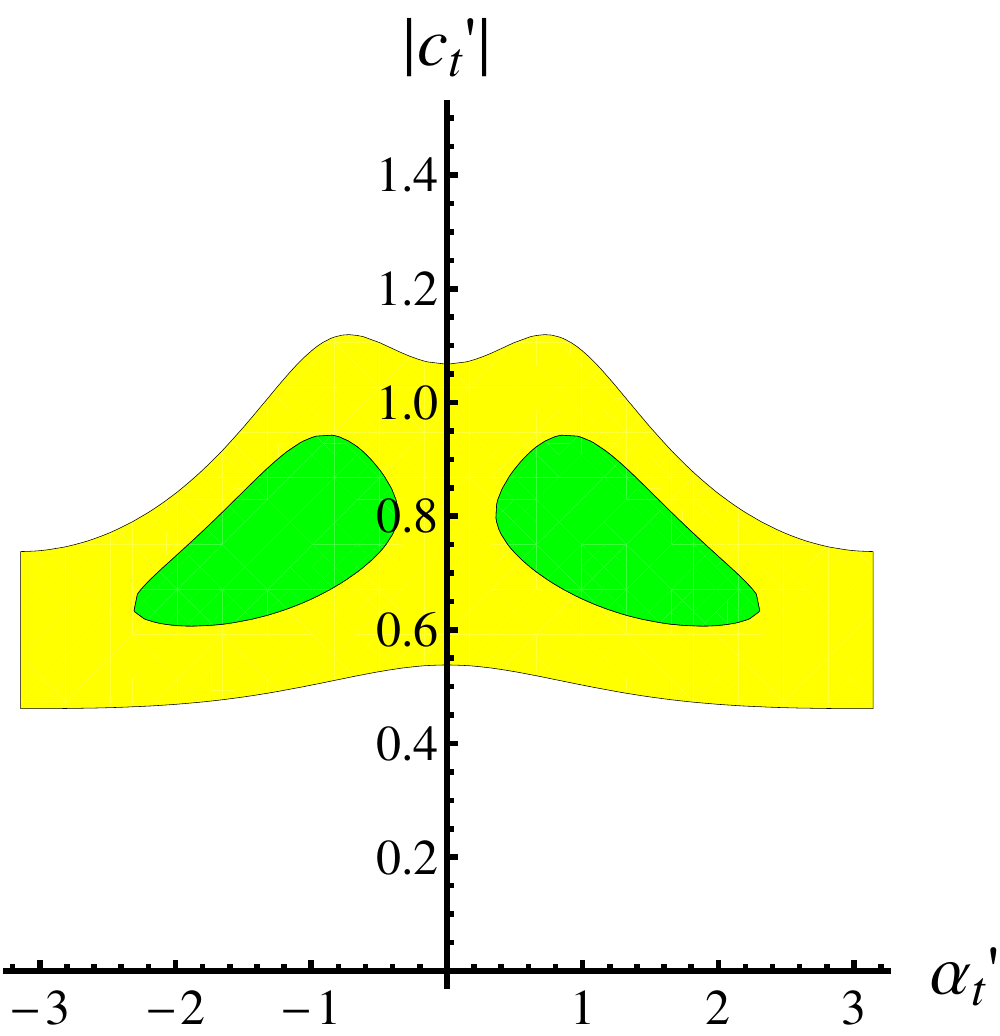}\quad\includegraphics[scale=0.7]{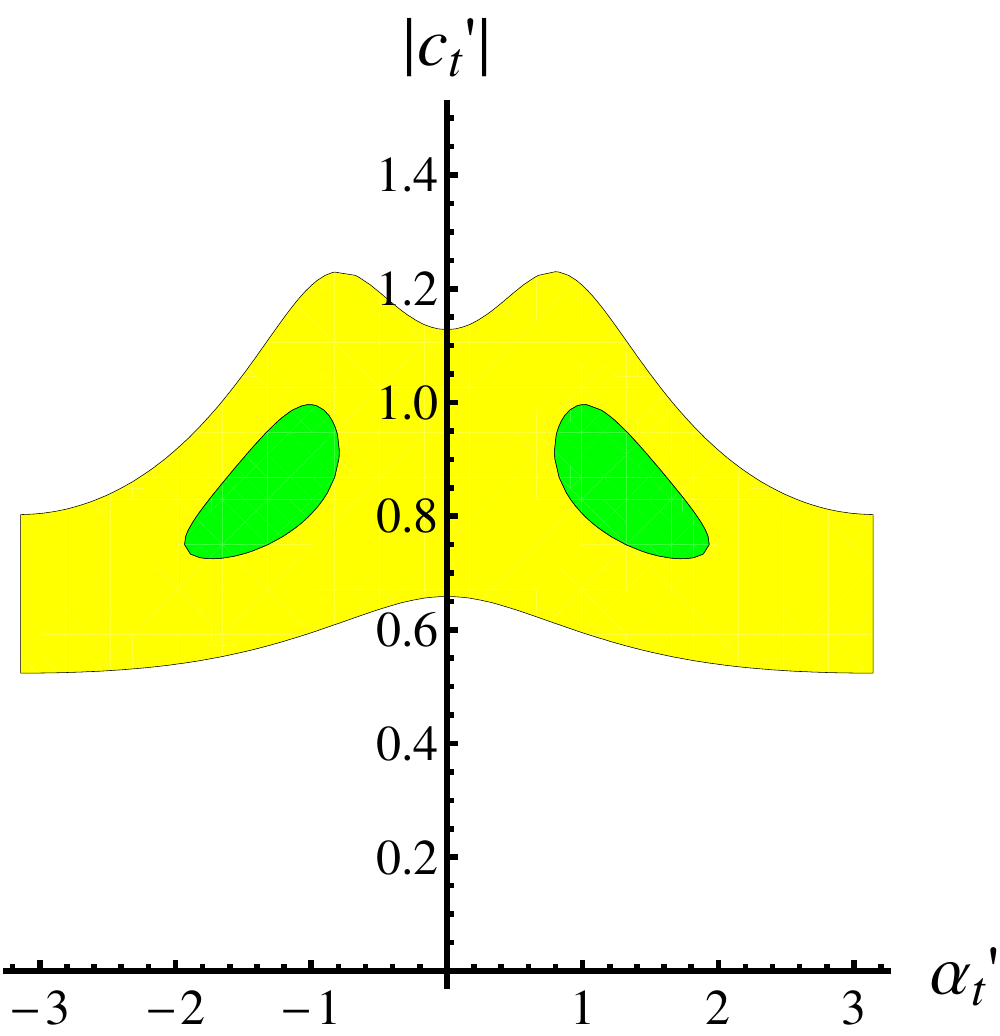}\\
\includegraphics[scale=0.7]{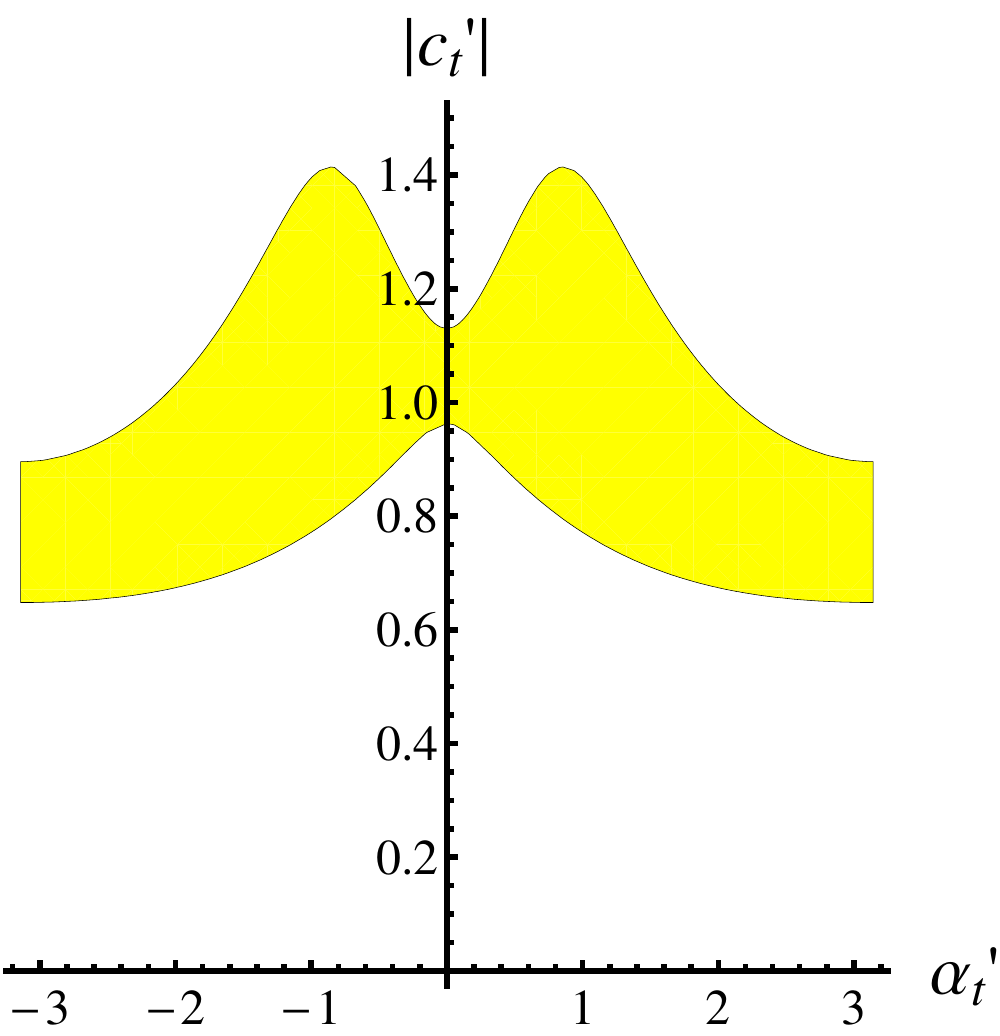}\quad\includegraphics[scale=0.7]{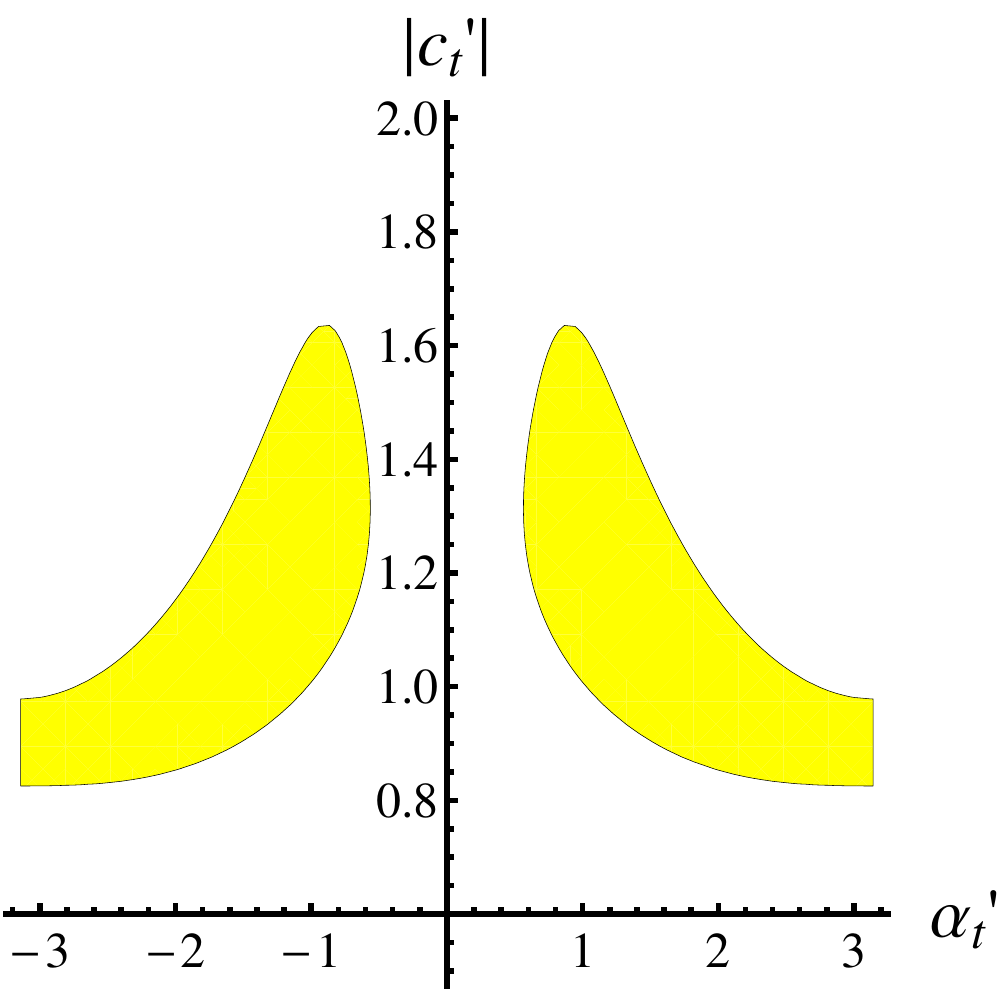}
\end{figure}

For the ATLAS data, when $c_V\sim0.5$, the $\chi^2_{\textrm{min}}\approx7$ which is near the $1\sigma$ allowed boundary.
The data favor smaller $|c_b|$ as well just like the CMS case.
\autoref{atlasct1}-\autoref{atlasct2} show ATLAS allowed $|c_t'|$ and $\alpha_t'\equiv\arg(c_t')$ for
some benchmark points.

In \autoref{atlasct1} we show the allowed regions for $c_t'$. Fixing $c_V=0.5$, $c_{\pm}=0.4$, and
$|c_{\tau}|=0.7$, choosing $|c_b|=0.2,0.4$, we have the two figures. In \autoref{atlasct2}, fixing $c_V=0.6$,
$c_{\pm}=0.4$, and $|c_{\tau}|=0.8$,
and taking $|c_b|=0.1,0.3,0.5,0.7$, we have the four figures. Usually $\alpha'_t\sim0$ is disfavored while for
smaller $|c_b|$ and larger $c_V$ any $\alpha'_t$ is allowed. The best fit points for $\alpha'_t$ are around $|\alpha'_t|\sim1.2$,
all these behaviors are similar as the results from CMS data.

For both CMS and ATLAS data, smaller $|c_b|$ is favored. In most case, the best fit points for
$|c_t'|, |c_{\tau}|$ are around $1$, and $c_{\pm}\sim\mathcal{O}(0.1)$. The fitting results for $\alpha'_t$ favor smaller
$|\alpha'_t|(\sim1.2)$ by both data for most $|c_b|$ input.
We also have the $\chi^2$ for SM as
\begin{equation}
\chi^2_{\textrm{SM,CMS}}=2.4,\quad\quad\chi^2_{\textrm{SM,ATLAS}}=3.7
\end{equation}
close to the minimal $\chi^2$ for Lee model we discussed in this paper. So the Lee model can fit the current data as well as that in the SM.

\subsection{Same Sign Top Production}
We put no additional symmetries in the Yukawa sector to avoid tree-level FCNC, thus the model must be
constrained by processes including flavor-changing interactions. The tree-level
FCNC for up type quarks will lead to same sign top quarks production at the LHC. An upper limit at $95\%$C.L.
was given as \cite{sst}
\begin{equation}
\label{tt}\sigma_{tt}<0.37\textrm{pb}
\end{equation}
by the CMS group with an integrated luminosity $19.5\textrm{fb}^{-1}$ at $\sqrt{s}=8\textrm{TeV}$.

In this model, we can write the interaction which can induce same sign top quark production at the LHC as
\begin{equation}
\label{hut}\mathcal{L}_{I,tuh}=-\frac{1}{\sqrt{2}}\bar{t}(\xi_{1tu}+\xi_{2tu}\gamma^5)uh+\textrm{h.c.}
\end{equation}
The lightest neutral boson gives the dominant contribution when the effect couplings are similar.
A direct calculation gives
\begin{equation}
\sigma_{tt}=\int dx_1dx_2f_u(x_1)f_u(x_2)\sigma_0
\end{equation}
in which
\begin{eqnarray}
\sigma_0&=&\frac{|\xi_{tu}|^4\beta_t}{64\pi s_0}\int_{-1}^1dc_{\theta}\left[\left(\frac{1-\beta_tc_{\theta}}
{1+\beta^2_t+4m^2_h/s_0-2\beta_tc_{\theta}}\right)^2\right.\nonumber\\
&&\left.+\left(\frac{1+\beta_tc_{\theta}}{1+\beta^2_t+4m^2_h/s_0+2\beta_tc_{\theta}}\right)^2
-\frac{1+\beta^2_tc^2_{\theta}-2\beta^2_t}{(1+\beta^2_t+4m^2_h/s_0)^2-4\beta^2_tc^2_{\theta}}\right],
\end{eqnarray}
if $\xi_{1tu}\xi_{2tu}^*+\xi_{2tu}\xi_{1tu}^*=0$ where $s_0$ is the square of energy in the frame
of momentum center of two partons (both $u$ quarks).
$\beta_t=\sqrt{1-4m^2_t/s_0}$ is the velocity of a top quark and $\theta$ is the radiative
angle in the same frame and $|\xi_{tu}|=\sqrt{|\xi_{1tu}|^2+|\xi_{2tu}|^2}$. Using the MSTW2008
PDF \cite{mstw} and comparing with (\ref{tt}), we can estimate that $|\xi_{tu}|\lesssim0.4$.
\subsection{Top Rare Decays}
\label{TRD}
In this model, the FCNC interactions including up type quarks will induce rare decay processes of top quark,
such as $t\rightarrow ch$ and $t\rightarrow uh$, usually with a larger rate than that in the SM. When the charged
Higgs boson is lighter than the top quark, there will be a new decay channel
$t\rightarrow H^+b$ as well. Direct search results at $\sqrt{s}=8\textrm{TeV}$ by CMS at LHC gave
the top pair production cross section\cite{ttexp} $\sigma_{t\bar{t}}=(237\pm13)\textrm{pb}$ assuming $m_t=173\textrm{GeV}$
and $Br(t\rightarrow bW)=1$, while theoretical calculation predicts that\cite{ttpre}
$\sigma_{t\bar{t},\textrm{pre}}=(246^{+9}_{-11})\textrm{pb}$. Assuming there is no effects beyond SM during
the production of top pair, these results can constrain the top rare decay (all channels except $bW$) branching ratio
\begin{equation}
\label{tr}Br_{t,\textrm{rare}}=1-Br(t\rightarrow bW)<7.4\%
\end{equation}
at $95\%$C.L.

For the rare decay processes above, the interactions can be written as
\begin{eqnarray}
\mathcal{L}_{I,tch}&=&-\frac{1}{\sqrt{2}}\bar{t}(\xi_{1tc}+\xi_{2tc}\gamma^5)ch+\textrm{h.c.}\\
\mathcal{L}_{I,tbH^+}&=&-\bar{t}(\xi_{1tb}+\xi_{2tb}\gamma^5)bH^++\textrm{h.c.}
\end{eqnarray}
together with (\ref{hut}). Direct calculations give the decay rates
\begin{eqnarray}
\Gamma_{hu(hc)}&=&\frac{|\xi_{tu(tc)}|^2m_t}{32\pi}\left(1-\frac{m^2_h}{m^2_t}\right)^2\\
\label{htb}
\Gamma_{H^+b}&=&\frac{|\xi_{tb}|^2m_t}{16\pi}\left(1-\frac{m^2_{H^{\pm}}}{m^2_t}\right)^2,
\end{eqnarray}
where $|\xi_{ti}|=\sqrt{|\xi_{1ti}|^2+|\xi_{2ti}|^2}$.

Direct search for $t\rightarrow c(u)h\rightarrow c(u)\gamma\gamma$ decays \cite{tqh} at ATLAS gives
the bound for branching ratios
\begin{equation}
Br(t\rightarrow ch)+Br(t\rightarrow uh)<0.79\%\cdot
\left(\frac{Br(h\rightarrow\gamma\gamma)_{\textrm{SM}}}{Br(h\rightarrow\gamma\gamma)}\right)
\end{equation}
at $95\%$ C.L. which leads to
\begin{equation}
\sqrt{|\xi_{tu}|^2+|\xi_{tc}|^2}<0.16\kappa,\quad\textrm{where}\quad
\kappa=\sqrt{\frac{Br(h\rightarrow\gamma\gamma)_{\textrm{SM}}}{Br(h\rightarrow\gamma\gamma)}}\sim\mathcal{O}(1)
\end{equation}
and $\kappa=1$ in the SM. For most cases it is a stronger constraint on $|\xi_{tu}|$ than that in the same sign
top production process, but they are of the same order. A similar measurement by CMS \cite{tqh2} gives a $95\%$ upper limit
$Br(t\rightarrow ch)<0.56\%$ hence $|\xi_{tc}|<0.14$ with the combination of Higgs decaying to diphoton
or multileptons assuming the SM decay branching ratios of Higgs boson. If we allow different branching
ratios to the SM, the constraints on this coupling is still of that order. Adopting the Cheng-Sher
ansatz \cite{CS}, we have
\begin{equation}
\label{tutc}
\frac{|\xi_{tc}|v}{\sqrt{2m_tm_c}}\lesssim1.5,\quad\quad\textrm{and}\quad\quad
\frac{|\xi_{tu}|v}{\sqrt{2m_tm_u}}\lesssim44
\end{equation}
assuming SM branching ratios of Higgs. For other branching ratio, the constraints are of the same order.

Direct searches for $t\rightarrow bH^+\rightarrow b\tau^+\nu_{\tau}(c\bar{s})$ at ATLAS \cite{tbh+1} for
$90\textrm{GeV}<m_{H^{\pm}}<160(150)\textrm{GeV}$ and at CMS \cite{tbh+2} for
$80\textrm{GeV}<m_{H^{\pm}}<160\textrm{GeV}$ gave the results in \autoref{charhig}.
\begin{table}
\caption{Constraints on the $t\rightarrow bH^+\rightarrow b\tau^+\nu(c\bar{s})$ from direct searches for
light charged Higgs boson (lighter than top quark).}\label{charhig}
\begin{tabular}{|c|c|c|}
\hline
\begin{tabular}{c}Process\\$(H^+\rightarrow f)$\end{tabular} & \begin{tabular}{c}Charged Higgs\\mass (GeV)\end{tabular} &
\begin{tabular}{c}$Br(t\rightarrow bH^+\rightarrow bf)$\\($95\%$C.L.)\end{tabular} \\
\hline
$H^+\rightarrow c\bar{s}$(ATLAS) & $90\sim150$ & $<(1.2\%\sim5.1\%)$ \\
\hline
$H^+\rightarrow \tau^+\nu$(ATLAS) & $90\sim160$ & $<(0.8\%\sim3.4\%)$ \\
\hline
$H^+\rightarrow \tau^+\nu$(CMS) & $80\sim160$ & $<(1.9\%\sim4.1\%)$ \\
\hline
$H^+\rightarrow c\bar{s}$(CMS) & $90-\sim160$ & $<(1.7\%\sim7.0\%)$ \\
\hline
\end{tabular}
\end{table}
These results lead to the upper limits region on $|\xi_{tb}|$ at $95\%$C.L. as
\begin{equation}
|\xi_{tb}|<\left\{\begin{array}{ll}(0.15\sim0.59)/\sqrt{Br(\tau\nu)},&(\textrm{CMS},\quad80\textrm{GeV}<m_{H^{\pm}}<160\textrm{GeV});\\
(0.15\sim1.12)/\sqrt{Br(c\bar{s})},&(\textrm{CMS},\quad90\textrm{GeV}<m_{H^{\pm}}<160\textrm{GeV});\\
(0.13\sim0.45)/\sqrt{Br(\tau\nu)},&(\textrm{ATLAS},\quad90\textrm{GeV}<m_{H^{\pm}}<160\textrm{GeV});\\
(0.19\sim0.27)/\sqrt{Br(c\bar{s})},&(\textrm{ATLAS},\quad90\textrm{GeV}<m_{H^{\pm}}<150\textrm{GeV}).
\end{array}\right.
\end{equation}
For some typical mass of charged Higgs boson (which are allowed for some cases in the S-T ellipse tests) we have the
upper limits of $|\xi_{tb}|$ in \autoref{charhig2}.
\begin{table}
\caption{Constraints on the $tbH^+$ vertex coupling $|\xi_{tb}|$ for some typical mass of the charged Higgs boson.}\label{charhig2}
\begin{tabular}{|c|c|c|c|}
\hline
Mass(GeV) & $100$ & $120$ & $150$ \\
\hline
CMS$(\tau\nu)$ & $0.17/\sqrt{Br(\tau\nu)}$ & $0.20/\sqrt{Br(\tau\nu)}$ & $0.38/\sqrt{Br(\tau\nu)}$ \\
\hline
CMS$(c\bar{s})$ & $0.15/\sqrt{Br(c\bar{s})}$ & $0.16/\sqrt{Br(c\bar{s})}$ & $0.43/\sqrt{Br(c\bar{s})}$ \\
\hline
ATLAS$(\tau\nu)$ & $0.16/\sqrt{Br(\tau\nu)}$ & $0.12/\sqrt{Br(\tau\nu)}$ & $0.25/\sqrt{Br(\tau\nu)}$ \\
\hline
ATLAS$(c\bar{s})$ & $0.17/\sqrt{Br(c\bar{s})}$ & $0.16/\sqrt{Br(c\bar{s})}$ & $0.27/\sqrt{Br(c\bar{s})}$ \\
\hline
\end{tabular}
\end{table}
From all the direct searches for top decays, we must have an relation
\begin{equation}
Br(t\rightarrow hc)+Br(t\rightarrow hu)+Br(t\rightarrow bH^+)<7.4\%
\end{equation}
according to (\ref{tr}) at $95\%$ C.L. as well.

\section{Constraints from Low Energy Phenomena}

The Lee model we discussed in this paper contains additional sources of CP
violation and tree-level FCNC interactions, therefore they will affect many kinds of low energy phenomena,
especially for the CP violation observables and the FCNC processes. For the CP violation observables, we will focus on the constraints from the electric dipole
moments(EDM) of electron and neutron \cite{EDM}. For the constraints on FCNC interactions, we will focus on the mesonic measurements.

\subsection{Constraints due to EDM and Strong CP Phase}

Direct searches of the electric dipole moment (EDM) for electron($d_e$) and neutron($d_n$) are given as \cite{dn}\cite{de}
\begin{equation}
d_e=(-2.1\pm4.5)\times10^{-29}e\cdot\textrm{cm},\quad\quad
d_n=(0.2\pm1.7)\times10^{-26}e\cdot\textrm{cm}
\end{equation}
which will constrain the corresponding CP-violation interactions.

The effective interaction for electron can be writen as \cite{EDM}
\begin{equation}
\mathcal{L}_{e,\textrm{EDM}}=-\frac{id_e}{2}\bar{e}\sigma^{\mu\nu}\gamma^5eF_{\mu\nu}
\end{equation}
where $d_e$ is the EDM for electron.
In our scenario, the dominant
contribution to electron EDM should be due to the two-loop
Barr-Zee type diagrams \cite{bz} \cite{bz2} involving the lightest scalar as follows
\begin{eqnarray}
\frac{d_e}{e}&=&\left(\frac{d_e}{e}\right)_{W^{\pm}}+\left(\frac{d_e}{e}\right)_t+\left(\frac{d_e}{e}\right)_{H^{\pm}}\nonumber\\
&=&\frac{2\sqrt{2}\alpha_{em}G_Fm_e}{(4\pi)^3}\left(-c_V\textrm{Im}(c_e)J_1(m_W,m_h)+\frac{8}{3}\textrm{Re}(c_e)\textrm{Im}(c_t)J_{1/2}(m_t,m_h)\right.\nonumber\\
&&\left.+\frac{8}{3}\textrm{Im}(c_e)\textrm{Re}(c_t)J_{1/2}'(m_t,m_h)-c_{\pm}\textrm{Im}(c_e)J_0(m_{H^{\pm}},m_h)\right)
\end{eqnarray}
in which the loop integration functions $J_1$ comes from the $W$ loop, $J_{1/2}(J_{1/2}')$ comes from the top loop
and $J_0$ comes from the charged scalar loop. The analytical expressions are \cite{bz2}
\begin{eqnarray}
J_1(m_W,m_h)&=&-\frac{m^2_W}{m^2_h}\left(\left(5-\frac{m^2_h}{2m^2_W}\right)I_1(m_W,m_h)\right.\nonumber\\
&&\left.+\left(3+\frac{m^2_h}{2m^2_W}\right)I_2(m_W,m_h)\right);\\
J_{1/2}(m_t,m_h)&=&-\frac{m^2_t}{m^2_h}I_1(m_t,m_h);\\
J_{1/2}'(m_t,m_h)&=&-\frac{m^2_t}{m^2_h}I_2(m_t,m_h);\\
J_0(m_{H^{\pm}},m_h)&=&-\frac{v^2}{2m^2_h}(I_1(m_{H^{\pm}},m_h)-I_2(m_{H^{\pm}},m_h));
\end{eqnarray}
where
\begin{eqnarray}
I_1(m_1,m_2)&=&\int_0^1dz\frac{m^2_2}{m_1^2-m_2^2z(1-z)}\ln\left(\frac{m_2^2z(1-z)}{m_1^2}\right);\nonumber\\
I_2(m_1,m_2)&=&\int_0^1dz\frac{m_2^2(1-2z(1-z))}{m_1^2-m_2^2z(1-z)}\ln\left(\frac{m_2^2z(1-z)}{m_1^2}\right).
\end{eqnarray}
Numerically, the contribution from charged Higgs loop is usually small comparing with the $W$ and top loop, especially
for heavy charged Higgs. As a benchmark point, take $m_{H^{\pm}}=150\textrm{GeV}$, we have
\begin{eqnarray}
d_e&=&[-(14.0c_V+1.28c_{\pm})\textrm{Im}(c_e)+6.53\textrm{Re}(c_t)\textrm{Im}(c_e)\nonumber\\
&&+9.32\textrm{Re}(c_e)\textrm{Im}(c_t)]\times10^{-27}e\cdot\textrm{cm}.
\end{eqnarray}

As benchmark points, take $c_V=c_{\pm}=0.5$, $|c_t|=1$. For both CMS and ATLAS data, small $\alpha(<\pi/2)$ is favored.
Take $\alpha_t'=1.2$ around the best fit point thus $\alpha_t\approx1.0$, the EDM data strongly constrains
the coupling $c_e$.
\begin{figure}
\caption{Constraints on $c_e$ taking $\alpha_t=1.0$.}\label{cmsedm}
\includegraphics[scale=0.7]{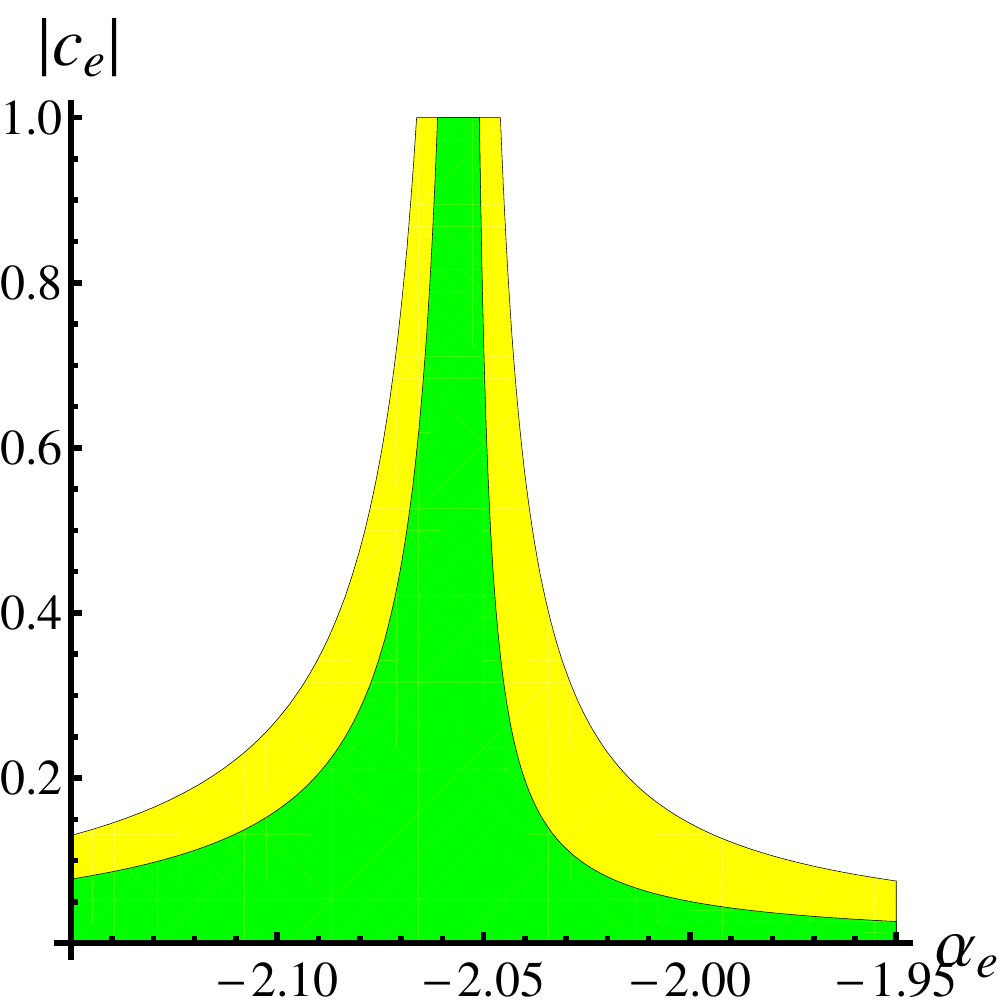}\quad\includegraphics[scale=0.7]{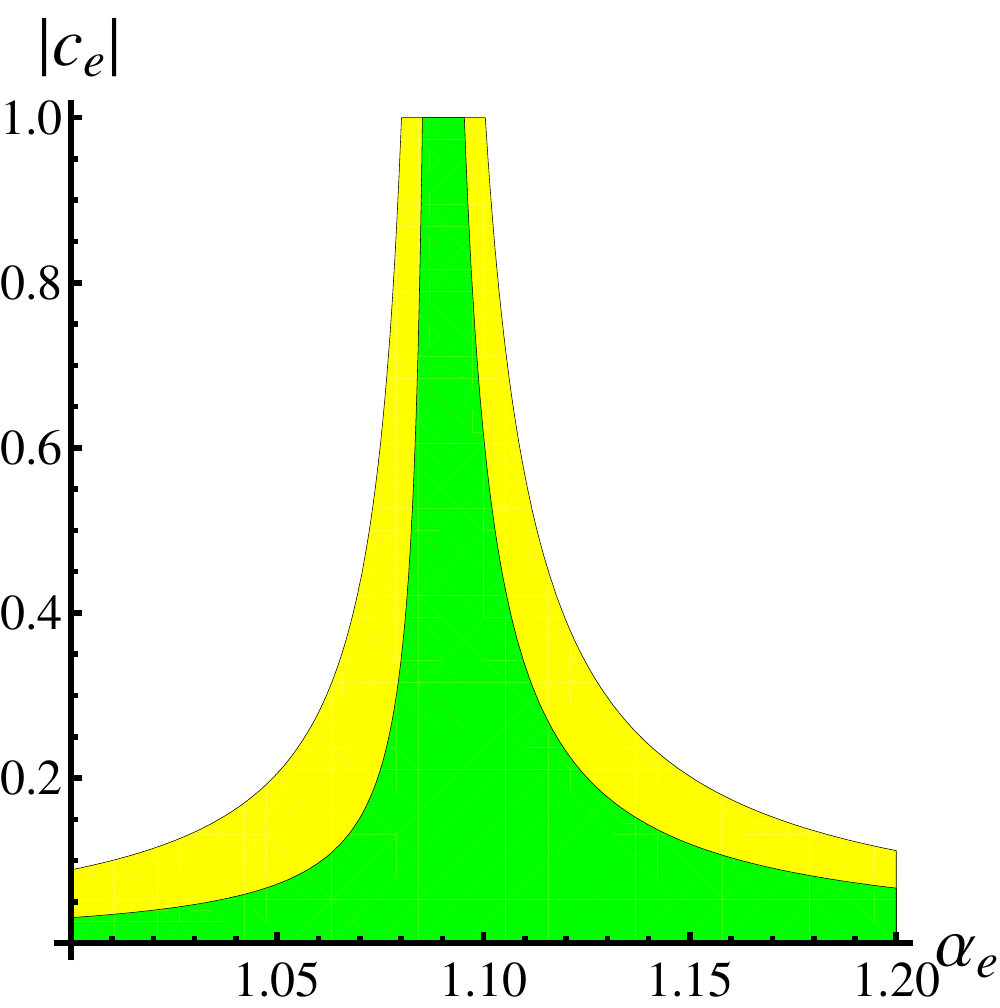}
\end{figure}
For most $\alpha_e\equiv\arg(c_e)$, the coupling strength $|c_e|$ is constrained to be as small as
$\mathcal{O}(10^{-2}-10^{-1})$. But for some special angles, as $\alpha_e\approx-2.04$ and $\alpha_e\approx1.09$,
$|c_e|$ may be as large as $\mathcal{O}(1)$. But the windows are very narrow,
in \autoref{cmsedm} we show the constraints close to the special angles.

\begin{figure}
\caption{An example of modified constraints by heavy neutral scalars.}\label{cmsedm2}
\includegraphics[scale=0.7]{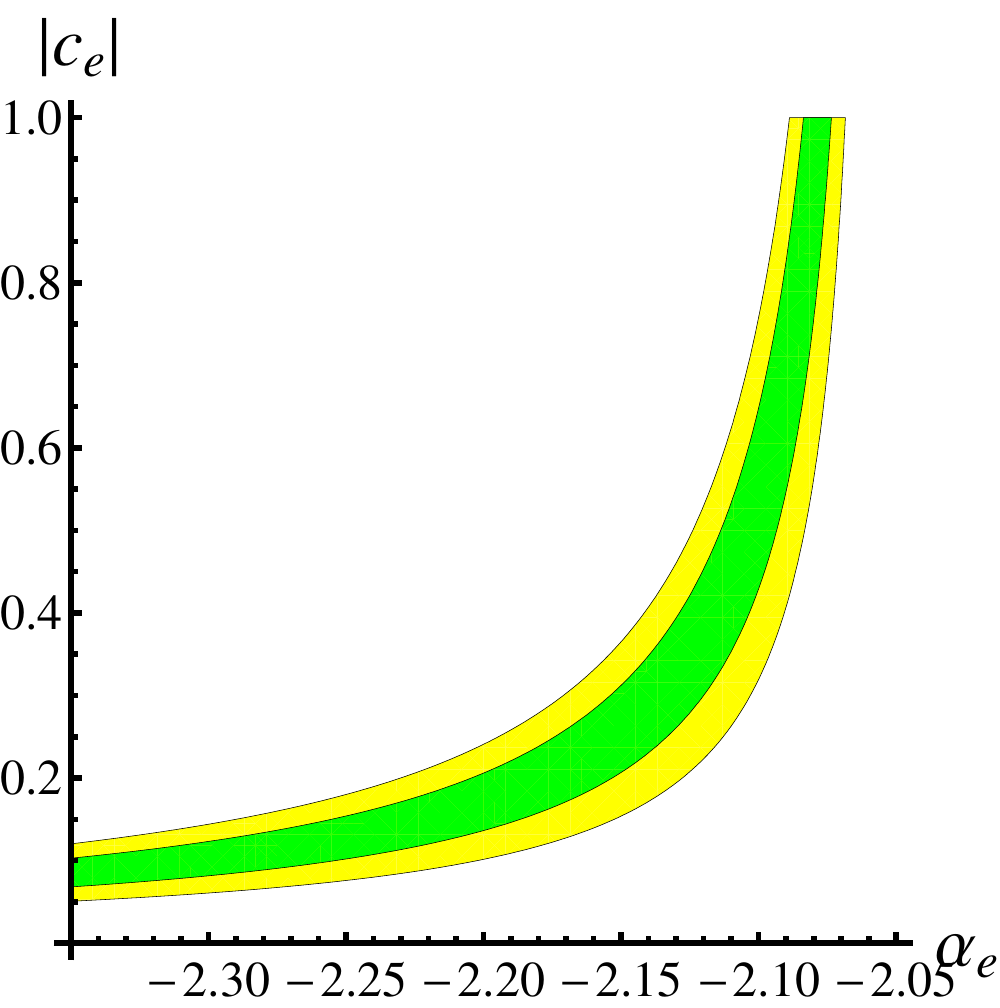}\quad\includegraphics[scale=0.7]{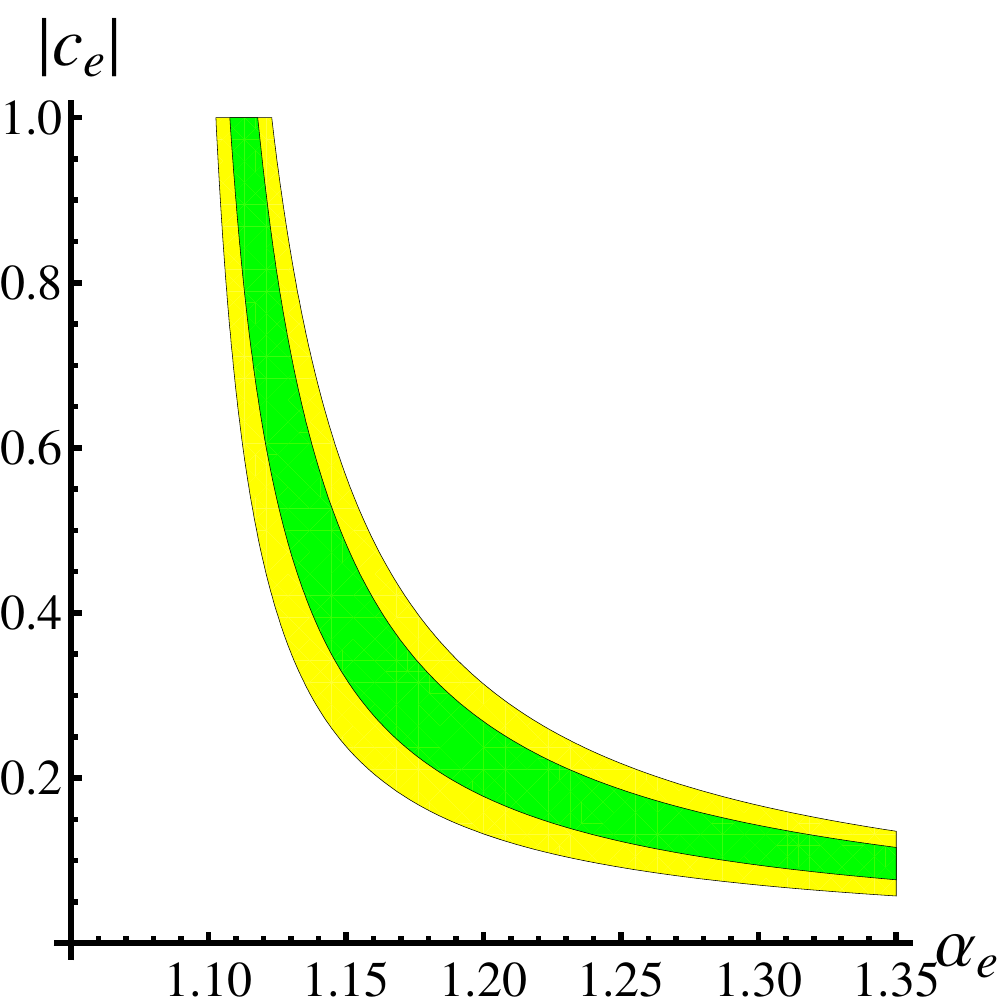}
\end{figure}

If adding the contributions from heavy neutral Higgs, the constraints on $c_e$ would be shifted.
Since both heavy scalars are CP-even dominant, we can estimate that
\begin{equation}
\arg(c_{e,2})\simeq\arg(c_{e,3})\simeq\arg(c_{t,2})\simeq\arg(c_{t,3})\sim\mathcal{O}(0.1)
\end{equation}
and for the two $|c_{e,i}|$, at least one of them is of $\mathcal{O}(1)$ because of its mass; which
is the same for $|c_{t,i}|$. For the couplings to gauge bosons, we can estimate
\begin{equation}
c_2^2+c_3^2=1-c_1^2\simeq0.7
\end{equation}
thus at least one of them must be large enough to be close to $\mathcal{O}(1)$.
For a neutral Higgs with mass $m_2\sim300\textrm{GeV}$ or $m_2\sim700\textrm{GeV}$,
the contributions can be estimated as
\begin{eqnarray}
d_{e,2}&\simeq&(1\sim5)\times10^{-28}e\cdot\textrm{cm};\\
d_{e,3}&\simeq&(0.5\sim3)\times10^{-28}e\cdot\textrm{cm}.
\end{eqnarray}
As an example, if the heavy scalars contribute a $d_e'=2\times10^{-28}e\cdot\textrm{cm}$,
\autoref{cmsedm} would be changed to \autoref{cmsedm2}.
It still imposes strict constraints on $c_e$ but the behaviors are different from that
without including the contributions from the heavy scalars.

For neutron, the effective interaction can be written as \cite{EDM,neu,neu2}
\begin{eqnarray}
\mathcal{L}_{n,\textrm{EDM}}&=&-\frac{i}{2}\mathop{\sum}_{q}(d_q\bar{q}\sigma^{\mu\nu}\gamma^5qF_{\mu\nu}+
\tilde{d}_qg_s\bar{q}\sigma^{\mu\nu}\gamma^5t^aqG^a_{\mu\nu})\nonumber\\
&&-\frac{w}{3}f^{abc}G_{\mu\nu}G_{\sigma}^{\nu,b}\tilde{G}^{\mu\sigma,c}+\frac{\theta\alpha_s}{8\pi}G_{\mu\nu}\tilde{G}^{\mu\nu}.
\end{eqnarray}
The first two operators correspond to the EDM($d_q$) and color EDM($\tilde{d}_q$) of light quarks; the third operator is the
Weinberg operator; and the last operator, in which $\theta=\arg(\det(M_u\cdot M_d))$ is the strong CP phase.
The EDM of neutron \cite{EDM,neu,neu2} is
\begin{eqnarray}
\frac{d_n}{e}&\simeq&1.4\left(\frac{d_d}{e}-0.25\frac{d_u}{e}\right)+1.1\left(\tilde{d}_d+0.5\tilde{d}_u\right)\nonumber\\
&&+(2.5\times10^{-16}\theta+4.3\times10^{-16}w(\textrm{GeV}^{-2}))\textrm{cm}
\end{eqnarray}
at the hadron scale with a theoretical uncertainty of about $50\%$. At weak scale the EDM and CEDM for quarks are given as \cite{neu}\cite{neu2}
\begin{eqnarray}
\frac{d_q}{e}&=&\frac{2\sqrt{2}\alpha_{em}Q_qG_Fm_q}{(4\pi)^3}\bigg(c_V\textrm{Im}(c_q)J_1(m_W,m_h)+
c_{\pm}\textrm{Im}(c_q)J_0(m_{H^{\pm}},m_h)\nonumber\\
&&-\frac{8}{3}\Big(\textrm{Re}(c_q)\textrm{Im}(c_t)J_{1/2}(m_t,m_h)+\textrm{Im}(c_q)\textrm{Re}(c_t)J'_{1/2}(m_t,m_h)\Big)\bigg);\\
\tilde{d}_q&=&-\frac{2\sqrt{2}\alpha_sG_Fm_q}{(4\pi)^3}\Big(\textrm{Re}(c_q)\textrm{Im}(c_t)J_{1/2}(m_t,m_h)\nonumber\\
&&+\textrm{Im}(c_q)\textrm{Re}(c_t)J'_{1/2}(m_t,m_h)\Big);
\end{eqnarray}
and the Weinberg operator
\begin{equation}
w=\frac{\sqrt{2}G_Fg_s\alpha_s}{4\cdot(4\pi)^3}\textrm{Re}(c_t)\textrm{Im}(c_t)g\left(\frac{m^2_t}{m^2_h}\right)
\end{equation}
with
\begin{equation}
g(x)=4x^2\int_0^1dv\int_0^1du\frac{u^3v^3(1-v)}{(xv(1-uv)+(1-u)(1-v))^2}.
\end{equation}

\begin{figure}
\caption{Plots on the allowed $\alpha_d-\alpha_u$, taking $\alpha_t=1.0$.}\label{neuedm1}
\includegraphics[scale=0.7]{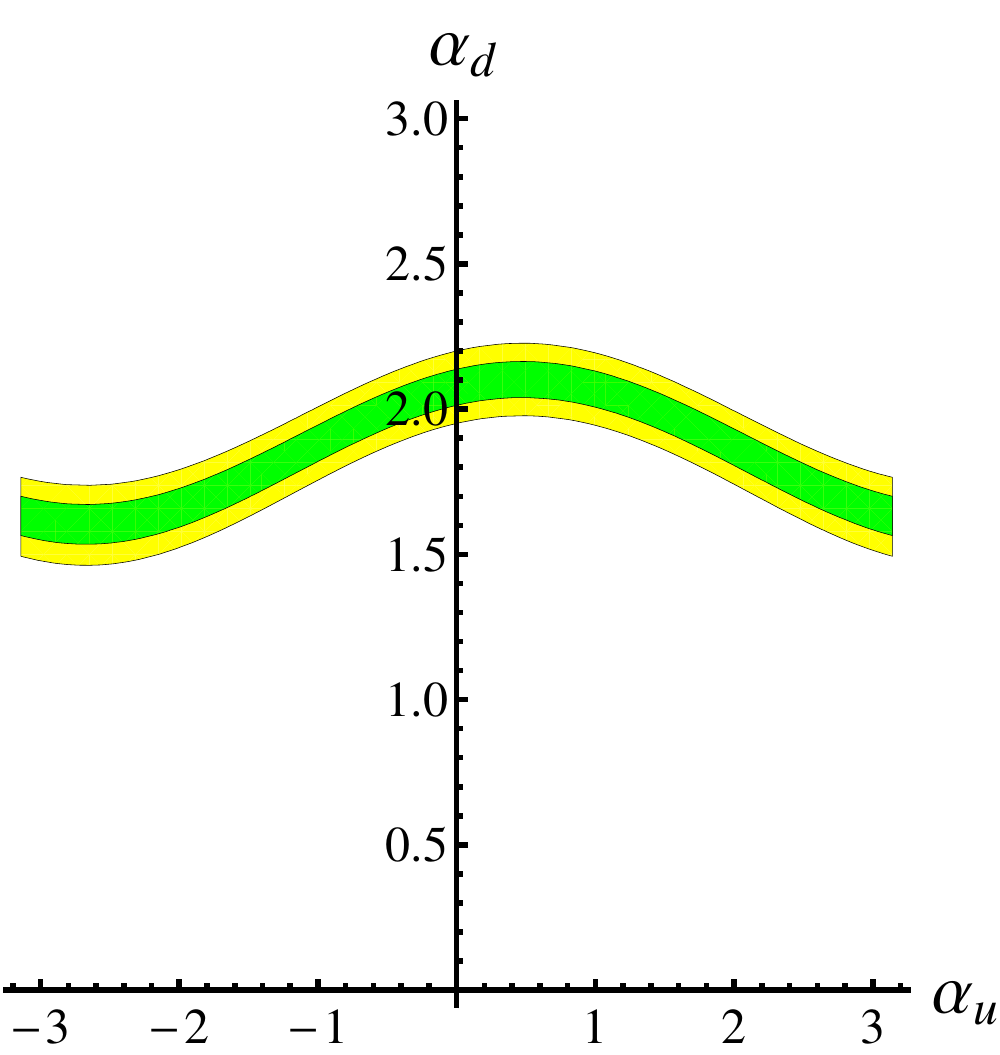}\quad\includegraphics[scale=0.7]{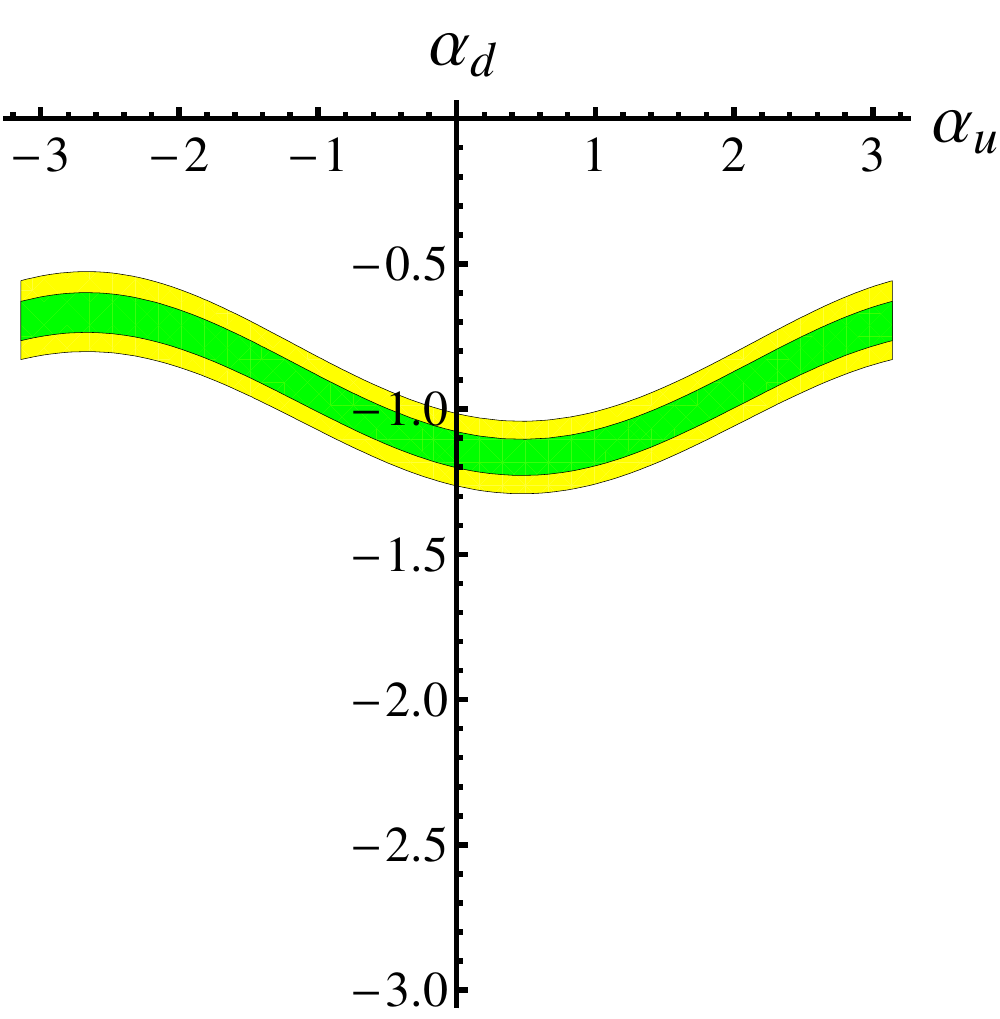}
\end{figure}

Following the appendix in \cite{neu}, with the input $m_u=2.3\textrm{MeV}$, $m_d=4.8\textrm{MeV}$
and $\alpha_s(m_t)=0.11$ \cite{PDG}, numerically the EDM for neutron is
\begin{eqnarray}
d_n&\simeq&(0.5\sim1.5)\times\Big(-(7.0\textrm{Re}(c_u)\textrm{Im}(c_t)+4.9\textrm{Im}(c_u)\textrm{Re}(c_t))\nonumber\\
&&-(29\textrm{Re}(c_d)\textrm{Im}(c_t)+20\textrm{Im}(c_d)\textrm{Re}(c_t))\nonumber\\
&&-(2.8c_V+0.25c_{\pm})\textrm{Im}(c_d)-(0.66c_V+0.06c_{\pm})\textrm{Im}(c_u)\nonumber\\
&&+2.5\times10^{10}\theta+2.3|c_t|^2\sin(2\alpha_t')\Big)\times10^{-26}e\cdot\textrm{cm}
\end{eqnarray}
Take benchmark points as usual, and fix $c_V=c_{\pm}=0.5$ and $|c_t|=1$, $\alpha_t=1.0$ as usual.
For $|c_u|\simeq|c_d|\sim\mathcal{O}(0.1)$, there is almost no constraints on $\alpha_u\equiv\arg(c_u)$
and $\alpha_d\equiv\arg(c_d)$. For $|c_u|\simeq|c_d|\sim\mathcal{O}(1)$, constraints  on $\alpha_d$
and $\alpha_u$ are shown in \autoref{neuedm1}.
Ignoring the $\theta$ term, for $|c_u|=|c_d|=1$, $\alpha_d$ is constrained in two bands with a width of
$\Delta\alpha_d\simeq(0.2\sim1)$ from the uncertainties in calculating $d_n$. And the width are more
sensitive to $c_d$, for example, if $|c_d|=0.5$, $\Delta\alpha_d\simeq(0.5\sim2)$. The constraints by
neutron EDM are less strict comparing with those by electron EDM in this model. Contributions from
heavy neutral Higgs bosons and nonzero $\theta(\lesssim10^{-10})$ would also change the location of the bands.

\subsection{Meson Mixing and CP Violation}
In SM the neutral mesons $K^0, D^0, B^0_d$ and $B^0_s$ mix with their corresponding anti-particles through weak interactions.
Usually BSM will give additional contributions to the mixing matrix elements
$\langle\bar{M}^0|\mathcal{H}_{\Delta F=2}|M^0\rangle$ thus they will modify the mass splitting
and mixing induced CP-violation observables. We can parameterize the new physics effects as \cite{CKMCPV}
\begin{equation}
\label{m12}
M_{12,M}\equiv\frac{1}{2m_M}\langle\bar{M}^0|\mathcal{H}_{\Delta F=2}|M^0\rangle=M_{12,M,\textrm{SM}}(1+\Delta_Me^{i\delta_M}).
\end{equation}
For mass splitting, we list the world averaging results \cite{PDG,hfag,hfag2} and SM
predictions \cite{smmix,smmix2,smmix3} for $\Delta m$ in \autoref{mixing} in \autoref{meson}.
The useful decay constants and bag parameters are from the lattice results \cite{Lat}. Only for the
$D^0-\bar{D}^0$ system it is difficult to predict $\Delta m_D$ since the long-distance effects are the
dominant contributions. Nonzero $\delta_M$ from new physics will modify the CP
violated effects from those in the SM, thus it will be constrained by CP-violated observable, as $\epsilon_K$ in
$K^0-\bar{K}^0$ mixing and $\sin(2\beta_{d(s)})$ in $B^0_{d(s)}-\bar{B}^0_{d(s)}$ mixing et. al.
They are defined as
\begin{equation}
\epsilon_K=\frac{1}{3}\left(\frac{\mathcal{M}(K_L\rightarrow2\pi^0)}{\mathcal{M}(K_S\rightarrow2\pi^0)}\right)+
\frac{2}{3}\left(\frac{\mathcal{M}(K_L\rightarrow\pi^+\pi^-)}{\mathcal{M}(K_S\rightarrow\pi^+\pi^-)}\right)
\end{equation}
where $K_{L(S)}$ is the long(short) lived neutral kaon and $\mathcal{M}$ is the amplitude for the process and
\begin{equation}
\beta=\arg\left(-\frac{V_{tb}V^*_{td}}{V_{cb}V^*_{cd}}\right),\quad\quad
\beta_s=\arg\left(\frac{V_{tb}V_{ts}^*}{V_{cb}V_{cs}^*}\right)
\end{equation}
where $V_{ij}$ are CKM matrix elements.

First assuming the charged Higgs is heavy and considering the contribution
only from the 126GeV Higgs Boson, we can write the flavor-changing effective interaction as
\begin{equation}
\mathcal{L}_{ij}=\bar{f}_i(\xi_{1ij}+\xi_{2ij}\gamma^5)f_jh+\textrm{h.c.}
\end{equation}
The lightest neutral Higgs boson contribution to matrix elements for meson mixing is \cite{Wells}\cite{Ope}
\begin{equation}
\label{m12*}
M_{12,M,\textrm{SM}}\Delta_Me^{i\delta_M}=\frac{f_M^2B_Mm_M}{6m^2_h}
\left(\xi^2_{1ij}-\xi^2_{2ij}+\frac{(\xi^2_{1ij}-11\xi^2_{2ij})m^2_M}{(m_i+m_j)^2}\right).
\end{equation}
The parameters $f_M,B_M$ and $m_M$ are
the decay constant, bag parameter and mass for meson $M^0$, and $m_{i(j)}$ are masses for the quark $f_{i(j)}$. For $B^0_{d(s)}-\bar{B}^0_{d(s)}$ mixing, according to fitting
results \cite{smcpv} (see the plots in \cite{future} for details), for different $\delta_{B_d(B_s)}$,
\begin{equation}
\Delta_{B_d}\lesssim(0.1\sim0.4)\quad\quad\textrm{and}\quad\quad\Delta_{B_s}\lesssim(0.1\sim0.3).
\end{equation}
For $\delta_{B_d(B_s)}=0$, the upper limit on $\Delta_{B_d(B_s)}$ is about $0.2$.
Comparing with (\ref{m12}), (\ref{m12*}) and adopting the Cheng-Sher ansatz \cite{CS}, the typical upper
limit on $\xi_{bs(bd)}$ have the order
\begin{equation}
\label{bsbd}
\frac{|\xi_{bs}|v}{\sqrt{2m_bm_s}}\lesssim2\times10^{-2}\quad\quad\textrm{and}\quad\quad
\frac{|\xi_{bd}|v}{\sqrt{2m_bm_d}}\lesssim6\times10^{-2}
\end{equation}
both of $\mathcal{O}(10^{-2}\sim10^{-1})$. For $D^0-\bar{D}^0$ mixing, we have the upper limit
\begin{equation}
\frac{|\xi_{cu}|v}{\sqrt{2m_cm_u}}\lesssim0.1
\end{equation}
For $K^0-\bar{K}^0$ mixing, when $\delta_K\approx0$ or $\pi$, we have
$\Delta_K\lesssim0.25$ which leads to
\begin{equation}
\frac{|\xi_{sd}|v}{\sqrt{2m_sm_d}}\lesssim2\times10^{-2}.
\end{equation}
While for a general $\delta_K$, $\Delta_K$ is strongly constrained to be less than $\mathcal{O}(10^{-3})$ because of
the smallness of $\epsilon_K$. New CP-violation effects must be very small in neutral $K$ system while they are allowed
or even favored \cite{smcpv} for other meson.

Next, consider the contribution to $B^0_{d(s)}-\bar{B}^0_{d(s)}$ mixing from charged Higgs boson. Box diagrams with
one or two charged Higgs boson instead of $W$ boson will contribute to $\Delta_{B_d(B_s)}\exp(i\delta_{B_d(B_s)})$
as \cite{chabox}\cite{chabox2}
\begin{equation}
\Delta_{B_d(B_s)}e^{i\delta_{B_d(B_s)}}=\frac{\mathcal{F}_1(x_{tW},x_{tH},x_{HW})+\mathcal{F}_2(x_{tH})}{\mathcal{F}_0(x_{tW})}
\end{equation}
where
\begin{eqnarray}
\mathcal{F}_0(x_{tW})&=&1+\frac{9}{1-x_{tW}}-\frac{6}{(1-x_{tW})^2}-\frac{6x_{tW}^2\ln x_{tW}}{(1-x_{tW})^3}\\
\mathcal{F}_1(x_{tW},x_{tH},x_{HW})&=&\eta_{d(s)}^2\frac{x_{tH}}{1-x_{HW}}\left(\frac{8-2x_{tW}}{1-x_{tH}}+\right.\nonumber\\
&&\left.\frac{(2x_{HW}-8)\ln x_{tH}}{(1-x_{tH})^2}+\frac{6x_{HW}\ln x_{tW}}{(1-x_{tW})^2}\right)\\
\label{F2}
\mathcal{F}_2(x_{tH})&=&\eta_{d(s)}^4x_{tH}\frac{1-x_{tH}^2+2x_{tH}\ln x_{tH}}{(1-x_{tH})^3}
\end{eqnarray}
at leading order in which $\eta_{d(s)}\approx(\xi_{1tb}\xi_{1td(s)}/2V_{tb}V_{ts(d)}^*)^{1/2}v/m_t$ and $x_{ij}=(m_i/m_j)^2$.
We can parameterize the interactions (\ref{htb}) as
\begin{equation}
\mathcal{L}_{I,tD_iH^+}=-\frac{V_{tD_i}}{v}\bar{t}(X_tm_tP_L+X_{D_i}m_{D_i}P_R)D_i+\textrm{h.c.}
\end{equation}
in which $P_{L(R)}=(1\mp\gamma^5)/2$. Thus $\eta_{d(s)}\approx |X_t|v/(\sqrt{2}m_t)$ and it is not sensitive to $X_{D_i}$
if they are of the same order as $X_t$.
According to the constraints in \autoref{TRD} for light charged Higgs $m_{H^{\pm}}<m_t$,
with a typical coupling $|X_t|\lesssim0.5$,
$\Delta_{B_d(B_s)}\lesssim0.2$ holds for $m_{H^{\pm}}\geq100\textrm{GeV}$ and additional CP-violation effects
induced by charged Higgs mediated loop are negligible. Thus
take a benchmark point $m_{H^{\pm}}=150\textrm{GeV}$ as usual, it is allowed by B meson mixing data. While for
heavy charged Higgs $m_{H^{\pm}}>m_t$, the coupling $X_t$ is not constrained by $t\rightarrow bH^+$ decay
process. We can give an upper limit $|X_t|\lesssim(0.6\sim1)$ when $200\textrm{GeV}<m_{H^{\pm}}<600\textrm{GeV}$.

In the
$D^0-\bar{D}^0$ mixing, another useful constraint comes from the neutral Higgs mediated box diagram. Its
contribution to $\Delta m_D$ is \cite{Dmix}
\begin{equation}
\Delta m_D^*=\frac{G^2_Fv^4|\xi_{tu}\xi_{tc}|^2}{12\pi^2m^2_t}f^2_Dm_DB_Br\mathcal{F}_2(x_{th})\approx4\times10^{-9}|\xi_{tu}\xi_{tc}|^2
\end{equation}
where $r=(\alpha_s(m_t)/\alpha_s(m_b))^{6/23}(\alpha_s(m_b)/\alpha_s(m_c))^{6/25}\approx0.8$ and loop
function $\mathcal{F}_2$ is the same as that in (\ref{F2}). For $\Delta m_D^*$ contributes less than
the order of measured $\Delta m_D$, we have $|\xi_{tu}\xi_{tc}|\lesssim1.5\times10^{-3}$ and hence we
can put a stronger constraint than (\ref{tutc}) on the flavor changing interactions including top as
\begin{equation}
\frac{|\xi_{tu}\xi_{tc}|v^2}{2m_t\sqrt{m_um_c}}\lesssim5
\end{equation}
which is of $\mathcal{O}(1)$.
\subsection{The B Leptonic Decays}
The rare decay process $B_{s,d}\rightarrow\mu^+\mu^-$ has been measured by
LHCb \cite{ralhcb} and CMS \cite{racms} Collaborations respectively with the results
\begin{equation}
\overline{Br}(B_s\rightarrow\mu^+\mu^-)=\left\{\begin{array}{ll}2.9^{+1.1}_{-1.0}\times10^{-9},&\quad(\textrm{LHCb},\quad4.0\sigma\textrm{significance}),\\
3.0^{+1.0}_{-0.9}\times10^{-9},&\quad(\textrm{CMS},\quad4.3\sigma\textrm{significance})\end{array}\right.;
\end{equation}
and
\begin{equation}
\overline{Br}(B_d\rightarrow\mu^+\mu^-)=\left\{\begin{array}{ll}3.7^{+2.5}_{-2.1}\times10^{-10},&\quad(\textrm{LHCb}),\\
3.5^{+2.5}_{-1.8}\times10^{-10},&\quad(\textrm{CMS})\end{array}\right..
\end{equation}
A combination result is $\overline{Br}(B_s\rightarrow\mu^+\mu^-)=(2.9\pm0.7)\times10^{-9}$ by CMS and LHCb Collaborations\cite{comb}.
There is no evidence for the process $B_d\rightarrow\mu^+\mu^-$.
The results correspond to the SM prediction \cite{rapre} (and updated results \cite{rapre2} in 2014)
\begin{eqnarray}
\overline{Br}(B_s\rightarrow\mu^+\mu^-)_{\textrm{SM}}&=&(3.65\pm0.23)\times10^{-9},\\
\overline{Br}(B_d\rightarrow\mu^+\mu^-)_{\textrm{SM}}&=&(1.06\pm0.09)\times10^{-10}.
\end{eqnarray}
Where the modified branching ratio $\overline{Br}$ means the averaged time-integrated branching ratio
and it has the relation with the branching ratio $Br$ as \cite{modif1}\cite{modif2}
\begin{equation}
Br(B_s\rightarrow\mu^+\mu^-)=\overline{Br}(B_s\rightarrow\mu^+\mu^-)\left(1+\mathcal{O}\left(\frac{\Delta \Gamma}{\Gamma}\right)\right).
\end{equation}
See \autoref{meson} for details.

Consider the neutral Higgs mediated flavor changing process first. Using the constraints in (\ref{bsbd}), we
can estimate the contributions to $Br(B_{s(d)}\rightarrow\mu^+\mu^-)$ as
\begin{eqnarray}
\delta Br(B_s\rightarrow\mu^+\mu^-)=\frac{m_{B_s}|c_{\mu}|^2}{8\pi\Gamma_{B_s,\textrm{tot}}}
\left(\frac{f_{B_s}m^2_{B_s}m_{\mu}|\xi_{bs}|}{(m_b+m_s)vm^2_h}\right)^2\lesssim4\times10^{-12}|c_{\mu}|^2;\\
\delta Br(B_d\rightarrow\mu^+\mu^-)=\frac{m_{B_d}|c_{\mu}|^2}{8\pi\Gamma_{B_d,\textrm{tot}}}
\left(\frac{f_{B_d}m^2_{B_d}m_{\mu}|\xi_{bd}|}{(m_b+m_d)vm^2_h}\right)^2\lesssim1\times10^{-12}|c_{\mu}|^2.
\end{eqnarray}
We cannot get stronger constraints through these processes on $|c_{\mu}|$ than direct search\cite{mu}
which gives $|c_{\mu}|\lesssim7$.

Next consider the charged Higgs contribution. For $|X_t|\sim|X_{b,s,\mu}|\sim\mathcal{O}(1)$, the charged Higgs
loop is sensitive to $X_t$ and $m_{H^{\pm}}$ only \cite{ali}. According to \cite{ali}, it is estimated that
\begin{equation}
\frac{\delta Br(B_s\rightarrow\mu^+\mu^-)}{Br(B_s\rightarrow\mu^+\mu^-)}\approx\left(1-\frac{|X_t^2|}{\eta}
\frac{Y_{\textrm{2HDM}}}{Y_{\textrm{SM}}}\right)^2
\end{equation}
where $\eta=0.987$ is the electro-weak and QCD correction factor and
\begin{eqnarray}
Y_{\textrm{SM}}&=&\frac{x_{tW}}{8}\left(\frac{x_{tW}-4}{x_{tW}-1}+\frac{3x_{tW}}{(x_{tW}-1)^2}\ln x_{tW}\right);\\
Y_{\textrm{2HDM}}&=&\frac{x_{tW}^2}{8}\left(\frac{1}{x_{HW}-x_{tW}}+\frac{x_{HW}}{(x_{HW}-x_{tW})^2}\ln\left(\frac{x_{tW}}{x_{HW}}\right)\right).
\end{eqnarray}
If the charged Higgs is light $(m_{H^{\pm}})<m_t$,
$|X_t|=0.5$ is allowed at $95\%$C.L. While for a heavy charged Higgs, when $200\textrm{GeV}<m_{H^{\pm}}<600\textrm{GeV}$,
we have the $95\%$C.L. upper limit on $|X_t|$ as $|X_t|\lesssim(0.6\sim1.1)$ with the combined experimental results or
$|X_t|\lesssim(0.8\sim1.4)$ with single experimental result.
\subsection{The B Radiative Decays}
The inclusive radiative decays branching ratio of $\bar{B}$ meson $\bar{B}\rightarrow X_s\gamma$ (or we say $b\rightarrow s\gamma$
at parton level) has the averaged value \cite{hfag}
\begin{equation}
Br(\bar{B}\rightarrow X_s\gamma)=(3.43\pm0.22)\times10^{-4}
\end{equation}
with the photon energy $E_{\gamma}>1.6\textrm{GeV}$. The SM prediction for that value is $(3.15\pm0.23)\times10^{-4}$
to $\mathcal{O}(\alpha_s^2)$ \cite{chm2}\cite{rad}. In a 2HDM, the dominant contribution to modify this decay rate is
from a loop containing a charged Higgs instead of the $W$ boson in SM. The neutral Higgs loop contribution is negligible
because of the suppression in $\xi_{bs}$ and $m_{b(s)}/v$.

\begin{figure}
\caption{Plots on allowed $\alpha_{bt}-|X_b|$. For the left figure, $|X_t|=0.5$ and $m_{H^{\pm}}=150\textrm{GeV}$; for the right
figure,  $|X_t|=0.8$ and $m_{H^{\pm}}=500\textrm{GeV}$.}\label{bsga}
\includegraphics[scale=0.7]{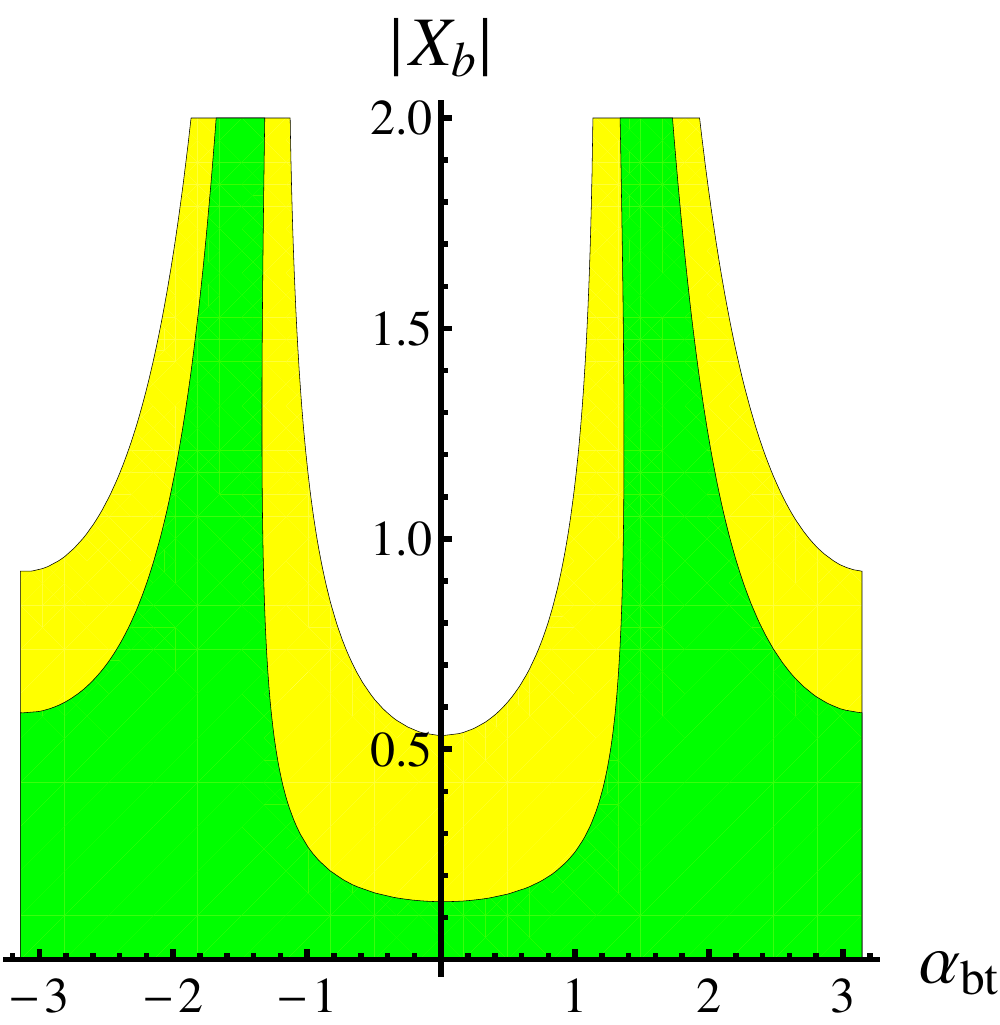}\quad\includegraphics[scale=0.7]{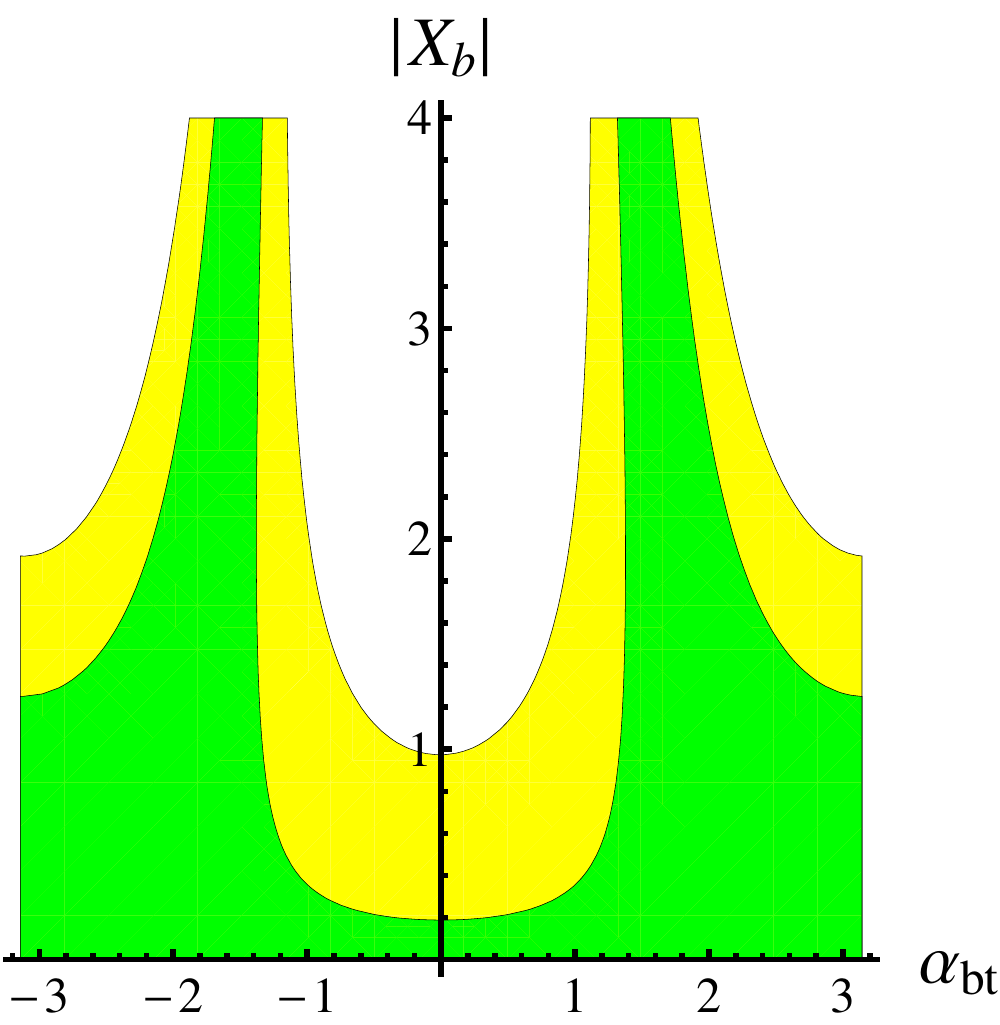}
\end{figure}

The charged Higgs loop is sensitive to both $X_t$ and $X_b$ that we should take some benchmark points. Define
$\alpha_{bt}\equiv\arg(X_b/X_t)$, for a light charged Higgs boson, take $|X_t|=0.5$ and $m_{H^{\pm}}=150\textrm{GeV}$
as before; while for a heavy charged Higgs boson, take $|X_t|=0.8$ and $m_{H^{\pm}}=500\textrm{GeV}$. We show the allowed
region for $\alpha_{bt}-|X_b|$  in \autoref{bsga} utilizing the calculations in \cite{chm2}\cite{rad2}.
From the figures, we can see that for most $\alpha_{bt}$ the coupling $|X_b|$ is constrained to be $\lesssim\mathcal{O}(1)$;
while for some angles it can be larger\footnote{That's because with merely the decay rate, we can only determine the absolute
value for the $b\rightarrow s\gamma$ amplitude. The largest allowed $X_b$ can reach $14$ for the left figure and $28$ for the
right figure, in which case the new physics contributes twice as large as the SM but with the opposite sign.}.

\section{Features of Lee model and its future perspectives }
\label{disco}
One of the main goals of this paper is the though phenomenological studies on the Lee model with spontaneous CP-violation \cite{Lee}. We can see from last two sections that Lee model is still viable confronting
the high and low energy experiments. The next natural question is how to confirm/exclude this model at future facilities.

In the scalar sector there are nine
free parameters $\mu_1^2,\mu_2^2$, and $\lambda_{1,2,\ldots,7}$, corresponding to nine observables:
\begin{itemize}
\item Four masses $m_h,m_2,m_3$ and $m_{H^{\pm}}$;
\item VEVs and a physical phase  $v_1,v_2,\xi$ (or equivalently $v,\tan\beta,\xi$);
\item Two neutral scalar mixing angles, equivalently we choose the ratios $c_1$ and $c_2$ of the couplings to gauge boson compared to
the corresponding ones in the SM.
\end{itemize}

We treat the discovered scalar with mass 126GeV as the lightest neutral Higgs boson. If Lee model is true, the extra neutral and charged Higgs bosons should be discovered at high energy colliders.
As the general rules, the lighter the extra Higgs bosons, the easier they can be produced.
In order to confirm the Lee model, another possible signal can be the FCNC decay of the neural Higgs bosons which are unobservable small in the SM.
Furthermore the CP properties of the Higgs boson are essential measurements, though it is a very challenging task.

As we have pointed out that there is no SM limit in this scenario, thus it is always testable at the future colliders,
such as LHC with $\sqrt{s}=14\textrm{TeV}$, CEPC,
ILC, or TLEP with $\sqrt{s}=(240\sim250)\textrm{TeV}$, even before the discovery of other neutral Higgs bosons and charged Higgs
boson.
The coupling between the lightest Higgs boson and
other particle(especially for massive gauge bosons $W^{\pm}$ and $Z^0$)
are usually suppressed by the factor of $\mathcal{O}(t_{\beta}s_{\xi})$.
In the $b\bar{b}$ decay channel or any VBF, V+H production
channel, a significant suppression can be the first sign of this scenario. On the contrary if the signals become even more SM-like,
this scenario will be disfavored.

For future LHC with $\sqrt{s}=14\textrm{TeV}$, the signal strengths will be measured with an uncertainty of about
$10\%$ at the luminosity $300\textrm{fb}^{-1}$ \cite{cms14}\cite{atl14}. Perform the same $\chi^2$ fit as in (\ref{chi2}),
and add the $b\bar{b}$ decay mode in. The value of $\chi^2$ is sensitive to $c_V$ and $c_b$, and the magnitude of $c_V$
is a criterion for this model. A Higgs boson with $c_V\gtrsim(0.6\sim0.7)$ is hardly to be pseudoscalar dominant thus if
$c_V\lesssim(0.6\sim0.7)$ is excluded, we can say this scenario is excluded. So we can test this scenario by fitting the
signal strengths. We list the estimating results in \autoref{excl}.
\begin{table}[h]
\caption{Abilities to test the scenario  at $\sqrt{s}=14\textrm{TeV}$ LHC. Lower limit for the allowed $c_V$ at
$2\sigma$ and $3\sigma$ level are listed in the tables. For the left/right tables we assume all signal strengths are consist
with SM at $1\sigma$/$2\sigma$ level respectively.}\label{excl}
\begin{tabular}{|c|c|c|}
\hline
Excluded level&$2\sigma$&$3\sigma$\\
\hline
$300\textrm{fb}^{-1}$&$0.62$&$0.55$\\
\hline
$3000\textrm{fb}^{-1}$&$0.77$&$0.72$\\
\hline
\end{tabular}
\quad\quad
\begin{tabular}{|c|c|c|}
\hline
Excluded level&$2\sigma$&$3\sigma$\\
\hline
$300\textrm{fb}^{-1}$&$0.53$&$0.45$\\
\hline
$3000\textrm{fb}^{-1}$&$0.7$&$0.65$\\
\hline
\end{tabular}
\end{table}

If all signal strengths and the overall $\chi^2$ are consist with SM at $1\sigma$ level, For the integrated
luminosity $300\textrm{fb}^{-1}$, all $c_V\lesssim0.62$ can be excluded at $95\%$C.L.($2\sigma$) while all
$c_V\lesssim0.55$ can be excluded at $99.7\%$C.L.($3\sigma$); For the integrated
luminosity $3000\textrm{fb}^{-1}$, all $c_V\lesssim0.77$ can be excluded at $95\%$C.L.($2\sigma$) while all
$c_V\lesssim0.72$ can be excluded at $99.7\%$C.L.($3\sigma$).
If all signal strengths and the overall $\chi^2$ are consistent with SM at $2\sigma$ level, For the integrated
luminosity $300\textrm{fb}^{-1}$, all $c_V\lesssim0.53$ can be excluded at $95\%$C.L.($2\sigma$) while all
$c_V\lesssim0.45$ can be excluded at $99.7\%$C.L.($3\sigma$); For the integrated
luminosity $3000\textrm{fb}^{-1}$, all $c_V\lesssim0.7$ can be excluded at $95\%$C.L.($2\sigma$) while all
$c_V\lesssim0.65$ can be excluded at $99.7\%$C.L.($3\sigma$). All the results are for the largest parameter
space in this scenario because the true ability to test this scenario by $\chi^2$ depends strongly on the
real signal strengths from future experiments.

Another useful observable is $f_{a3}$ defined in (\ref{fa3}). For $\sqrt{s}=14\textrm{TeV}$, the $95\%$C.L.
upper limit on $f_{a3}$ will reach about $0.14(0.04)$ for the luminosity $300(3000)\textrm{fb}^{-1}$\cite{cms14}
\cite{atl14}\footnote{Almost the same for CMS and ATLAS detector, with $300(3000)\textrm{fb}^{-1}$ luminosity,
the upper limit can reach $0.15(0.037)$ for ATLAS and $0.13(0.04)$ for CMS, see details in the references.} which
leads to the constrains $|a_3/a_1|<1.0(0.5)\sim\mathcal{O}(1)$ separately. For $|c_t|\sim\mathcal{O}(1)$, it is
still too large to give direct constrains on $\alpha_t\equiv\arg(c_t)$.

At a Higgs factory with the $e^+e^-$ initial state at $\sqrt{s}=(240\sim250)\textrm{GeV}$, the dominant production process
for a Higgs boson is associated with a $Z^0$ boson. Another important production process is through VBF. In this scenario
it is suppressed by a factor $c_1^2$ thus we can exclude this scenario if the total cross section favors SM. For the total
cross section, a measurement with $\mathcal{O}(10\%)$ uncertainty is accurate enough to distinguish the scenario we discussed
in this paper and SM at $3\sigma$ or even $5\sigma$ significance.
Such accuracy can be achieved at CEPC/ILC/TLEP.
 At $\sqrt{s}=240\textrm{GeV}$ TLEP, the total cross
section can be measured with an uncertainty $0.4\%$ for the integrated
luminosity $500\textrm{fb}^{-1}$ \cite{TLEP}\cite{TLEP2}, while that value is about $3\%$ for the integrated
luminosity $250\textrm{fb}^{-1}$ ILC at $\sqrt{s}=500\textrm{GeV}$ \cite{CLIC}.

\section{Conclusions and discussions}

In this paper we proposed a scenario in which the smallness of CP-violation
and the lightness of Higgs boson are correlated through small $t_{\beta}s_{\xi}$,
based on the Lee model, namely the 2HDM with spontaneous CP-violation.
The basic assumption is that CP, which spontaneously broken by the complex vacuum,
is an approximate symmetry. We found that $m_h$ as well as the quantities $K$
and $J$ are $\propto t_{\beta}s_{\xi}$
in the limit $t_{\beta}s_{\xi}\rightarrow0$. Here $K$
and $J$ are the measures for CP-violation effects in scalar and Yukawa sectors respectively.
It is a new way to
understand why the Higgs boson discovered at LHC is light. In this scenario, all the three neutral physical degrees of freedom
mix with each other thus none of them is a CP eigenstate.

We then investigated the phenomenological constraints from both high energy and low energy experiments and found the scenario still alive.
The lightest Higgs boson usually couples with SM gauge and fermion particles with a smaller strength than in the SM, thus the total width must be narrower
than that in SM. Such choice of the parameters makes Lee model still allowed by the
CMS or ATLAS data. The LHC search for heavy neutral bosons implies the masses of other two neutral bosons
should be away from the region $300-700\textrm{GeV}$. The S-T ellipse also strictly constrains the mass relation between the charged
and neutral bosons as can be seen in \autoref{I}-\autoref{III}. We also fitted the CMS and ATLAS data respectively, for example, see
\autoref{cmsct1}-\autoref{atlasct2}. We found that this scenario is still allowed for either data. It does not sensitive to
the charged Higgs contribution. After considering all the data, a light charged Higgs with the mass about $100\textrm{GeV}$ is still
allowed. Small $h\bar{b}b$ vertex is favored for both CMS and ATLAS data. The minimal $\chi^2$ is close to
the $\chi^2$ in SM, thus we cannot conclude that SM is better than Lee model.

We forbid the explicit CP-violation in the whole lagrangian including the Yukawa sector, thus we must tolerate the tree-level FCNC.
The flavor-changed couplings including top quark are constrained by same sign top production process and
the top quark rare decay, besides the constraints by B physics processes. The tree-level
FCNC vertices including five light quarks are strongly constrained to be less than $\mathcal{O}(10^{-2}\sim10^{-1})\sqrt{2m_im_j}/v$
while for the vertices including top quark it should be less than $\mathcal{O}(1)\sqrt{2m_tm_q}/v$. The coupling $X_t$ for $tbH^+$ vertex
are constrained to be less than $\mathcal{O}(0.1\sim1)$ for different $m_{H^{\pm}}$, while $X_b\sim\mathcal{O}(1)$ are usually
allowed by $b\rightarrow s\gamma$ data.

The constraints by EDMs are usually very important in discussing a model with CP-violation, because new sources of CP-violation
may modify the theoretical prediction of EDMs from the SM by several orders of magnitude, and maybe testable by the experiments now.
The EDM for electron gave very strict constraints on the $h\bar{e}e$ vertex as shown in \autoref{cmsedm}-\autoref{cmsedm2}. While
the EDM for neutron gave weaker constraints on $h\bar{d}d$ and $h\bar{u}u$ vertices, see \autoref{neuedm1}.

There is no SM limit for the lightest Higgs boson in this scenario, thus it is testable at future colliders.
At $\sqrt{s}=14\textrm{TeV}$, besides discovering the extra neutral and charged Higgs bosons,
the ability to
test this scenario depends on how far the signal strengths for the 126 GeV Higgs boson differ from the SM predictions, as listed in \autoref{excl}. From the discovery point of view, if any
suppression in the VBF, VH production channel or $b\bar{b}$ decay channel are confirmed, this scenario would be favored. On the contrary, if all signal
are SM-like more and more at future colliders, this scenario would be disfavored by data. For most cases
$300\textrm{fb}^{-1}$ luminosity is not enough to exclude this scenario, while $3000\textrm{fb}^{-1}$ luminosity is
better. At $\sqrt{s}=(240\sim250)\textrm{GeV}$ $e^+e^-$ colliders, several $\textrm{fb}^{-1}$ luminosity is enough to
distinguish this scenario and SM at $(3\sim5)\sigma$ level by accurately measuring the total cross section. We emphasize that measuring
the CP properties and the flavor-changing decay of the Higgs bosons are essential to pin down Lee model.

We did not build the model for flavor sector in details thus we did not solve the natural FCNC and strong CP problems.
It is possible to solve the FCNC and strong CP problems together, for example, see the model proposed
by Liao \cite{Liao}. We also did not discuss the constraints from flavor changing processes in lepton sector.
As a model with CP-violation, there may also
some new CP-violation effects, especially in top, $\tau$ and neutral D sector where no CP-violation has been discovered.
We did not study the cosmological effects in this paper, like the domain wall and electro-weak baryogenesis in this model.
All these consequences will be further scrutinized in the future.

\section*{Acknowledgement}
We thank J.-J. Cao, Q.-H. Cao, S.-L. Chen, L. Dai, W. Liao, and C. Zhang et. al. for helpful discussions.
This work was supported in part by the Natural Science Foundation of China (Nos. 11135003 and 11375014).

\numberwithin{equation}{section}
\appendix
\section{Vacuum Stability Conditions}
\label{vs}
For the potential (\ref{pot}), when $|\phi_i|\equiv\sqrt{\phi_i^{\dag}\phi_i}\rightarrow\infty$, $V\geq0$ must
hold to keep the vacuum stable. Write $|\phi_1|=r_1, |\phi_2|=r_2$ and $\phi_1^{\dag}\phi_2=r\exp(i\alpha)$, we have
\begin{equation}
|\phi_1^{\dag}\phi_2|=r\leq|\phi_1|\cdot|\phi_2|=r_1r_2.
\end{equation}
The $R_{ij}$ and $I_{ij}$ can be expressed as
\begin{equation}
R_{11}=r_1^2,\quad R_{22}=r^2_2,\quad R_{12}=r\cos\alpha,\quad I_{12}=r\sin\alpha.
\end{equation}
Thus we have the equation
\begin{eqnarray}
V&=&\lambda_1r_1^4+\lambda_3r_1^2r_2^2+\lambda_6r^4_2\nonumber\\
&&+rc_{\alpha}(\lambda_2r_1^2+\lambda_5r^2_2)+r^2(\lambda_4c^2_{\alpha}+\lambda_7s^2_{\alpha})
\end{eqnarray}
holds for any $\alpha$ and $0\leq r\leq r_1r_2$.

Another type of conditions is that the potential should be minimized when
\begin{equation}
\langle\phi_1\rangle=\frac{1}{\sqrt{2}}\left(\begin{array}{c}0\\v_1\end{array}\right),\quad\quad
\langle\phi_2\rangle=\frac{1}{\sqrt{2}}\left(\begin{array}{c}0\\v_2e^{i\xi}\end{array}\right).
\end{equation}
It equivalents to the conditions that the mass matrix for neutral Higgs, $\tilde{m}$ in (\ref{higmass}),
must be positive definite. We write the conditions as
\begin{equation}
\textrm{tr} \tilde{m}>0,\quad(\textrm{tr} \tilde{m})^2-\textrm{tr} (\tilde{m})^2>0,\quad\det\tilde{m}>0.
\end{equation}
If there exist more than one local minimal points for the potential, the physical vacuum should
be chosen at the global minimum if we want to forbid a meta-stable vacuum.
\section{Scalar Spectra and Small $t_{\beta}s_{\xi}$ Expansion}
\label{mass}
In the unitary gauge the mass square matrix for charged scalars reads
\begin{equation}
\label{mcha}M^2_{\pm}=-\frac{\lambda_7}{2}\left(\begin{array}{cc}v_2^2&-v_1v_2e^{-i\xi}\\-v_1v_2e^{i\xi}&v_1^2\end{array}\right)
\end{equation}
The eigenvalues are
\begin{equation}
m^2_{G^{\pm}}=0;\quad\quad\quad m^2_{H^{\pm}}=-\frac{\lambda_7v^2}{2};
\end{equation}
where the zero eigenvalue corresponds to the charged goldstones which will be eaten by the longitudinal
part of W bosons. Diagonalize (\ref{mcha}) by performing a rotation
\begin{equation}
\label{rot}\left(\begin{array}{c}G^+\\H^+\end{array}\right)=\left(\begin{array}{cc}\cos\beta&e^{-i\xi}\sin\beta\\
-e^{i\xi}\sin\beta&\cos\beta\end{array}\right)\left(\begin{array}{c}\phi_1^+\\ \phi_2^+\end{array}\right)
\end{equation}
For the neutral parts, in the basis $(I_1,I_2,R_1,R_2)^T$, for any angle $\alpha$ to appear below,
we have the mass square matrix $M^2_0=(v^2/2)m_{ij}$, where
\begin{eqnarray*}
m_{11}&=&(\lambda_4-\lambda_7)s^2_{\beta}s^2_{\xi};\\
m_{12}&=&\lambda_5s^2_{\beta}s^2_{\xi};\\
m_{13}&=&(\lambda_2c_{\beta}+(\lambda_4-\lambda_7)s_{\beta}c_{\xi})s_{\beta}s_{\xi};\\
m_{14}&=&((\lambda_4-\lambda_7)c_{\beta}+\lambda_5s_{\beta}c_{\xi})s_{\beta}s_{\xi};\\
m_{22}&=&4\lambda_6s^2_{\beta}s^2_{\xi};\\
m_{23}&=&2((\lambda_3+\lambda_7)c_{\beta}+\lambda_5s_{\beta}c_{\xi})s_{\beta}s_{\xi};\\
m_{24}&=&(\lambda_5c_{\beta}+4\lambda_6s_{\beta}c_{\xi})s_{\beta}s_{\xi};\\
m_{33}&=&4\lambda_1c^2_{\beta}+2\lambda_2c_{\beta}s_{\beta}c_{\xi}+(\lambda_4-\lambda_7)c^2_{\xi}s^2_{\beta};\\
m_{34}&=&\lambda_2c^2_{\beta}+(2\lambda_3+\lambda_4+\lambda_7)s_{\beta}c_{\beta}c_{\xi}+\lambda_5s^2_{\beta}c^2_{\xi};\\
m_{44}&=&(\lambda_4-\lambda_7)c^2_{\beta}+2\lambda_5c_{\beta}s_{\beta}c_{\xi}+4\lambda_6s^2_{\beta}c^2_{\xi}.
\end{eqnarray*}
Perform the same rotation as (\ref{rot}) between $\phi_1$ and $\phi_2$, which in the basis above can be written as
\begin{equation}
R=R_1R_2=\left(\begin{array}{cccc}c_{\beta}&s_{\beta}&0&0\\-s_{\beta}&c_{\beta}&0&0\\0&0&1&0\\0&0&0&1\end{array}\right)
\left(\begin{array}{cccc}1&0&0&0\\0&c_{\xi}&0&-s_{\xi}\\0&0&1&0\\0&s_{\xi}&0&c_{\xi}\end{array}\right)
\end{equation}
we have
\begin{equation}
\widetilde{M}^2_0=RM^2_0R^{-1}=\frac{v^2}{2}\left(\begin{array}{cc}0&\\&\left(\tilde{m}\right)_{3\times3}\end{array}\right)
\end{equation}
where the zero eigenvalue corresponds to the neutral Goldstone,
\begin{equation}
G^0=c_{\beta}I_1+s_{\beta}c_{\xi}I_2-s_{\beta}s_{\xi}R_2,
\end{equation}
which will be eaten by the longitudinal part of Z boson. The matrix elements for $\tilde{m}$ in the basis
$(-s_{\beta}I_1+c_{\beta}c_{\xi}I_2-c_{\beta}s_{\xi}R_2,R_1,s_{\xi}I_2+c_{\xi}R_2)^T$ should be
\begin{eqnarray}
\tilde{m}_{11}&=&(\lambda_4-\lambda_7)s^2_{\xi};\nonumber\\
\tilde{m}_{12}&=&-(\lambda_2c_{\beta}+(\lambda_4-\lambda_7)s_{\beta}c_{\xi})s_{\xi};\nonumber\\
\tilde{m}_{13}&=&-(\lambda_5s_{\beta}+(\lambda_4-\lambda_7)c_{\beta}c_{\xi})s_{\xi};\nonumber\\
\tilde{m}_{22}&=&4\lambda_1c^2_{\beta}+2\lambda_2c_{\beta}s_{\beta}c_{\xi}+(\lambda_4-\lambda_7)s^2_{\beta}c^2_{\xi};\nonumber\\
\tilde{m}_{23}&=&\lambda_2c^2_{\beta}c_{\xi}+(2(\lambda_3+\lambda_7)+(\lambda_4-\lambda_7)c^2_{\xi})s_{\beta}c_{\beta}+\lambda_5s^2_{\beta}c_{\xi};\nonumber\\
\tilde{m}_{33}&=&(\lambda_4-\lambda_7)c^2_{\beta}c^2_{\xi}+2\lambda_5s_{\beta}c_{\beta}c_{\xi}+4\lambda_6s^2_{\beta}.
\end{eqnarray}

We can expand $\tilde{m}$ in powers of $t_{\beta}s_{\xi}$ as follows,
\begin{equation}
\tilde{m}=\tilde{m}_0+(t_{\beta}s_{\xi})\tilde{m}_1+(t_{\beta}s_{\xi})^2\tilde{m}_2+\cdots
\end{equation}
In the basis $(-s_{\beta}I_1+c_{\beta}c_{\xi}I_2-c_{\beta}s_{\xi}R_2,R_1,s_{\xi}I_2+c_{\xi}R_2)^T$ the matrix $\tilde{m}_0$
can be written as
\begin{equation}
\tilde{m}_0=\left(\begin{array}{ccc}(\lambda_4-\lambda_7)s^2_{\xi}&-\lambda_2s_{\xi}&-(\lambda_4-\lambda_7)s_{\xi}c_{\xi}\\
-\lambda_2s_{\xi}&4\lambda_1&\lambda_2c_{\xi}\\
-(\lambda_4-\lambda_7)s_{\xi}c_{\xi}&\lambda_2c_{\xi}&(\lambda_4-\lambda_7)c^2_{\xi}\end{array}\right)
\end{equation}
Diagonalize it with a $3\times3$ matrix
\begin{equation}
r=r_1r_2=\left(\begin{array}{ccc}1&0&0\\0&c_{\theta}&s_{\theta}\\0&-s_{\theta}&c_{\theta}\end{array}\right)
\left(\begin{array}{ccc}c_{\xi}&0&s_{\xi}\\0&1&0\\-s_{\xi}&0&c_{\xi}\end{array}\right)
\end{equation}
we have
\begin{equation}
r\tilde{m}_0r^{-1}=\left(\begin{array}{ccc}0&&\\
&(\tilde{m}_0)_{22}&\\
&&(\tilde{m}_0)_{33}\end{array}\right),
\end{equation}
in which
\begin{eqnarray}
(\tilde{m}_0)_{22(33)}&=&\frac{4\lambda_1+\lambda_4-\lambda_7}{2}\pm
\left(\frac{4\lambda_1-(\lambda_4-\lambda_7)}{2}c_{2\theta}+\lambda_2s_{2\theta}\right);\\
\theta&=&\frac{1}{2}\arctan\left(\frac{2\lambda_2}{4\lambda_1-(\lambda_4-\lambda_7)}\right).
\end{eqnarray}
The two heavy scalars have their masses
\begin{equation}
m^2_{2(3)}=\frac{v^2}{2}((\tilde{m}_0)_{22(33)}+\mathcal{O}(t_{\beta}s_{\xi})).
\end{equation}
The new basis is then
\begin{equation}
r\left(\begin{array}{c}c_{\xi}I_2-s_{\xi}R_2\\R_1\\s_{\xi}I_2+c_{\xi}R_2\end{array}\right)=
\left(\begin{array}{c}I_2\\c_{\theta}R_1+s_{\theta}R_2\\-s_{\theta}R_1+c_{\theta}R_2\end{array}\right)
\end{equation}
in which the useful matrix elements for $\tilde{m}$ are
\begin{eqnarray}
(\tilde{m}_1)_{11}&=&0;\\
(\tilde{m}_1)_{12}&=&(2(\lambda_3+\lambda_7)c_{\theta}+\lambda_5s_{\theta});\\
(\tilde{m}_1)_{13}&=&(\lambda_5c_{\theta}-2(\lambda_3+\lambda_7)s_{\theta});\\
(\tilde{m}_2)_{11}&=&4\lambda_6.
\end{eqnarray}
Thus to the leading order of $t_{\beta}s_{\xi}$, for the lightest scalar $h$ we have
\begin{eqnarray}
m^2_{h}&=&\frac{v^2t^2_{\beta}s^2_{\xi}}{2}\left((\tilde{m}_2)_{11}-\frac{(\tilde{m}_1)^2_{12}}{(\tilde{m}_0)_{22}}
-\frac{(\tilde{m}_1)^2_{13}}{(\tilde{m}_0)_{33}}\right)\nonumber\\
&=&\frac{v^2t^2_{\beta}s^2_{\xi}}{2}\bigg[4\lambda_6+2\lambda_5(\lambda_3+\lambda_7)s_{2\theta}
\left(\frac{1}{(\tilde{m}_0)_{22}}-\frac{1}{(\tilde{m}_0)_{33}}\right)\nonumber\\
&&-4(\lambda_3+\lambda_7)^2\left(\frac{c^2_{\theta}}{(\tilde{m}_0)_{22}}+
\frac{s^2_{\theta}}{(\tilde{m}_0)_{33}}\right)-\lambda^2_5\bigg(\frac{s^2_{\theta}}{(\tilde{m}_0)_{22}}
+\frac{c^2_{\theta}}{(\tilde{m}_0)_{33}}\bigg)\bigg];\\
h&=&I_2-t_{\beta}s_{\xi}\left(\frac{(\tilde{m}_1)_{12}}{(\tilde{m}_0)_{22}}(c_{\theta}R_1+s_{\theta}R_2)
+\frac{(\tilde{m}_1)_{13}}{(\tilde{m}_0)_{33}}(c_{\theta}R_2-s_{\theta}R_1)+\frac{I_1}{t_{\xi}}\right)\nonumber\\
&=&I_2-t_{\beta}s_{\xi}\bigg[\bigg(2(\lambda_3+\lambda_7)\bigg(\frac{c^2_{\theta}}{(\tilde{m}_0)_{22}}+
\frac{s^2_{\theta}}{(\tilde{m}_0)_{33}}\bigg)+\frac{\lambda_5s_{2\theta}}{2}\bigg(\frac{1}{(\tilde{m}_0)_{22}}
-\frac{1}{(\tilde{m}_0)_{33}}\bigg)\bigg)R_1\nonumber\\
&&+\bigg((\lambda_3+\lambda_7)s_{2\theta}\bigg(\frac{1}{(\tilde{m}_0)_{22}}
-\frac{1}{(\tilde{m}_0)_{33}}\bigg)+\lambda_5\bigg(\frac{s^2_{\theta}}{(\tilde{m}_0)_{22}}+
\frac{c^2_{\theta}}{(\tilde{m}_0)_{33}}\bigg)\bigg)R_2+\frac{I_1}{t_{\xi}}\bigg].
\end{eqnarray}
\section{Some Useful Feynman-Rules in this Model}
\label{Fr}
From the lagrangian we have some useful coupling vertexes directly,
\begin{eqnarray}
\label{c1}
\mathcal{L}_{hVV}&=&\left(\frac{2m^2_W}{v}W^+_{\mu}W^{\mu-}+\frac{m^2_Z}{v}Z_{\mu}Z^{\mu}\right)\nonumber\\
&&(c_{\beta}R_1+s_{\beta}c_{\xi}R_2+s_{\beta}s_{\xi}I_2);\\
\mathcal{L}_{hH^+H^-}&=&-vH^+H^-\bigg[\frac{\lambda_2+\lambda_5}{2}s_{\beta}s_{\xi}I_1\nonumber\\
&&+\bigg(\lambda_3c^2_{\beta}+\left(-\lambda_2+\frac{\lambda_5}{2}\right)c^2_{\beta}s_{\beta}c_{\xi}\nonumber\\
&&+(2\lambda_1-\lambda_4c^2_{\xi}-\lambda_7s^2_{\xi})c_{\beta}s^2_{\beta}+\frac{\lambda_2}{2}s^3_{\beta}c_{\xi}\bigg)R_1\nonumber\\
&&+((2\lambda_6-\lambda_7)c^2_{\beta}-\lambda_5c_{\beta}s_{\beta}c_{\xi}+\lambda_3s^2_{\beta})I_2\nonumber\\
&&+\bigg(\frac{\lambda_5}{2}c^3_{\beta}-(\lambda_4-2\lambda_6)c^2_{\beta}s_{\beta}c_{\xi}\nonumber\\
&&+\left(\frac{\lambda_2}{2}-\lambda_5c^2_{\xi}\right)c_{\beta}s^2_{\beta}+\lambda_3s^3_{\beta}c_{\xi}\bigg)R_2\bigg];\\
\mathcal{L}_{hDD}&=&-\frac{1}{\sqrt{2}}\bar{D}_{Li}(Y'_{1d}(R_1+iI_1)+Y'_{2d}(R_2+iI_2))_{ij}D_{Rj}+\textrm{h.c.};\\
\label{c4}
\mathcal{L}_{hUU}&=&-\frac{1}{\sqrt{2}}\bar{U}_{Li}(Y'_{1u}(R_1-iI_1)+Y'_{2u}(R_2-iI_2))_{ij}U_{Rj}+\textrm{h.c.};\\
\mathcal{L}_{Ch}&=&-\frac{1}{\sqrt{2}}\bar{U}_{Li}(V_{CKM})_{ij}(-Y'_{1d}s_{\beta}e^{-i\xi}+Y'_{2d}c_{\beta})_{jk}D_{Rk}H^+\nonumber\\
&&-\frac{1}{\sqrt{2}}\bar{D}_{Li}(V^{\dag}_{CKM})_{ij}(Y'_{1u}s_{\beta}e^{i\xi}-Y'_{2u}c_{\beta})_{ji}U_{Rk}H^-+\textrm{h.c.}
\end{eqnarray}
The $Y'$ in Yukawa couplings means the couplings in the mass eigenstates. For neutral Higgs triple vertex, the Feynman rules are all from
\begin{equation}
-i\lambda_{ijk}=-\frac{i\partial^3V}{\partial h_i\partial h_j \partial h_k}
\end{equation}
\section{Formalism for Neutral Meson}
\label{meson}
The for meson $K^0,D^0,B^0_d$ and $B^0_s$ can mix with their charged conjugate particles, through weak
interaction in SM. We begin with the Schr$\ddot{o}$dinger equation
\begin{equation}
i\frac{\partial}{\partial t}\left(\begin{array}{c}|M_0\rangle\\|\bar{M}_0\rangle\end{array}\right)=
\left(\mathbf{m}-\frac{i}{2}\mathbf{\Gamma}\right)
\left(\begin{array}{c}|M_0\rangle\\|\bar{M}_0\rangle\end{array}\right)
\end{equation}
where $\mathbf{m}$ and $\mathbf{\Gamma}$ are $2\times2$ matrix. Write the hamiltonian as
\begin{equation}
\mathcal{H}=\mathcal{H}_0+\mathcal{H}_{\Delta F=1}+\mathcal{H}_{\Delta F=2},
\end{equation}
we have the matrix elements
\begin{eqnarray}
\label{mael}
\left(\mathbf{m}-\frac{i}{2}\mathbf{\Gamma}\right)_{ij}&=&m_M\delta_{ij}+\frac{1}{2m_M}\langle\psi_i|\mathcal{H}_{\Delta F=2}|\psi_j\rangle\nonumber\\
&&+\frac{1}{2m_M}\int d\Pi_f \frac{\langle\psi_i|\mathcal{H}_{\Delta F=1}|f\rangle\langle f|\mathcal{H}_{\Delta F=1}|\psi_j\rangle}{m_M-E(f)+i\epsilon}
\end{eqnarray}
with the normalized condition $\langle\psi_i|\psi_j\rangle=2m_M\delta_{ij}$ where $\psi_{i,j}=|M^0\rangle$ or $\bar{M}^0$.
The second and third terms come from from short-distance and long-distance effects separately and according to (\ref{mael})
\begin{equation}
\mathbf{\Gamma}_{ij}=\frac{1}{2m_M}\int d\Pi_f\langle\psi_i|\mathcal{H}_{\Delta F=1}|f\rangle\langle f|\mathcal{H}_{\Delta F=1}|\psi_j\rangle2\pi\delta(E(f)-m_M)
\end{equation}
The solutions for the eigenvalues are
\begin{eqnarray}
m_{H(L)}&=&m_M\pm\textrm{Re}\left(\sqrt{\left(\mathbf{m}_{12}-\frac{i}{2}\mathbf{\Gamma}_{12}\right)
\left(\mathbf{m}_{12}^*-\frac{i}{2}\mathbf{\Gamma}_{12}^*\right)}\right);\\
\Gamma_{H(L)}&=&\Gamma\mp\textrm{Im}\left(\sqrt{\left(\mathbf{m}_{12}-\frac{i}{2}\mathbf{\Gamma}_{12}\right)
\left(\mathbf{m}_{12}^*-\frac{i}{2}\mathbf{\Gamma}_{12}^*\right)}\right).
\end{eqnarray}
The H(L) means the heavy(light) mass eigenstate
\begin{equation}
|M_{H(L)}\rangle=p|M^0\rangle\mp q|\bar{M}^0\rangle
\end{equation}
where
\begin{equation}
|p|^2+|q|^2=1;\quad\quad\textrm{and}\quad\quad\left(\frac{p}{q}\right)^2=\frac{\mathbf{m}_{12}-i\mathbf{\Gamma}_{12}/2}
{\mathbf{m}_{12}^*-i\mathbf{\Gamma}_{12}^*/2}.
\end{equation}
The time-dependent solution
\begin{equation}
\left(\begin{array}{c}|M_0(t)\rangle\\|\bar{M}_0(t)\rangle\end{array}\right)=\left(\begin{array}{cc}g_+(t)&-(q/p)g_(t)\\
g_+(t)&-(p/q)g_(t)\end{array}\right)\left(\begin{array}{c}|M_0(0)\rangle\\|\bar{M}_0(0)\rangle\end{array}\right)
\end{equation}
where
\begin{equation}
g_{\pm}(t)=\frac{1}{2}\left(e^{-im_Ht-\frac{\Gamma_H}{2}t}\pm e^{-im_Lt-\frac{\Gamma_L}{2}t}\right).
\end{equation}

For $\mathbf{\Gamma}_{12}\sim\mathbf{m}_{12}$ and $\mathbf{m}_{12}$ is almost real like $K^0$ system,
$\Delta m\approx2\textrm{Re}\mathbf{m}_{12}$; while for $\mathbf{\Gamma}_{12}\ll\mathbf{m}_{12}$ like
$B^0_{d(s)}$ system, $\Delta m\approx2|\mathbf{m}_{12}|$. All the measurements and SM predictions are
listed here. It is difficult to estimate the long-distance effects which give the dominant contribution
in $D^0$ system.
\begin{table}
\caption{SM predictions and Experimental values for mass difference in meson mixing.}\label{mixing}
\begin{tabular}{|c|c|c|}
\hline
Meson&$\Delta m_{\textrm{exp}}$(GeV)&$\Delta m_{\textrm{SM}}$(GeV)\\
\hline
$K^0(d\bar{s})$&$(3.474\pm0.006)\times10^{-15}$&$(3.30\pm0.34)\times10^{-15}$\\
\hline
$D^0(c\bar{u})$&$(1.0\pm0.3)\times10^{-14}$&$-$\\
\hline
$B^0_d(d\bar{b})$&$(3.33\pm0.03)\times10^{-13}$&$(3.3\pm0.4)\times10^{-13}$\\
\hline
$B^0_s(s\bar{b})$&$(1.1663\pm0.0015)\times10^{-11}$&$(1.14\pm0.17)\times10^{-11}$\\
\hline
\end{tabular}
\end{table}

For decay processes to CP eigenstate $f$, for example, $B^0\rightarrow\mu^+\mu^-$, the
direct observable is time integrated averaged branching ratio which has an relation
\begin{equation}
\overline{Br}(M\rightarrow f)\equiv\frac{1}{2}\mathop{\int}_0^{\infty}(\Gamma(M(t)\rightarrow f)+\Gamma(\bar{M}(t)\rightarrow f))
\end{equation}
which leads to
\begin{equation}
\overline{Br}(M\rightarrow f)=\frac{1+A\Delta\Gamma/\Gamma}{1-(\Delta\Gamma/\Gamma)^2}Br(M\rightarrow f)
\end{equation}
where $-1\leq A\leq1$ and in SM $A=1$.

\clearpage\end{CJK*}

\end{document}